\documentclass[11pt]{article}
\usepackage{jheppub}
\usepackage{nicefrac}
\usepackage{slashed} 
\usepackage[titletoc, title]{appendix}

\graphicspath{{./Artwork/}}

\hypersetup{pdftitle={Yang-Mills amplitude relations at loop level from non-adjacent BCFW shifts}} 
\hypersetup{pdfauthor={Rutger H. Boels and Reinke Sven Isermann}} 
\hypersetup{pdfsubject={Amplitudes}} 
\hypersetup{pdfkeywords={Amplitudes}} 

\def\braket#1{\mathinner{\langle{#1}\rangle}}
\newcommand{\sbraket}[1]{\lbrack #1\rbrack}

\newcommand{\ii}{\mathrm{i}}
\newcommand{\Tr}{\textrm{Tr}}

\newcommand{\da}{\dot{\alpha}}

\newtheorem{suspicion}{Suspicion}[section]

\newcommand{\boxit}[1]{%
  \[\fbox{%
      \addtolength{\linewidth}{-2\fboxsep}%
      \addtolength{\linewidth}{-2\fboxrule}%
      \begin{minipage}{\linewidth}%
      #1%
      \end{minipage}%
    } \nonumber \]%
}
\title{Yang-Mills amplitude relations at loop level from non-adjacent BCFW shifts}
\author{Rutger H. Boels}\emailAdd{Rutger.Boels@desy.de}
\author{and Reinke Sven Isermann}\emailAdd{Reinke.Sven.Isermann@desy.de}
\affiliation{II. Institut f\"ur Theoretische Physik, Universit\"at Hamburg\\ Luruper Chaussee 149, D- 22761 Hamburg, Germany }

\keywords{Amplitudes}

\abstract{This article studies methods to obtain relations for scattering amplitudes at the loop level, with concrete examples at one loop. These methods originate in the analysis of large so-called Britto-Cachazo-Feng-Witten shifts of tree level amplitudes and loop level integrands. In particular BCFW shifts for particles which are not color adjacent and some particular generalizations of this situation are analyzed in some detail in four and higher dimensions. For generic non-adjacent shifts our results are independent of loop order for integrands and hold for generic minimally coupled gauge theories with possible scalar potential and Yukawa terms. By a standard argument this result indicates a generalization of the Bern-Carrasco-Johansson relations for tree level amplitudes exists to the integrand at all loop levels. A concrete relation is presented at one loop. Furthermore, inspired by results in QED it is shown that the results on generalized BCFW shifts of tree level amplitudes imply relations for the so-called rational, bubble and triangle terms of one loop amplitudes in pure Yang-Mills theory. Bubble and triangle terms for instance are shown to obey a five photon decoupling identity, while a three photon decoupling identity is demonstrated for the rational terms. Along the same lines recently conjectured relations for helicity equal amplitudes at one loop are shown to generalize to helicity independent relations for the massive box coefficient of the rational terms.}

\begin{document}
\maketitle

\section{Introduction}
Most of our knowledge of the standard model of particle physics comes from collider experiments such as those at the Large Hadron Collider (LHC). A crucial issue at hadron colliders is that the main research interest is in the electroweak sector of the standard model (which contains the Higgs particle for instance), while the scattered particles primarily interact through the strong nuclear force, Quantum ChromoDynamics (QCD). As the words already suggest, the strong force dominates over the weak force. Hence strong quantitative control over the strong sector backgrounds is of fundamental importance for the ability of the LHC to distinguish new from known physics. QCD is an example of a Yang-Mills theory coupled to massive quarks. As a step towards experiment this is the main phenomenological motivation to study scattering amplitudes in Yang-Mills theories.

There is also a theoretical motivation to study scattering amplitudes: there are many cases known in which the outcome of a calculation displays unexpected simplicity. Whenever this happens a symmetry not manifest in the calculation is expected to be at work. The benchmark result is the expression for the tree level color ordered MHV amplitude by Parke and Taylor \cite{Parke:1986gb} in Yang-Mills theory, which manages to express an all multiplicity result in one line for a particular choice of external helicities. The Parke-Taylor result is instrumental in the many recent developments in scattering amplitude technology triggered by Witten's twistor string \cite{Witten:2003nn} insights into the underlying symmetries of this result. 

An example of such a development  where the two different motivations cross is the question  how much work is involved in calculating cross sections from tree level Yang-Mills amplitudes. Naively, one would expect in the amplitudes all $n!$ different color structures for  the $n$ gluons in the amplitude. The color information can however be factorized from the amplitudes at tree level as  \cite{Berends:1987cv} \cite{Mangano:1987xk}
\begin{equation}
A^{\textrm{full}}_{0} =  g^{n-2} \sum_{P_n/\mathbb{Z}_n} \Tr\left(T^{a_{\sigma(1)}} \ldots T^{a_{\sigma(n)}}\right) A_0\left(\sigma(1)\ldots \sigma(n) \right)
\end{equation}
where the sum ranges over all non-cyclic permutations of the external legs and the $T^a$ are
matrices in the fundamental representation of the gauge group. The component amplitudes on the right hand side are known as
color-ordered amplitudes. This reduces the complexity of the calculation from $n!$ to $(n-1)!/2$. The extra factor of a half comes from inversion symmetry of the color
trace which leads to the identity
\begin{equation}\label{eq:reflprop}
A_0(1,\ldots, n) = (-1)^n A_0(n,\ldots, 1) 
\end{equation}
Interestingly, this organization of the amplitude is natural in string theory  \cite{Mangano:1987xk} where the color traces are known as Chan-Paton factors \cite{Paton:1969je}. Further relations for tree level amplitudes were formulated by Kleiss and Kuijf \cite{Kleiss:1988ne} which reduce the number $(n-1)!/2$ further down to $(n-2)!$. Only very recently more relations at tree level were found by Bern, Carrasco and Johansson (BCJ) \cite{Bern:2008qj} and subsequently proven in string theory \cite{BjerrumBohr:2009rd} \cite{Stieberger:2009hq} as well as in field theory \cite{Feng:2010my}. The BCJ relations reduce the number of independent tree level amplitudes down to $(n-3)!$. 

In light of these recent developments one can ask if there is an extension of these relations to the  loop level. Considering the  complexity of even one loop calculations any reduction in workload is welcome. An analog of the Kleiss-Kuijf relations exists at one loop \cite{Bern:1994zx} where it relates non-planar to planar amplitudes. However, not much is known beyond this apart from some relations for the leading color part of the finite one loop amplitudes  \cite{Bern:1993qk} \cite{BjerrumBohr:2011xe}. These amplitudes involve either all helicities equal or one unequal of the participating particles and are known to be given by rational functions of polarizations and momenta. The helicity equal amplitudes for instance obey a `three photon decoupling relation' \cite{Bern:1993qk}. More relations for the finite loop amplitudes were conjectured very recently in \cite{BjerrumBohr:2011xe}. At two loops some results have been obtained in \cite{Feng:2011fja} and for four point all loop relations have appeared very recently in \cite{Naculich:2011ep}. In this article several new, helicity-blind relations at one loop will be proven for the coefficients in the standard scalar integral basis of pure Yang-Mills generalizing both  \cite{Bern:1993qk} and  \cite{BjerrumBohr:2011xe}, as well as a generalization of the BCJ relations to the one loop level. These results have been announced in a companion paper, \cite{Boels:2011tp}; their proof is in this article. 

The main technical result needed to prove these relations originates in yet another development triggered by  Witten's twistor string \cite{Witten:2003nn} insights: the derivation of on-shell recursion relations by Britto, Cachazo, Feng and Witten (BCFW) \cite{Britto:2004ap} \cite{Britto:2005fq}. These relations allow one to express tree amplitudes in terms of three point tree amplitudes only. The relations have been extended to $D\geq4$ dimensions in  \cite{ArkaniHamed:2008yf} and explicitly supersymmetrized in \cite{Brandhuber:2008pf}. Recently their extension to the integrand at loop level has been discussed in \cite{ArkaniHamed:2010kv} and \cite{Boels:2010nw}. Public packages for evaluating the recursion in Mathematica exist \cite{Dixon:2010ik}, \cite{Bourjaily:2010wh}.

The derivation of the on-shell recursion relations singles out two legs of an amplitude for which the on-shell momenta are shifted by a certain vector $q$,
\begin{equation}\label{eq:BCFWshift}
k_i \rightarrow k_i + q \, z \qquad \qquad k_j \rightarrow k_j - q \, z
\end{equation}
Crucial in the derivation of on-shell recursion relations is the behavior of the amplitude or integrand as $z \rightarrow \infty$: if this vanishes as  $\sim\!(z^{-1})$ or better on-shell recursion relations exist. For Yang-Mills amplitudes for instance there are choices of helicities of the shifted legs for which this behavior is realized. However, cases are known at tree level for which the shift behavior is more suppressed than $\sim\!(z^{-1})$. 

One example of this is QED \cite{Badger:2008rn}, \cite{Badger:2010eq} where the improved behavior is intimately tied in with the vanishing of certain integral coefficients at one loop. Moreover, in Yang-Mills theory it is known that the shift of non-color adjacent pairs of particles scales better than a shift of color adjacent particles. This result has been used in \cite{Feng:2010my} at tree level to prove the BCJ relations  \cite{Bern:2008qj}. Investigating non-adjacent shifts at loop level for the integrand therefore can yield evidence BCJ-type relations exist for this object. In addition, any all-loop information on the integrand of pure Yang-Mills amplitudes is of course always welcome. 

This article is structured as follows. Section \ref{sec:intro} contains a lightning review of  BCFW paying particular attention to the adjacent shift case. In section \ref{sec:nonadjshftintegr} it is established using Feynman graph techniques that non-adjacent shifts of gluons  are better behaved than the adjacent case for  the integrand of Yang-Mills coupled to scalar or spin-$\frac{1}{2}$ matter at any loop level.  In section \ref{sec:BCJatoneloop} a concrete generalization of the BCJ relations for tree level amplitudes to the one loop integrand of quite general gauge theories is presented. A particular generalization of the non-adjacent shift, its physical interpretation and proofs of some of its properties are presented in section \ref{sec:gennonadj}. In section \ref{sec:amplreloneloop} the results of the previous sections are used to study relations for pure Yang-Mills one loop amplitudes inspired by similar work in QED \cite{Badger:2008rn}. In particular it is shown that the coefficients of the standard basis for one loop amplitudes for pure Yang-Mills obey novel relations which are remarkably similar to known effects in (maximally) supersymmetric field theories.  The main section of the paper ends with a discussion and conclusions. Appendix \ref{app:feynrules} and \ref{app:diagrams} contain an overview over Feynman rules and graphs used for the analysis in section \ref{sec:nonadjshftintegr}. Appendix \ref{app:othermatters} explains how to obtain some shifts of other fields than gluons using on-shell supersymmetry. Appendix \ref{sec:shiftoneloop} contains results and conjectures on BCFW shifts of non-planar (integrated) one loop amplitudes.


\section{Lightning review}\label{sec:intro}
This section contains a lightning review of various techniques and concepts used throughout the article thereby establishing our conventions. The focus will be on amplitudes in $U(N)$ Yang-Mills theory and its supersymmetric cousins in four and higher dimensions, except where indicated otherwise.

\subsection{Color ordering at tree and loop level}
All amplitudes in Yang-Mills theory can be written in terms of permutation sums over so-called 
color-ordered amplitudes multiplied by certain color factors. At tree level this reads
\begin{equation}
A^{\textrm{full}}_{0} =  g^{n-2} \sum_{P_n/Z_n} \Tr\left(T^{a_{\sigma(1)}} \ldots T^{a_{\sigma(n)}}\right) A_0\left(\sigma(1)\ldots \sigma(n) \right)
\end{equation}
while at the one loop level this decomposition takes the form
\begin{multline}
A^{\textrm{full}}_{1} = N g^{n} \sum_{P_n/Z_n} \Tr\left(T^{a_{\sigma(1)}} \ldots T^{a_{\sigma(n)}}\right) A_1\left(\sigma(1)\ldots \sigma(n) \right) \\ 
+  g^{n} \sum_{c=1}^{[n]-1} \sum_{P_{n,c}} \Tr\left(T^{a_{\sigma(1)}} \ldots T^{a_{\sigma(c)}}\right) \Tr\left(T^{a_{\sigma(c+1)}} \ldots T^{a_{\sigma(c)}}\right) \\
A^{n|c}_1\left(\sigma(1)\ldots \sigma(c)| \sigma(c+1)\ldots \sigma(n) \right). 
\end{multline}
Here $P_{n,c}$ are those permutations which leave the  double trace structures invariant. The order of the gluons on the color-ordered amplitudes is fixed. As an immediate benefit of this representation, note that there are $(n-1)!/2$ partial amplitudes at tree level, while there are $(n)!$ full amplitudes. The color-ordered amplitudes can be calculated using color-ordered perturbation theory
\cite{Mangano:1988kk}, \cite{Bern:1990ux}.
\begin{figure}[!ht]
  \begin{center}
  \includegraphics[scale=0.6, angle=90, trim = 530 0 0 0 ]{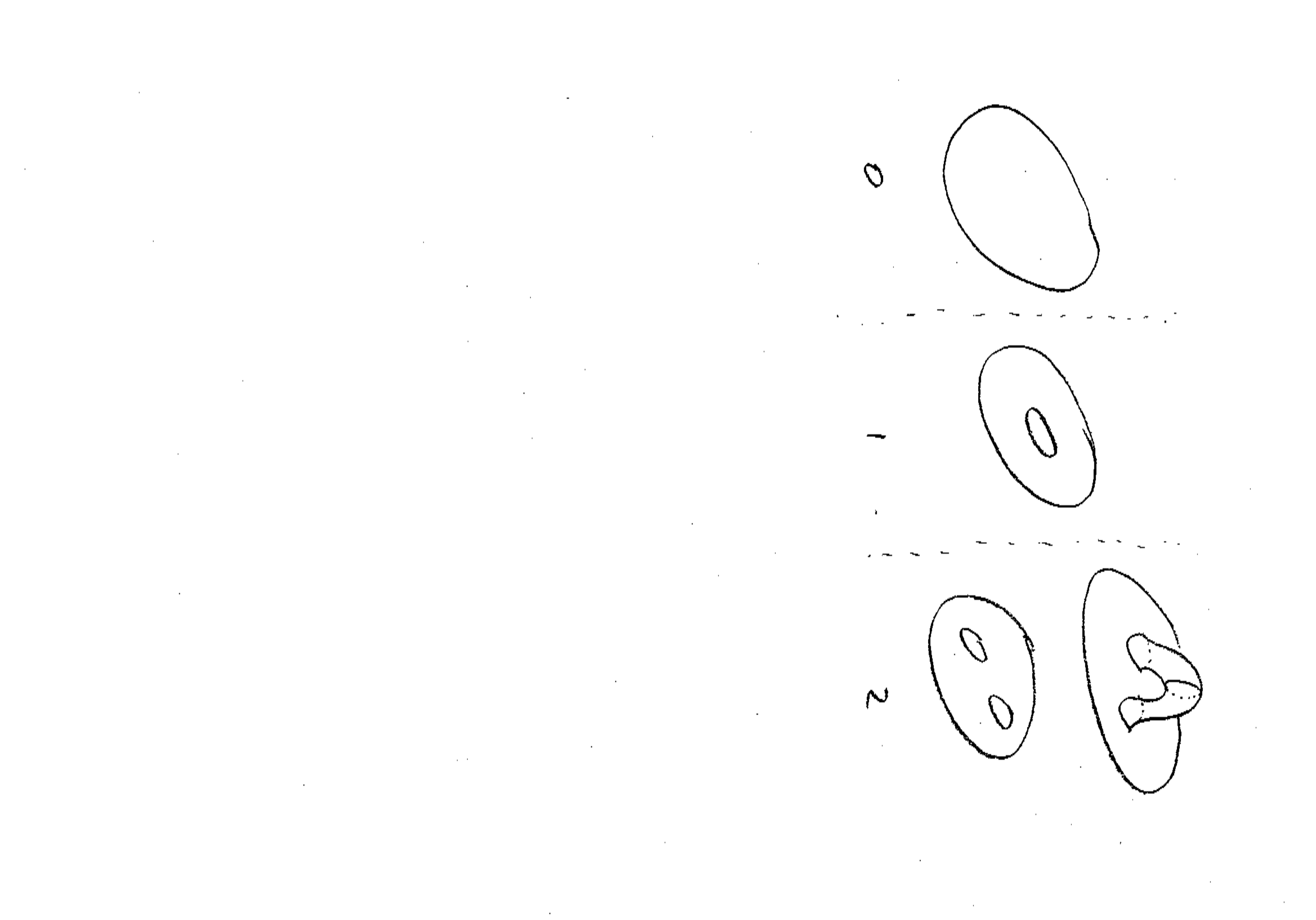}
   \label{fig:stringtop}
 \caption{Topologies for open string perturbation theory at the tree, one and two loop level}
   \end{center}
\end{figure}

For a string theorist this expansion is easily understood in terms of open string perturbation theory. At the first three loop levels of the open string the diagrams are given in figure \ref{fig:stringtop}. For open string color ordered amplitudes open string vertex operators have to be inserted on the available boundaries, in a particular order, keeping the order invariant after integration. The full amplitude is a sum over these partial amplitudes multiplied by Chan-Paton factors.

There is a close link of color-ordered perturbation theory to 't Hooft's \cite{'tHooft:1973jz} large $N$ limit,
\begin{equation}
 N\rightarrow \infty \quad \text{while} \quad g^2\;N = \lambda=\text{fixed}
\end{equation}
This is also called the planar limit. In the planar limit  the single trace terms contain the leading terms. Note however that the single trace terms can have non-planar corrections, starting at two loops where one obtains sketchily
\begin{multline}
A^{\textrm{full}}_{2} =  g^{n+2} \sum \left(N^2 f_1 + f_2 \right) \Tr\left(\phantom{P} \right)  + \\ g^{n+2}  \sum \left(N f_3 \right) \Tr\left(\phantom{P} \right) \Tr\left(\phantom{P} \right) + g^{n+2} \sum \left(f_4 \right) \Tr\left(\phantom{P} \right) \Tr\left(\phantom{P} \right) \Tr\left(\phantom{P} \right) 
\end{multline}

At tree level there has been some discussion in the literature about the number of independent  partial amplitudes in both field and string theory. In string theory it was argued early on in
\cite{Plahte:1970wy} that the correct number was $(n-3)!$.  In field theory in \cite{Kleiss:1988ne} relations with a concrete realization of $(n-2)!$ were presented by Kleiss and Kuijf (see also \cite{Berends:1988zn}). The Kleiss-Kuijf relations read
\begin{equation}\label{eq:KKrel}
A (1,\alpha,2,\beta) = \sum_{\omega \in OP(\alpha^T \cup \beta)} A(1, 2, \omega)
\end{equation}
where the ordered product $OP(\alpha^{T} \cup \beta)$ is the set of all permutations of the set $(\alpha^{T} \cup \beta$ which leave the order of the ordered subsets invariant. The set $\alpha^T$ is the inverse of the set $\alpha$.

Very recently new relations were conjectured in \cite{Bern:2008qj} and subsequently proven first through string theory methods in \cite{BjerrumBohr:2009rd} \cite{Stieberger:2009hq}. These new relations allow one to reduce the number of independent amplitudes further to $(n-3)!$, with an explicit expression of all amplitudes in terms of a basis available. 

At one loop one can in general calculate the double trace parts of the color ordered amplitudes as  certain permutation sums over the leading color single trace amplitude \cite{Bern:1994zx},
\begin{equation}\label{eq:sub-leadingfromleading}
A^{nlc}_1(\alpha | \beta) = \sum_{\omega \in COP(\alpha^{T} \cup \beta)} A_{LC} (\omega)
\end{equation}
The sum ranges over all ordered products such that the cyclic order of the sets $\beta$ and $\alpha$ are preserved. This formula can be used to calculate all sub-leading-in-color corrections at one loop, given the leading color answer. It should be noted that the naive extension of the CFT based derivation of the Kleiss-Kuijf relation in string theory in \cite{Boels:2010bv} to the loop level also yields this result. As explained in the introduction, only some relations have been explored for one-loop single trace color ordered amplitudes.

\subsection{BCFW on-shell recursion relations and shifts}
As mentioned in the introduction the derivation of on-shell recursion relation involves the notion of a BCFW shift (\ref{eq:BCFWshift}). This shift is designed to introduce a single complex variable into amplitudes by shifting two momenta 
\begin{equation}\nonumber
\begin{array}{cc}k_i \rightarrow & k_i + q\, z \\
k_{j} \rightarrow  &\;\; k_j - q\, z \ ,
\end{array}
\end{equation}
which conserve momentum conservation while the vector $q$ is constructed to obey
\begin{equation}\label{eq:BCFWconstraint} 
q \cdot k_i = q\cdot k_j = q \cdot q = 0
\end{equation}
so that the masses of both legs are invariant. There are two complex solutions for $q$, as can be checked in a lightcone frame for instance. If legs $i$ and $j$ are adjacent on a trace on a color ordered amplitude the shifts are called adjacent shifts. All other possibilities, including shifts of amplitudes on different traces will be referred to as non-adjacent. The shift turns an amplitude into a function of a complex variable $z$. The original amplitude, $A(0)$, can be obtained from $A(z)$ by a contour integral around the origin
\begin{equation}
 A(0)=\oint_{z=0}\frac{A(z)}{z}
\label{r_recursion}
\end{equation}
assuming $z=0$ is an isolated singularity.  The contour integral can be deformed to infinity which yields
\begin{equation}
\label{r_bcfw}
 A(0)=\oint_{z=0}\frac{A(z)}{z}=\sum(\text{residues})\Big|_{z=\text{finite}} + \sum(\text{residues})\Big|_{z = \infty}
\end{equation}
At tree level, the finite $z$ residues are products of tree level amplitudes, summed over all intermediate states (simply consider the Feynman graphs for instance). There is no such physical interpretation for the residues at infinity. If those residues can be shown to be absent equation \eqref{r_bcfw} constitutes an on-shell recursion relation for tree level amplitudes. If they are not absent progress can still be made if the boundary contributions can be calculated \cite{Feng:2010ku}, \cite{Feng:2011tw}, but this case will not be considered here. To investigate whether a theory obeys on-shell recursion relations it is crucial to know the behavior of its scattering amplitudes for $z \rightarrow \infty$. In all known cases absence of the residue at infinity follows from fall-off of the amplitudes at infinity of the form  $\sim\! (z^{-1})$ or better.

Better than  $\sim\! (z^{-1})$ falloff at infinity can be used to generalize equation \eqref{r_bcfw}. For fall-off of $A(z)$ of the form  $\sim\!(z^{-2})$ or better one consider for instance
\begin{equation}\label{eq:bonusrelfromBCFW}
 A(0)=\oint_{z=0}\frac{\alpha - z}{\alpha z}A(z)=\sum(\text{residues})\Big|_{z=\text{finite}}\cdot
f(p_i)
\end{equation} 
for some constant $\alpha$. The residues on the right hand side are the same products of known amplitudes as before, but now multiplied by an additional factor. This has two known uses. First, by tuning $\alpha$ one can eliminate particular terms in the recursion relations, see \cite{Badger:2010eq} for an example in QED.  Second and more important for this article, the above implies certain relations between amplitudes referred to as bonus relations. This has been used for instance in gravity in \cite{Spradlin:2008bu}. As was shown in  \cite{Feng:2010my} and will be re-derived in detail in section \ref{sec:BCJatoneloop} this can be used to prove the BCJ relations at tree level. 

\subsection{Deriving large shift scaling for color-adjacent shifts}
The large-$z$ scaling behavior of tree level amplitudes can be analyzed in several ways all of which are variants of power counting: tracing explicit factors of $z$ in Feynman diagrams. In standard Feynman-'t Hooft gauge for instance, if they contain legs with shifted momenta the three-vertex will scale as $\sim\! (z)$ , the four-vertex as $\sim\! (z^0)$  and the propagator as $\sim\! (z^{-1})$ . Hence the leading scaling behavior in Yang-Mills theory is $\sim\!(z)$ coming from graphs with only three vertices. As will be shown below, this can be improved to yield the result for a color-adjacent shift in the form
\begin{equation}\label{large_z_r}
 A(z)\sim \epsilon^\mu_1(\hat{p}_1)\epsilon^\nu_2(\hat{p}_2)M_{\mu\nu}\sim \epsilon^\mu_1(\hat{p}_1)\epsilon^\nu_2(\hat{p}_2) \Big(z\eta_{\mu\nu}f_1(1/z)+f_{2,\mu\nu}(1/z)+\mathcal{O}(1/z)\Big) 
\end{equation}
where $f_i$ are polynomial functions in $z^{-1}$ and $f_2$ is antisymmetric in its indices and $ \epsilon^\mu_i$ are the z-dependent polarization vectors of the shifted legs. This scaling result holds for all  Yang-Mills theories minimally coupled to fermionic and scalar matter with possible scalar potential or Yukawa terms \cite{Cheung:2008dn} in four and higher dimensions. The scaling of the amplitudes depends on the little group index of the shifted $D$ dimensional gluon legs. Using four-dimensional notation derived from the space spanned by the vectors $k_1, k_2, q$ and $q^*$ this yields  \cite{ArkaniHamed:2008yf} table \ref{tab:r_tab1}. Since the amplitude $A(z)$ in equation \eqref{large_z_r} vanishes as $z$ approaches to $\infty$ for polarization combinations $(-,\pm)$ on-shell recursion relations hold in these cases.

\begin{table}[h!]
\centering
\begin{tabular}{l|lll}
$\epsilon_1 \backslash \epsilon_2 $& $-$ & $+$ & T\\
\hline
$-$ & $\nicefrac{1}{z}$ & $\nicefrac{1}{z}$ & $\nicefrac{1}{z}$\\
$+$ & $z^3$ & $\nicefrac{1}{z}$ & $z$\\
T & $z$ & $\nicefrac{1}{z}$ & $z$\\
T' & $z$ &$\nicefrac{1}{z}$ & $z^0$\\

\end{tabular}
\caption{\label{tab:r_tab1} Large $z$ behavior of a tree amplitude / loop integrand in 
YM for an adjacent gluonic shift and all possible polarizations of these
gluons for a special choice of $q$. T and T' indicate if contractions between the 
polarization vectors vanish or not, i.e $\epsilon_1^T \cdot
\epsilon_1^{T'}=\delta^{TT'}$.}
\end{table} 

An efficient way to obtain equation \eqref{large_z_r} for tree level amplitudes is \cite{ArkaniHamed:2008yf}  to split the Yang-Mills fields in the Lagrangian into $z$-dependent 'hard' and $z$-independent `soft' fields. Furthermore, the soft fields can be treated as a background. This can be done neatly in terms of the background method with the result 
\begin{equation}
\mathcal{L}=-\frac{1}{4} \Tr \left( D_\nu a_\mu D^\nu a^\mu + \frac{i}{2}[a_\mu, a_\nu] F^{\mu\nu}[A] \right)
\label{bgflag}
\end{equation}
for the quadratic part of the Lagrangian for the hard fields $a_{\mu}$. Here the background gauge version of the Feynman-'t Hooft gauge has been implemented for the hard fields $a^{\mu}$. Since only $a_{\mu}$ have $z$-dependent momenta the only three vertex which depends on $z$ comes from the first term which contains a metric contraction between the hard fields. All hard propagators scale as $\sim\!(z)^{-1}$. Every insertion of a graph from the second term in \eqref{bgflag} will be suppressed by one order of $\sim\!(z^{-1})$ as well as be anti-symmetric in the hard fields. Combining these observations yields equation \eqref{large_z_r}.  For future reference, note this derivation of \eqref{large_z_r} is confined to tree level since the equations of motions have been used in the derivation. 

Further simplifications arise if the gauge freedom of the background fields in equation \eqref{bgflag} is used to impose the natural `spacecone' \cite{Chalmers:1998jb} gauge,
\begin{equation}
q \cdot A = 0 \label{ahkgauge}
\end{equation}
with $q$ the BCFW shift (\ref{eq:BCFWconstraint}). This will be referred to as AHK gauge. This gauge choice eliminates most $z$-dependence from the three vertices, leaving only few diagrams for the leading terms. More on the background field method can be found in section \ref{sec:gennonadj}.

A variant of the above powercounting-based approach is to investigate Feynman graphs directly in AHK gauge \cite{Boels:2010nw}, dispensing with the distinction between hard and soft fields. This method has the distinct advantage that no on-shell conditions are necessary. The proof of the large-$z$ scaling for adjacent shifts \eqref{large_z_r} using this approach will be repeated here briefly as a warm-up: the same approach will be used in the following section to obtain the large-$z$ behavior for shifts of non-color-adjacent particles.

The AHK gauge \eqref{ahkgauge} can be chosen for all fields including all polarization vectors except those of the shifted legs. Since $q$ is orthogonal to the momentum in these legs the AHK gauge is not a valid gauge choice here. For these shifted legs one has instead
\begin{equation}
\label{tradeitoff}
 \hat{p}\cdot \epsilon(\hat{p})=(p\pm zq )\cdot \epsilon(\hat{p})=0 \quad \Rightarrow \quad q\cdot \epsilon(\hat{p})=\mp \frac{p\cdot \epsilon(\hat{p})}{z}
\end{equation}
The propagator in AHK gauge reads
\begin{equation}
 G(p)_{\mu\nu}=-\frac{i}{p^2}\Big(\eta_{\mu\nu}-\frac{q_\mu p_\nu+p_\mu q_\nu }{p\cdot q}\Big)
\label{AHKpropagator}
\end{equation}
This propagator is orthogonal to $q$ by construction and collapses if contracted into its momentum
\begin{subequations}
\begin{equation}
\label{pprop1}
 q^{\mu} G(p) _{\mu\nu}= 0
\end{equation}
\begin{equation}
\label{pprop2}  p^{\mu} G(p) _{\mu\nu} = \frac{i q_{\nu}}{p \cdot q}
\end{equation}
\end{subequations}

The powercounting argument now requires one to identify which parts of the Feynman graphs depend on the shifted momenta. At tree level there is a unique line in the diagram connecting the shifted legs, but at loop level for the integrand of amplitudes this is no longer true. One can however always choose a routing of the loop momenta such that only the shortest path through the diagram depends on the shifted momenta. For color-adjacent shifts this path is along the edge of the color-ordered graphs. This path will be referred to as the hard line. The complete set of Feynman graphs follows from the graphs containing only the hard line by contracting off-shell currents onto this. Note that there is no canonical routing of a hard line if the shifted legs belong to different traces of a non-planar integrand. 

In the following the focus will be almost exclusively on the scaling of the hard-line graphs, leaving the external lines which are not the shifted legs arbitrary. In this way the scaling results obtained hold for integrands of Yang-Mills scattering amplitudes to arbitrary loop order. Phrased differently, our results are for tree level correlation  functions calculated in AHK gauge. When combined into a gauge invariant object the scaling result holds for this object. An example application of this beyond scattering amplitudes are form-factors (see e.g. \cite{Brandhuber:2011tv} \cite{Bork:2011cj} and references therein) at both tree and loop level. 

A propagator along the hard line scales as $z^0$
\begin{equation}
G(\hat{p})_{\mu\nu}^{\text{hard}}\sim \frac{2z}{p^2\pm2z q\cdot p}\Big(\frac{q_\mu q_\nu}{q\cdot
p}\Big)+\mathcal{O}(1/z)
\label{ahkhardline}
\end{equation}
Due to its dependence on two $q$'s the $z^0$ part of the propagator hardly ever contributes since $q$ vanishes when contracted into any unshifted external leg. An exception is when it contract into the momentum on a three vertex. Apart from this effect, additional propagators in hard-line graphs lower the $z$ scaling. Note further that the three vertices in AHK gauge can be taken to be $z$-independent. Hence for the leading scaling behavior under a BCFW shift there are only a few graphs to be drawn. 

The leading diagram to consider for an adjacent shift is a three-vertex with two shifted legs and one off-shell leg. It can be shown using the Feynman rules of appendix \ref{app:feynrules} by power counting to scale like equation \eqref{large_z_r}. A subtlety occurs in this derivation:  the momentum in the off-shell leg is proportional to $\hat{p}_i+\hat{p}_j$ which is orthogonal to $q$ and hence the AHK gauge is singular for this class of graphs. This divergence can be circumvented by imposing an auxiliary gauge \cite{Boels:2010nw} which yields the structure of equation \eqref{large_z_r}.

\begin{figure}[h!]
\centering
\includegraphics[scale=0.08]{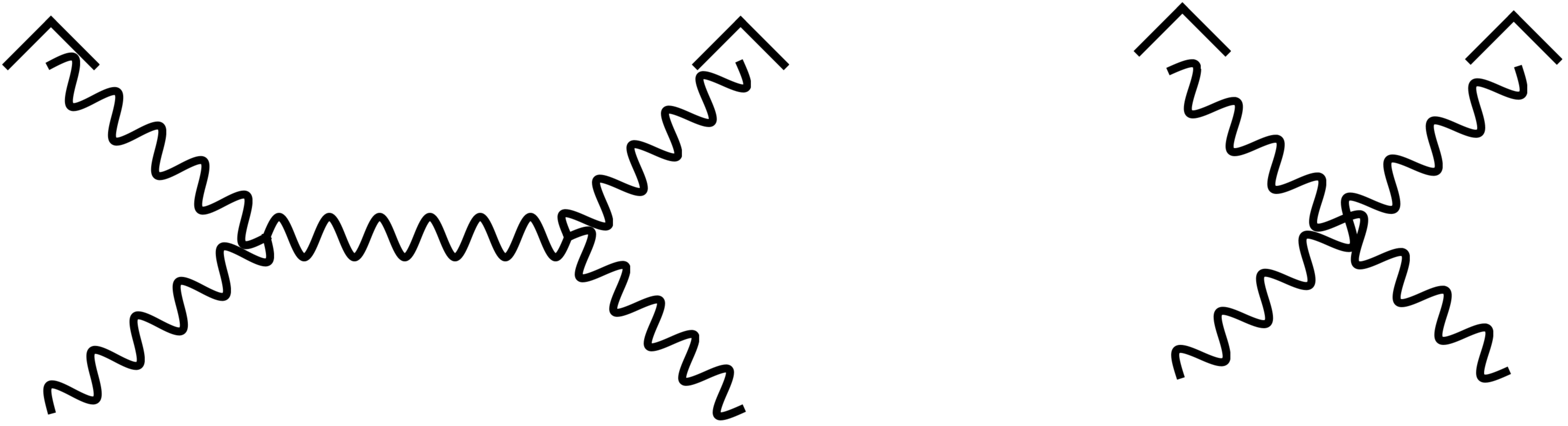}
\caption{\label{fig:sublea} $\mathcal{O}(z^0)$ contributions for adjacent shifts.}
\end{figure}
The sub-leading $\sim(z^0)$ behavior come from the graphs depicted in figure \ref{fig:sublea}. The second contributes at order $\sim(z^0)$ due to equation \eqref{ahkhardline} which leads to 
\begin{equation}
\epsilon_1^\mu(\hat{p}_1)V_{\rho\mu\kappa}G^{\kappa\lambda}V_{\lambda\nu\sigma}\epsilon_2^\nu(\hat{p
}_2) \sim z^0 \epsilon_1^\mu \epsilon_2^\nu
\eta_{\mu\sigma}\eta_{\nu\rho}+\mathcal{O}(1/z)
\end{equation}
Combining it with the ight diagram in figure \ref{fig:sublea} obtains an effective four-vertex for adjacent shifts antisymmetric in the shifted legs is obtained, 
\begin{equation}
\label{effecad}
 {}^4 V_{\mu\nu\sigma\rho} = i
z^0\Big(\eta_{\mu\rho}\eta_{\nu\sigma}-\eta_{\mu\sigma}\eta_{\nu\rho}-\frac{1}{2}\eta_{\mu\nu}
\eta_{\rho\sigma}    
\Big)+\mathcal{O}(1/z).
\end{equation}
The last term proportional to the metric contraction between the shifted legs can be thought of as a part of the function $f_1$ of \eqref{large_z_r}. Hard line graphs with more hard line propagators will be suppressed in $z$ and so do not contribute at this order. Combing results the large-$z$ scaling of the integrand of a scattering amplitude under an adjacent BCFW shift is given by equation \eqref{large_z_r}. Even more generally, the above shows that the BCFW shift of two color adjacent legs on a tree level color ordered correlation function in AHK gauge scales as equation \eqref{large_z_r}.

\subsection*{Some notation for sets}
In the following sums over certain sets will play a central role. Ordered sets will be indicated by round brackets, e.g. $(1,2,\ldots,n)$, while unordered sets will have curly brackets, e.g. $\{1,2,\ldots,n\}$.
An overview over the notation for other sets follows. 

\begin{enumerate}
\item $P_n$: the set of permutations of a set with $n$ elements. 
\item $\mathbb{Z}_n$: the set of cyclic permutations of a set with $n$ elements, assuming canonical order.
\item $P_n / \mathbb{Z}_n$: the set of non-cyclic permutations of a set with $n$ elements.
\item $OP(\alpha^T \cup \beta):$ the ordered product, i.e. sum over all unions of the two sets
$\alpha$ and $\beta$, leaving the order of $\beta$ and the inverse of $\alpha$ intact. 
\item $POP({\alpha}\cup{\beta}):$ the partially ordered product, i.e. sum over all unions of the
two sets $\alpha$ and $\beta$, leaving the order of $\beta$ preserved.
\item $COP({\alpha}\cup{\beta}):$ the cyclicly ordered product, i.e. sum over all unions of the
two sets $\alpha$ and $\beta$, leaving the cyclic order of $\beta$ and $\alpha $preserved.
\end{enumerate}

\noindent Examples for all these sets follow:
\begin{enumerate}
\item E.g. for $n=3$: $\{(123), (132), (231), (213), (312), (321)\}$.
\item E.g. for $n=3$: $\{(123), (231), (312)\}$ .
\item E.g. for $n=3$: $\{(123), (132)\}$.
\item E.g. for the sets $\alpha=(1,2)$ and $\beta=(3,4)$: $\{(2134), (3214), (3421), (2314), (2341), (3241)\}$.
\item E.g. for the sets  $\alpha=\{1,2\}$ and $\beta=(3,4)$: $\{(2134), (3214), (3421), (2314), (2341)$, $(3241),(1234), (3124), (3412), (1324), (1342), (3142) \}$. 
\item E.g. for the sets  $\alpha=(1,2)$ and $\beta=(3,4)$: $\{(2134), (3214), (3421), (2314), (2341)$, $(3241),(1234), (3124), (3412), (1324), (1342), (3142) \}$. 
\end{enumerate}
Note that the difference between $POP$ and $COP$ only sets in for bigger sets which  defeat the purpose of writing simple examples here.


\section{Generic non-adjacent BCFW shifts for integrands}\label{sec:nonadjshftintegr}

In this section the scaling of integrands of Yang-Mills theory under BCFW shifts of two legs which are not color adjacent will be studied, generalizing the analysis for adjacent shifts reviewed in the previous section. At tree level it is a folk theorem these shifts scale one power of $z$ better than their adjacent counterparts\footnote{Although widely known, we have been unable to find a general proof of this improved scaling for non-adjacent shifts in the literature.}. For orientation, take a tree level $n$-point MHV amplitude for the
helicity configuration $(--+\dots +)$. In spinor language it is given up to unimportant numerical constants by
\begin{equation}
 A^{\text{MHV}}_n(--+++\dots+)=\frac{\langle 1 2 \rangle^4}{\langle 12 \rangle \langle 23 \rangle
\langle 34 \rangle \langle 45 \rangle\dots \langle n 1 \rangle}
\end{equation}
A non-adjacent shift of particles one and three implies for the holomorphic spinors
\begin{equation}
 |1\rangle \rightarrow |1\rangle \quad |3\rangle \rightarrow |3\rangle + z|1\rangle
\end{equation}
which implies for the amplitude
\begin{equation}
 A^{\text{MHV}}_n(--+++\dots+)=\frac{\langle 1 2 \rangle^4}{\langle 12 \rangle \langle 2(3 +z 1)
\rangle \langle (3+z1)4
\rangle \langle 45 \rangle\dots \langle n 1 \rangle}.
\end{equation}
which indeed hows an $z^{-2}$ scaling behavior.

Since for adjacent shifts integrands show the same scaling behavior as tree amplitudes it is natural to suspect that the same holds for non-adjacent shifts. Hence integrands are expected to show the same improved scaling behavior under non-adjacent shifts as their tree level counterparts, i.e.
\boxit{\begin{equation}
 A(z)\sim \epsilon_i^{\mu}\epsilon_j^{\rho}\Bigg(z^0 \eta_{\mu\rho}f_1(1/z)+\frac{1}{z} B_{\mu\rho}(1/z)+\mathcal{O}\left(\frac{1}{z^2}\right)\Bigg)
\label{r_gennonadscale}
\end{equation}}
where $f_1$ is a polynomial in 1/z, $B$ is an antisymmetric matrix. $i$ and $j$ denote the shifted legs and $\mu$ and $\rho$ are the space-time indices of these non-adjacently shifted legs. This is structurally the same formula as equation \eqref{large_z_r} for the adjacent shift only one power down in $z$. The central result of this section is that this suspicion is true for all minimally coupled gauge theories with possible scalar potential and Yukawa terms

The analysis of the large-$z$ behavior of the integrand for non-adjacent shifts presented below proceeds via powercounting in AHK gauge, just as in the adjacent case. The main difference is that in this case hard line graphs up and including six points will have to be considered to evaluate the scaling up to and including order $\sim\!(z^{-1})$ as required to prove equation \eqref{r_gennonadscale}. All higher point hard line graphs will start at $\sim\!(z^{-2})$ and do not need to be considered here. The calculations presented below have been performed with the aid of FeynCalc \cite{Kublbeck:1992mt}. The Feynman rules can be found in appendix \ref{app:feynrules} and an overview of the graphs used can be found in the appendix \ref{app:diagrams}.  The Mathematica files are available on request.

A consistency check of the calculations we have found useful in intermediate stages is to put all external legs on-shell in four dimensions as the scaling of these tree level amplitudes is known.

\subsection{Gluonic contributions}
\subsection*{Four point graphs}
\begin{figure}[h!]
\centering
\includegraphics[scale=0.5]{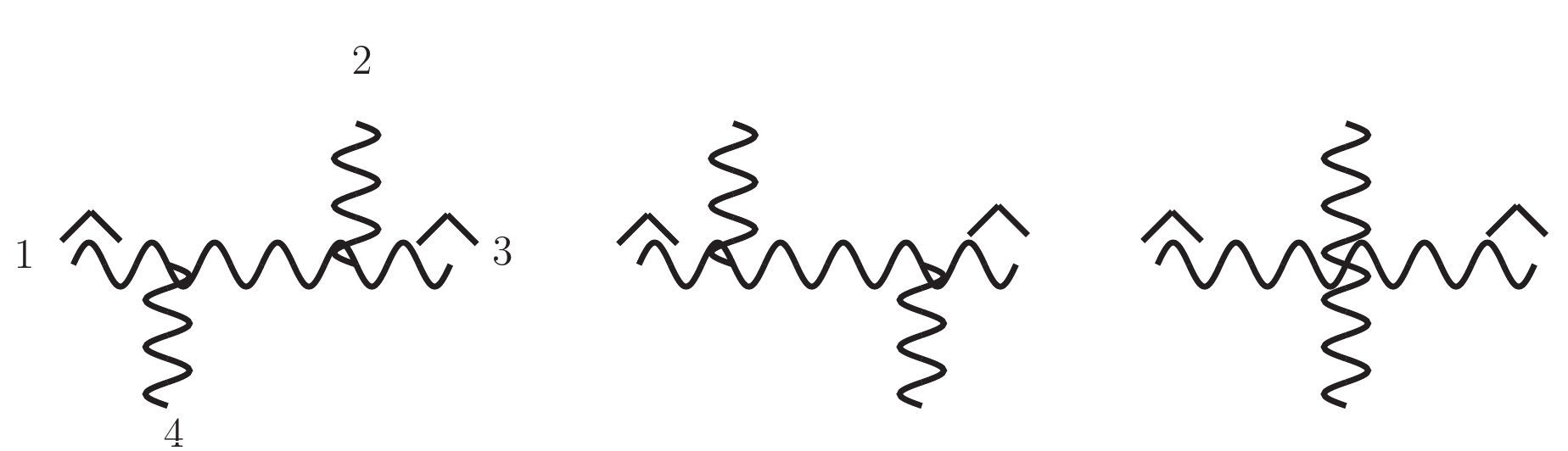}
\caption{\label{fig:4ptglue} All color-ordered glue 4-point diagrams. Legs 1 and 3 have been
shifted.}
\end{figure}

At four points there are only three diagrams to consider: the YM four-vertex and the $s$-channel and $t$-channel graph depicted in figure (figure \ref{fig:4ptglue}). Note the $u$-graph does not appear in color-ordered perturbation theory. 
\noindent Labeling the legs clockwise from one to four, shift for instance legs 1 and 3. The sum of the three graphs in the large-$z$ limit is
\begin{equation}
\begin{split}
\label{4ptresult}
 A(z)_{4pt} &= \epsilon_1^\mu(\hat{p}_1)\epsilon_3^\rho(\hat{p}_3)M_{\mu\nu\rho\sigma}\\
	    &= i\epsilon_1^\mu(\hat{p}_1)\epsilon_3^\rho(\hat{p}_3)\Bigg(z^0 \eta_{\mu\rho}\eta_{\nu\sigma}(1+\mathcal{O}(1/z))+\frac{1}{z \, p_2\cdot q}\Big[p_1\cdot p_3(\eta_{\mu\nu}\eta_{\rho\sigma}-\eta_{\mu\sigma}\eta_{\rho\nu})\\
            &+\eta_{\mu\sigma}(p_{4\nu}p_{2\rho}+p_{2\rho}p_{4\nu}+\frac{1}{2}p_{2\rho}p_{2\nu}) + \eta_{\nu\rho}(p_{2\sigma}p_{2\mu}+p_{2\sigma}p_{4\mu}+\frac{1}{2}p_{4\sigma}p_{4\mu}) \\
            &- \mu \leftrightarrow \rho\Big]+\mathcal{O}\left(\frac{1}{z^2}\right)\Bigg)
\end{split}
\end{equation}
Here the indices $\nu$ and $\sigma$ belong to the unshifted legs $2$ and $4$respectively. Sub-leading terms proportional to the metric have been dropped since they appear in the expansion of the function $f_1(1/z)$ as already encountered in \eqref{large_z_r}. This will be done throughout this section without further warning. This result shows the scaling behavior of equation \eqref{r_gennonadscale}: the leading part is $\sim\!(z^0)$ and proportional to the metric while the sub-leading part not proportional to the metric is antisymmetric in the shifted legs. 

\subsection*{Five point graphs}
Powercounting suggests that the class of diagrams with five gluons (figure \ref{fig:5ptglue}) will contribute only up to order $\sim\!(z^{-1})$. The shifted non-adjacent legs will be labelled $1$ and $3$. As expected the large-$z$ behavior scales like $\sim\!(z^{-1})$ but the result is not antisymmetric at this order i.e. a symmetric $\sim\!(z^{-1})$ part of the sum of these diagrams remains. If the legs are labeled by $(\hat{p}_1,\mu)$, $({p}_2,\nu)$, $(\hat{p}_3,\rho)$, $({p}_4,\sigma)$, $({p}_5,\tau)$ the result of the symmetric part of the sum of diagrams not proportional to $\eta_{\mu\rho}$ is given by
\begin{multline}
\label{135pointsym}
 A_{5, \textrm{symmetric}}=
\frac{-i\epsilon_1^\mu(\hat{p}_1)\epsilon_3^\rho(\hat{p}_3)}{2
\sqrt{2} \,z} \Bigg(
p_{2\nu}\Big(\frac{p_2 \cdot q
(\eta_{\mu\tau}\eta_{\rho\sigma}+\eta_{\rho\tau}\eta_{\mu\sigma})}{p_4\cdot q \, p_5 \cdot
q}\Big)\\+
p_{4\sigma}\Big(\frac{p_4\cdot q
(\eta_{\mu\tau}\eta_{\rho\nu}+\eta_{\rho\tau}\eta_{\mu\nu})}{p_2\cdot q \, p_5 \cdot
q}\Big)+
p_{5\tau}\Big(\frac{p_5\cdot
q(\eta_{\mu\sigma}\eta_{\rho\nu}+\eta_{\rho\sigma}\eta_{\mu\nu})}{p_2\cdot q \, p_4\cdot
q}\Big)\Bigg)+\mathcal{O}\left(\frac{1}{z^2}\right)
\end{multline}
This symmetric part consists of terms proportional to the momentum in one of each off-shell legs. To obtain the result of a shift of particles (1, 4) from the previous result replace $p_4$ by $p_3$, interchange $\sigma$ and $\rho$ and multiply the whole expression by minus one.

This symmetric part seems to be in conflict with the scaling of equation \eqref{r_gennonadscale}. This can be resolved as follows. If the unshifted legs are put on-shell the symmetric part written in equation \eqref{135pointsym} will vanish. For more general cases one contracts currents into the off-shell legs. As shown before in equation \eqref{pprop2} the propagator will collapse and one obtains (indices suppressed)
\begin{equation}
 p\cdot J(p) \sim p\cdot G(p)\sum(\text{graphs})\sim q\sum(\text{graphs})
\end{equation}
By the choice of gauge, this $q$ can only contract into the momentum of a three point vertex to give a non-vanishing result. Hence this particular symmetric part contributes to six point graphs. As will be shown explicitly below they combine with the six point hard line graphs to ensure the scaling behavior of equation \eqref{r_gennonadscale}.

\subsection*{Six point graphs}
This class of hard line diagrams scales maximally as $\sim\!(z^{-1})$ as follows from powercounting the graphs (figure \ref{fig:6ptglue}). For six points the number of graphs increases significantly. Furthermore, there are several possibilities for the choice of a non-adjacent shift. The shift $(1, 3)$ or $(1, 5)$ shifts involve $15$ graphs and while a $(1, 4)$ shift  involves $21$ graphs. To verify equation \eqref{r_gennonadscale} only the symmetric part of the sum of these graphs needs to be calculated. One finds this is nonzero. For instance with the labeling $(\hat{p}_1,\mu)$,
$({p}_2,\nu)$, $(\hat{p}_3,\rho)$, $({p}_4,\sigma)$, $({p}_5,\tau)$, $({p}_6,\lambda)$ the symmetric part of the result of the $(1, 3)$ shift is given by
\begin{multline}
 A_{6,\textrm{sym}}=\epsilon_1^\mu(\hat{p}_1)\epsilon_3^\rho(\hat{p}_3)M_{\mu\nu\rho\sigma\tau\lambda}
=- \frac{i \epsilon_1^\mu\epsilon_3^\rho }{4 \, z}\Bigg( \frac{\eta_{\sigma\tau}(p_4\cdot q - p_5 \cdot q)(\eta_{\mu\nu}\eta_{\rho\lambda}+\eta_{\rho\nu}\eta_{\mu\lambda})}{p_2\cdot q\, p_6 \cdot q} \\
+ \frac{\eta_{\lambda\tau}(p_5\cdot q - p_6 \cdot q)(\eta_{\mu\sigma}\eta_{\rho\nu}+\eta_{\rho\sigma}\eta_{\mu\nu})}{p_2\cdot q \,p_4 \cdot q} \Bigg) + \mathcal{O}\left(\frac{1}{z^2}\right)
\label{6ptsym}
\end{multline}
Taking into account contributions arising from the symmetric part of the five gluon graphs, the symmetric parts will cancel. 

\begin{figure}[h!]
\centering
\includegraphics[scale=0.18]{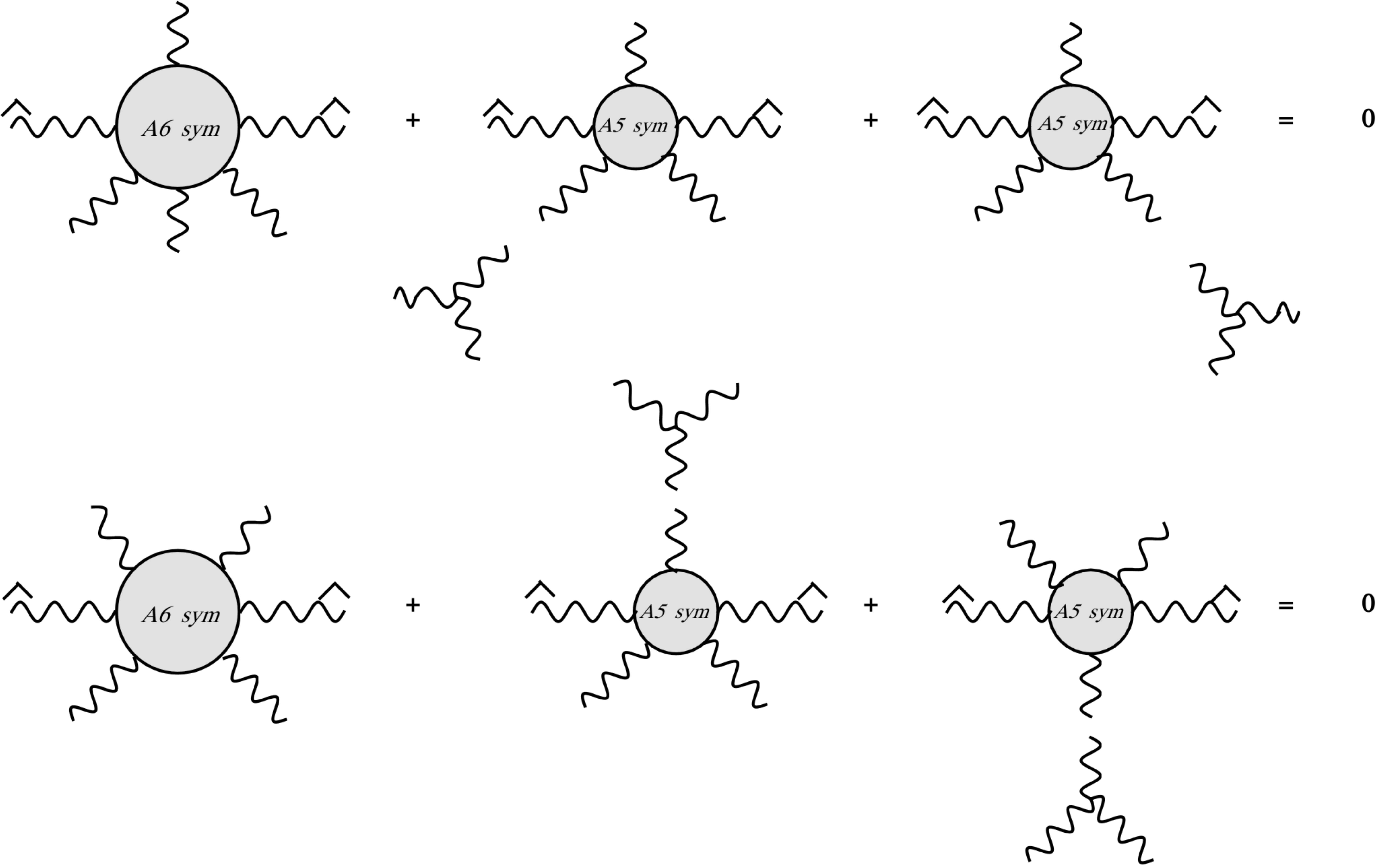}
\caption{\label{fig:5to6} Diagrammatics of the cancellation of the symmetric parts of the five- and
six point hard line graphs for different shifts at six
points. The case of three particle above the hard line is not shown.}
\end{figure}

It is instructive to study this in somewhat more detail. Consider the result for the five point case, equation \eqref{135pointsym}. If one contracts a three-gluon vertex into one of the off-shell legs, the connecting propagator collapses (due to $p\cdot G(p)$) and the result looks effectively like one of the six point diagrams under consideration. To give an example: contract a three-vertex $V$ into leg $5$ of equation \eqref{135pointsym} or to be more precise into the term that is proportional to the momentum in leg $5$.  Upon replacing $p_5\rightarrow p_5+p_6$ and $\tau
\rightarrow \alpha$ one obtains
\begin{equation}
\begin{split}
 &  \frac{-i\epsilon_1^\mu\epsilon_3^\rho}{2\sqrt{2} \, z} V_{\lambda\tau \beta}G^{\alpha\beta}(p_5+p_6)(p_5+p_6)_{\alpha} \Big(\frac{((p_5+p_6)\cdot q) (\eta_{\mu\sigma}\eta_{\rho\nu} +\eta_{\rho\sigma}\eta_{\mu\nu})}{p_2\cdot q \, p_4\cdot q}\Big)\\
&=\frac{-i\epsilon_1^\mu\epsilon_3^\rho}{2\sqrt{2} \, z}  V_{\lambda\tau \beta}\frac{i q^{\beta}}{(p_5+p_6)\cdot q}\Big(\frac{((p_5+p_6)\cdot q)(\eta_{\mu\sigma}\eta_{\rho\nu}+\eta_{\rho\sigma}\eta_{\mu\nu})}{p_2\cdot q \, p_4\cdot q}\Big)\\
&=\frac{i\epsilon_1^\mu\epsilon_3^\rho}{4 \, z} \Big(\eta_{\tau \lambda}(p_5-p_6)_{\beta}-\eta_{\tau \beta}(2p_5+p_6)_{\lambda}+\eta_{\lambda \beta}(2p_6+p_5)_{\tau}\Big)\\& \quad\frac{ q^{\beta}}{(p_5+p_6)\cdot q}\Big(\frac{((p_5+p_6)\cdot q)(\eta_{\mu\sigma}\eta_{\rho\nu}+\eta_{\rho\sigma}\eta_{\mu\nu})}{p_2\cdot q \, p_4\cdot q}\Big)\\
&=\frac{i\epsilon_1^\mu\epsilon_3^\rho((p_5-p_6)\cdot q)}{4\, z} \frac{(\eta_{\mu\sigma}\eta_{\rho\nu}+\eta_{\rho\sigma}\eta_{\mu\nu})}{p_2\cdot q \, p_4\cdot q} +  \mathcal{O}\left(\frac{1}{z^2}\right)
\end{split}
\end{equation}
The two other terms of the three-vertex did not survive the last line because they are proportional to a $q$ contracted into an on or off-shell leg in AHK gauge. The result of this exercise is up to sign the second term of \eqref{6ptsym}. Of course one gets the same topology by contracting a current into leg $4$ and both possibilities have to be taken into account:

\begin{equation}
\begin{split}\begin{matrix}\text{Influence of five} \\ \text{points on six points}\\ \text{for shift
} (1, 3)\end{matrix}
 =A_{5,4,sym}^{(1, 3)}+A_{5,5,sym}^{(1, 3)}&=\frac{i}{4\, z}\epsilon_1^\mu\epsilon_3^\rho \Bigg(
\frac{\eta_{\sigma\tau}(p_4\cdot q - p_5 \cdot
q)(\eta_{\mu\nu}\eta_{\rho\lambda}+\eta_{\rho\nu}\eta_{\mu\lambda})}{p_2\cdot q \, p_6 \cdot q}\\
&+
\frac{\eta_{\lambda\tau}(p_5\cdot q - p_6 \cdot
q)(\eta_{\mu\sigma}\eta_{\rho\nu}+\eta_{\rho\sigma}\eta_{\mu\nu})}{p_2\cdot q \, p_4 \cdot q}
\Bigg) + \mathcal{O}\left(\frac{1}{z^2}\right)
\end{split}
\end{equation}
where $A_{5,x}$ means that a three-particle vertex has been contracted into the term proportional to the momentum of leg $x$ of the symmetric part of the five-particle graphs of shift $(1, 3)$ (see equation \eqref{135pointsym}). Comparing this with equation \eqref{6ptsym} one sees that the expressions are identical up to sign and their sum vanishes. Connecting a current to the second leg of the five point symmetric part will result in canceling terms of the symmetric part of the $(1, 4)$ six leg graphs. The other shifts work along the same lines. This is represented in figure \ref{fig:5to6}. 

To repeat this once more: each term of the symmetric five point gluon graphs has to be treated separately since each of these terms will give rise to contributions for different color orders/shifts of six point graphs. Remember that contractions into leg 4 and 5 gave rise to terms at the color ordering ($\hat{1}2\hat{3}456$) whereas a contraction into leg 2 yielded contributions at ($\hat{1}23\hat{4}56$) each canceling symmetric terms at six points. This is generic since all legs are kept off-shell, so it follows that the symmetric parts will cancel each other at higher points too. Therefore the large-$z$ behavior of purely gluonic integrands subjected to non-adjacent shifts is given by equation \eqref{r_gennonadscale}.

\subsection{Minimally coupled scalar contributions}
In this subsection it will be shown that the scaling behavior of the integrand under a non-adjacent BCFW shift of two gluons does not change if minimal scalar-gluon couplings are included. This involves analyzing all scalar contributions to hard-line graphs up to order $\sim\!(z^{-1})$. The scalar Feynman rules are given in appendix \ref{app:feynrules}. Note that these rules are for adjoint matter, for fundamental matter one simply restricts to the diagrams where the scalar legs are adjacent.  

At four points the three new graphs to be considered of two gluons and two scalars are drawn in figure (fig. \ref{fig:4ptscalar}). Let $1$, $3$ denote the gluons and $2$,$4$ denote the scalars. The sum of
the three graphs evaluates to
\begin{equation}
 A_{4pt} \sim \epsilon_1^\mu\epsilon_3^\rho\Bigg(i \eta_{\mu \rho }+\frac{i}{z \, p_4 \cdot
q}\Big(-P_{\rho } p_{2\mu }+ P_{\mu
} p_{2\rho }\Big)\Bigg)+\mathcal{O}\left(\frac{1}{z^2}\right)
\end{equation}
where $P= p_1 + p_3$. This shows the scaling behavior of equation \eqref{r_gennonadscale}.

The analysis of the large-$z$ behavior of five legged scalar/gluon graphs is similar for all possibilities of particle combinations. Take for example particles $1$ to $3$ to be gluons and particle $4$ and $5$ to be scalars. The Feynman graphs for this choice are depicted in figure \ref{fig:5ptscalar}. The result of the sum under the non-adjacent shift is given by
\begin{equation}
A_{5pt} \sim \epsilon_1^\mu\epsilon_3^\rho\Bigg(
\frac{i \left(\eta_{\mu \nu } \left(p_{5}-p_{4}\right)_{\rho }-\eta_{\nu \rho }
\left(p_{5}-p_{4}\right)_{\mu }\right)}{2
\sqrt{2} \, z \, p_{2}\cdot q}
\Bigg)+\mathcal{O}\left(\frac{1}{z^2}\right)
\end{equation}
This result is antisymmetric in the indices of the shifted legs at order $\sim\!(z^{-1})$. Note that in contrast to the five point gluon diagrams no symmetric piece remains. The other possibilities of choosing particles yield the same result. 

At six points the number of graphs increases significantly. One can pick either two gluons and four scalars or vice versa. The first case is unimportant because the graphs one would have to consider scale as $\sim\!(z^{-2})$. Hence the only diagrams to be considered here have four gauge bosons and two scalars. They are depicted in figure \ref{fig:6pt2scalar} for a particular distribution of particles with the shift (1, 4). Summing all the graphs one finds a non-vanishing symmetric part at order $\sim\!(z^{-1})$ given by
\begin{equation}
\label{6ptscasym}
 A_{6pt sym} \sim \epsilon_1^\mu\epsilon_3^\rho\Bigg(
\frac{i (p_{5}\cdot q-p_{6}\cdot q) \left(\eta_{\mu \nu } \eta_{\rho \sigma }+\eta_{\rho \mu }
\eta_{\sigma \nu
}\right)}{4 z \, p_{2}\cdot q \, p_{3}\cdot q}\Bigg) + \mathcal{O}\left(\frac{1}{z^2}\right)
\end{equation}
This parallels the case of six gluons, where the five point gluonic graphs have to be taken into account. Since the five-scalar-gluon graphs are already antisymmetric at order $\sim\!(z^{-1})$ the missing contributions can only come from the five gluon graphs. The five point gluon diagrams leading to the sought-for cancellation are shift $(1, 4)$ diagrams with a scalar-scalar-gluon vertex contracted into leg five. The result of these graphs is given by \eqref{6ptscasym} with opposite sign and consequently the symmetric part vanishes when summed.

\begin{figure}[h!]
\centering
\includegraphics[scale=0.15]{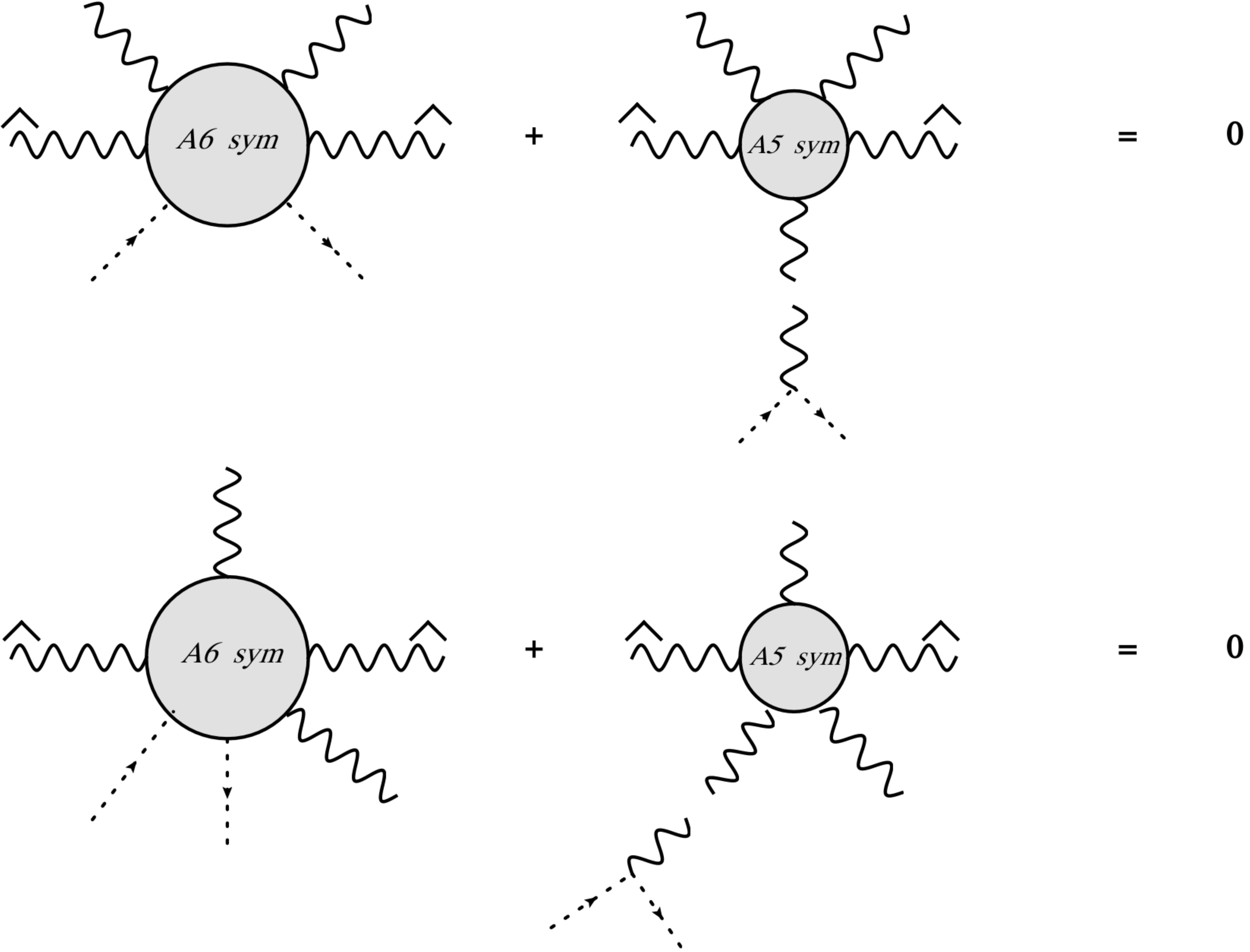}
\caption{\label{fig:5to6sca} Diagrammatics of the cancellation of the symmetric parts of the five
and six point hard line graphs for different shifts at six
points for scalar/gluon graphs.}
\end{figure}

Up to now the scalar legs have been adjacent. For non-adjacent scalar legs the sum of all diagrams has been checked explicitly to be antisymmetric at order $\sim\!(z^{-1})$ by itself at both five and six points.  Note that the all-gluon five point graphs symmetric parts uncovered above do not influence any six point graph that has two non-adjacent scalar legs. 

Therefore the large-$z$ behavior of integrands of minimally coupled scalar theories subjected a to non-adjacent shift of two gluons is given by equation \eqref{r_gennonadscale}.

\subsection{Minimally coupled fermion contributions}
In this subsection it will be shown that the scaling behavior of the integrand under a non-adjacent BCFW shift of two gluons does not change if minimal fermion-gluon couplings are included. This involves analyzing all fermion contributions to hard-line graphs up to order $\sim\!(z^{-1})$. The fermion Feynman rules are given in appendix \ref{app:feynrules}.   Note that these rules are for adjoint matter, for fundamental matter one simply restricts to the diagrams where the fermionic legs are adjacent.  

The fermion propagator scales as $\sim\!(z^0)$ along the hard line due to the occurrence of $z\,\slashed{q}$ in the numerator. On the other hand this scaling is hard to realize since $\slashed{q}$ squares to zero and $\slashed{q}$ anti-commutes with any external gluon leg. At four points there are two diagrams depicted in figure \ref{fig:4ptferm}. In the large-$z$ limit they sum to
\begin{equation}
 A_{4pt}\sim  i\epsilon_1^\mu\epsilon_3^\rho\Big(\frac{ \gamma _{\rho } p_{4\mu }-\gamma_{\mu } p_{4\rho }}{2 z \, p_4 \cdot q} \Big)+\mathcal{O}\left(\frac{1}{z^2}\right)
\end{equation}
and obey the scaling scheme of equation \eqref{r_gennonadscale}. Terms proportional to $\gamma^\mu\gamma^\rho$ at sub-leading order have been neglected since they can be rewritten as an antisymmetric tensor plus a metric since
\begin{equation}
 \gamma^\mu\gamma^\rho = \frac{1}{2} \{\gamma^\mu,\gamma^\rho\}+  \frac{1}{2} [\gamma^\mu,\gamma^\rho] =  \eta^{\mu\rho} +  \frac{1}{2} [\gamma^\mu,\gamma^\rho]
\end{equation}
by the usual Clifford algebra. 

At five points there are three diagrams depicted in figure \ref{fig:5ptferm}. Let $1$ to $3$ denote the gluons and $4$, $5$ the (anti-) fermions. Similar to the all-gluon case there is a symmetric piece left at  order $\sim\!(z^{-1})$ in the large-$z$ limit given by
\begin{equation}
 A_{5pt,ferm,sym} \sim  \epsilon_1^\mu\epsilon_3^\rho\frac{i}{4\sqrt{2} z p_2\cdot
q}\Bigg(\eta_{\mu\nu}\Big(\frac{\gamma_\rho\slashed{q}\slashed{p}_5}{p_4\cdot
q}+\frac{\slashed{p}_4\gamma_\rho\slashed{q}}{p_5\cdot q}\Big)+\mu \leftrightarrow \rho
\Bigg)\label{5ptferm_r}
\end{equation}
For a five point amplitude the symmetric part vanishes on-shell because of the Dirac equation
\begin{equation}
 \bar{u}(p_4)\slashed{p}_4=0 \quad \text{and}\quad \slashed{p}_5 u(p_5)=0.
 \end{equation} 
For off-shell legs the fermion propagator connecting to these legs collapse by
\begin{equation}
\frac{\slashed{p}_i  \slashed{p}_i }{p_i^2} = 1
\end{equation}
Hence these terms contribute to six point graphs with two fermionic legs. 

As before at six points the analysis becomes more intricate because of various possibilities for shifts and distribution of external particles. Graphs contain either two gluons and four fermions or four gluons and two fermions. The former case scales as $\sim\!(z^{-2})$ and does not need to be considered here. To understand how the five point result is needed in order to make the six point result scale correctly, take for instance the configuration particle $1$ to $4$ glue and $5$ and $6$ fermions with shift $(1, 3)$ as seen in figure \ref{fig:6ptfermother}.
\begin{figure}[h!]
\centering
\includegraphics[scale=0.15]{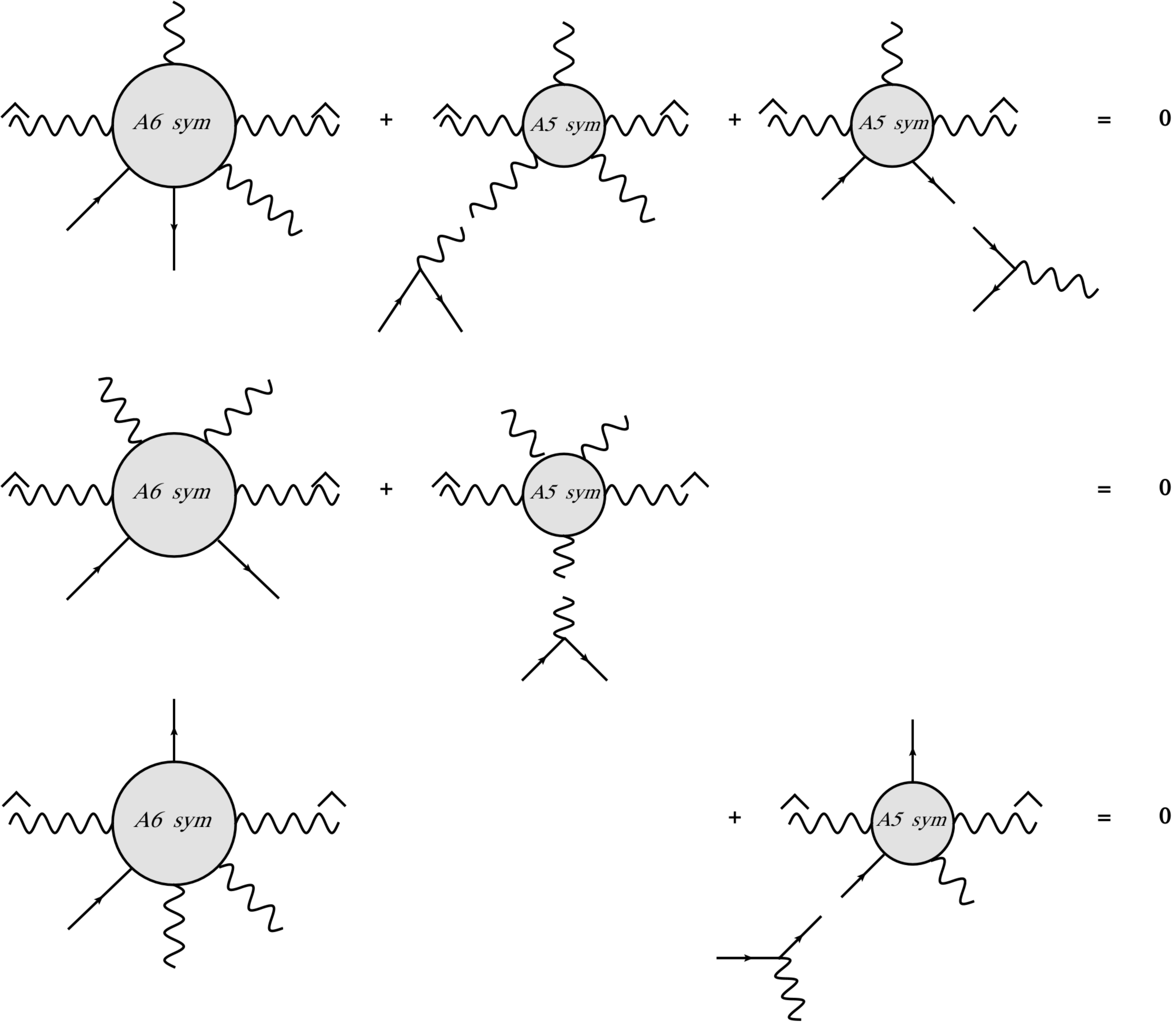}
\caption{\label{fig:5to6ferm} Diagrammatics of the cancellation of the symmetric parts of the five and six point hard line graphs for different shifts at six points for fermion/gluon graphs.}
\end{figure}

The symmetric part of this set of six point graphs is given by
\begin{equation}
 A_{6pt,ferm,sym}\sim \epsilon_1^\mu\epsilon_3^\rho\Big(\frac{i \eta_{\mu \sigma } \eta_{\rho \nu } \slashed{q}}{4 z \, {p_2}\cdot q \,{p_4}\cdot q}-\frac{i \eta_{\nu \mu } \gamma_{\sigma }\gamma_{\rho }\slashed{q}}{8 \, z\, p_2\cdot q\, p_6\cdot q} +\mu \leftrightarrow \rho \Big)
\end{equation}
Investigating the terms more closely one can already guess how they will be canceled: the first term consists purely of metrics and will therefore be canceled by a contribution from the symmetric five point gluon part of the $(1, 3)$ shift \eqref{135pointsym} upon contraction with a fermion-gluon vertex into leg five. The second term consists of three Dirac-matrices. Two of them are already present in the five point result \eqref{5ptferm_r} and a third matrix can be obtained if one adds another fermion-gluon three-vertex to the diagram as this vertex is basically only a Dirac matrix. Therefore this term will be canceled by the symmetric part of the (1, 3) fermion graphs discussed above upon contraction with a fermion-gluon vertex on leg 4 (see figure \ref{fig:5to6ferm}). The propagator will collapse since
\begin{equation}
\frac{\slashed{p}_i}{p_i^2}\,\slashed{p}_i=1
\end{equation}
and one obtains the desired result. The term in equation \eqref{5ptferm_r} proportional to $\slashed{p}_5$ does not play a role here but at other diagrams and can therefore be neglected. This mimics once more the logic behind the all-gluon graphs discussed above. The computations for the same particle configuration with shift $(1, 4)$ are similar except that there is no contribution of the symmetric fermion five point hard line graphs. The symmetric part is canceled by the terms coming from the five point gluon graphs alone depicted in figure \ref{fig:5to6ferm}).

As in the case of the scalars, there is the possibility that the fermions might be chosen non-adjacent. In this situation there will not be a contribution from the symmetric part of the purely gluonic hard line graphs because, as said previously, these kind of diagrams cannot be constructed from the gluon diagrams.

 In conclusion the large-$z$ behavior of integrands of minimally coupled fermion theories subjected a to non-adjacent shift of two gluons is given by equation \eqref{r_gennonadscale}.

\subsection{Scalar potential and Yukawa terms}
The above discussion can be generalized further to include scalar potential and Yukawa terms. For $\phi^3$ and $\phi^4$ type couplings for example the scaling behavior of equation \eqref{r_gennonadscale} can be easily checked by powercounting. The inclusion of Yukawa couplings follow simply by the observation that Yukawa couplings can be generated by considering a Yang-Mills theory minimally coupled to fermions in one dimension higher than the one under study. Then the momenta in this extra direction are all restricted to vanish. All Feynman graphs have been analyzed already in one dimension higher in the text above, leading to the scaling displayed in equation \eqref{r_gennonadscale}. The extra graphs in the number of dimensions under study are exactly those given by Yukawa couplings.

\begin{figure}[t]
\centering
 \includegraphics[scale=0.5]{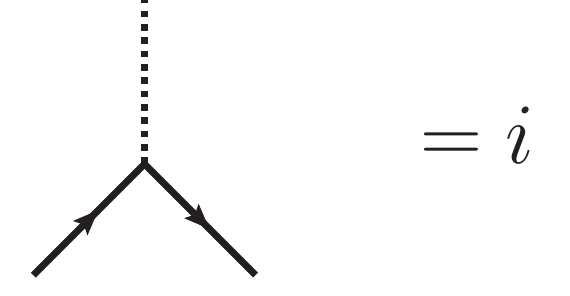}
\caption{The Feynman rule for Yukawa coupling. The black arrow lines represent fermions, the dotted line a scalar. \label{fig:yukawarule}}
\end{figure}
\begin{figure}[h!]
\centering
 \includegraphics[scale=0.5]{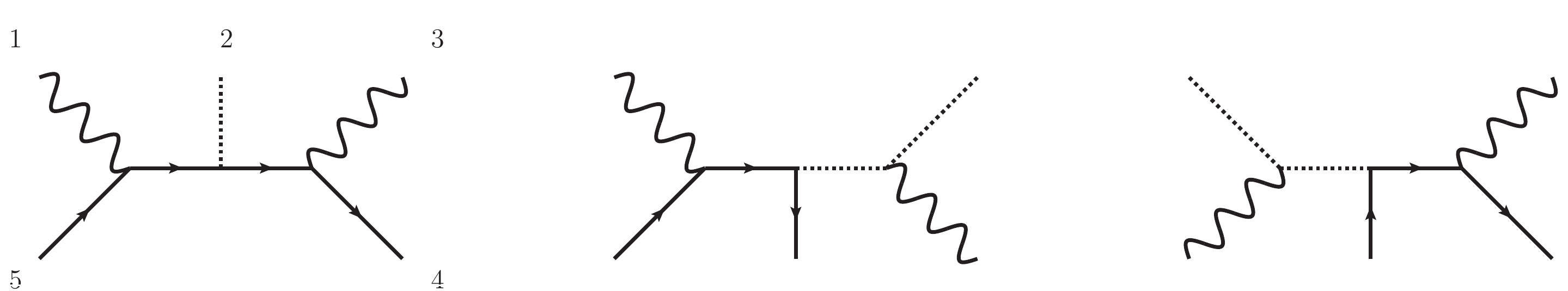}
\caption{Feynman diagrams at five points for Yukawa couplings. \label{fig:yukacou}}
\end{figure}

This can of course also be verified directly. Yukawa couplings couple a pair of fermion lines to a scalar particle. The corresponding color ordered Feynman rule is given in figure
\ref{fig:yukawarule}. As this section deals with gluon shifts only, there are no diagrams at four points to consider. The graphs at five points consist of two gluons, a fermion anti-fermion pair and a scalar. An example choice of ordering the external particles is depicted in figure \ref{fig:yukacou}, other choices will lead to basically the same computation.
When summed these graphs  give an antisymmetric expression at order $\sim\!(z^{-1})$ given by
\begin{equation}
 A_{\text{Yukawa, 5pt}}=-i\epsilon_1^\mu\epsilon_3^\rho\frac{\Big(p_2^\mu\gamma^\rho-p_2^\rho\gamma^\mu \Big)\slashed{q}}{4 z \,  p_2\cdot q \,  p_5\cdot q}
\end{equation}
This leaves the six point graphs. For instance the graphs which arise by adding a gluon directly on the scalar vertex in figure \eqref{fig:yukacou} get cancelled by adding a gluonic three vertex graph to the shifted gluon legs.

\subsection*{Conclusion}

It has been proven in this section that BCFW shifts of two non-color adjacent gluons on integrands of Yang-Mills theories minimally coupled to scalar or spin $1/2$ matter with possible scalar or Yukawa terms scale as given in equation \eqref{r_gennonadscale}. The calculation although conceptually straightforward is more intricate compared to the color adjacent case. As explained in appendix \ref{app:othermatters}, the above results on shifts of two gluons can be used in a supersymmetric field theory to obtain shifts of certain fermions and scalar pairs.


\section{BCJ relations for the one loop integrand}\label{sec:BCJatoneloop}

One of the main motivations to study non-adjacent shifts for loop level integrands is that the improved BCFW shift behavior found in section \ref{sec:nonadjshftintegr} immediately implies the existence of `bonus relations'  for the integrand through equation \eqref{eq:bonusrelfromBCFW}. In this section a particular generalization to the one loop integrand is proposed. 

\subsection{Review of improved BCFW shifts and BCJ relations at tree level}\label{eq:treelevelBCJfromBCFW}
First the argument of \cite{Feng:2010my} for the tree level derivation of the BCJ relations will be reviewed in a slightly different setup\footnote{The structure of the following tree level derivation was pointed out to us by Michael Kiermaier.}. Consider the following sum of tree level amplitudes,
\begin{equation}\label{eq:BCJtreelevel}
R = \sum_{i=1}^{n-1} k_\alpha \cdot (k_{n-1} + \sum_{j=1}^{i-1} k_j) A(1\ldots i-1,\alpha,i,\ldots, n-1)
\end{equation}
with the convention $k_0=0$. This can be disentangled into the $U(1)$ decoupling relation for the coefficient of $k_\alpha \cdot k_{n-1}$ plus the BCJ relation,
\begin{equation}
\sum_{i=2}^{n-1} k_\alpha \cdot (\sum_{j=1}^{i-1} k_j) A(1\ldots i-1,\alpha,i,\ldots, n-1) = 0
\end{equation}
This particular form arises naturally in superstring theory as the order $\alpha'$ terms in the relation
\begin{equation}\label{eq:stringrelBCJ}
\sum_{i=1}^{n-1} e^{2 \ii \alpha' k_\alpha \cdot (\sum_{j=1}^i k_j)} A(1\ldots i-1,\alpha,i,\ldots, n-1) =0 
\end{equation} 
derived in various places \cite{Plahte:1970wy, BjerrumBohr:2009rd, Stieberger:2009hq, Boels:2010bv}. Hence the quantity $R$ in equation \eqref{eq:BCJtreelevel} vanishes. The point of writing the BCJ relation in this form is that $R$ manifestly has good BCFW shifts under a shift of $1$ and $n-1$. The only possible spoiler of this is the $z$-dependent coefficient of the first term, 
\begin{equation}
\sim k_{\alpha} \cdot k_{n-1}  A(\alpha, 1, \ldots, n-1)
\end{equation}
but this multiplies the one and only non-adjacently shifted amplitude. Hence for every helicity combination of particles $1$ and $n-1$, a BCFW shift exists such that on-shell recursion holds and the quantity $R$ defined in equation \eqref{eq:BCJtreelevel} can be reconstructed from its singularities by on-shell recursion.

To proof $R=0$ and hence the BCJ relations a recursive argument can now be set up. Note first that the base step for three particle amplitudes holds trivially as all Lorentz kinematic invariants vanish in this case. Next note that despite appearances $R$ is cyclic, hence to study all of its kinematic singularities it is enough to consider all subsets $1,\ldots m$ for some value of $m$. The pole at this singularity can be written,
\begin{equation}
\lim_{(k_1 + \ldots k_m)^2 \rightarrow 0} R \sim \frac{A_L(1,\ldots,m, X)}{(k_1 + \ldots k_m)^2} \left[\sum_{\sigma \in OP({\alpha}\cup \{m+1, \ldots n-1 \})} (k_\alpha \cdot (k_X + k_{<\alpha})) A_R(X,\sigma)  \right]
\end{equation}
here a sum over helicities has been suppressed. The momentum $k_{<\alpha}$ is the sum over all momenta between the particle $\alpha$ and particle $X$. This of course simply spells the quantity $R$ from equation \eqref{eq:BCJtreelevel} for a restricted set of tree amplitudes with a strictly lower number of particles than the original case. Note that the structurally the proof can be applied if all momenta in the coefficients would be set to unity: this proves the $U(1)$-decoupling relation. Hence by the existence of on-shell recursion relations in $D \geq 4$, this furnishes a proof of the BCJ relations in these numbers of dimensions. Since the improved large $z$ behavior has been shown to hold in gauge theories coupled to various forms of matter in the adjoint, the BCJ relation also holds for all amplitudes in these theories.  

\subsection*{Extension of the BCJ relations to the integrand at one loop}
In section \ref{sec:nonadjshftintegr} it was shown the integrand of any minimally coupled gauge theory with possible scalar potential and Yukawa terms scales under a non-adjacent shift one order of $z$ better than an adjacent shift of the same quantity. It can therefore be expected BCJ-type relations exists at any loop order. A problem for a direct derivation is that a pretty clear idea is needed what these relations look like before they can be proven. Leaving the higher loop case to future work, here the focus will be on one loop. Some experimentation with on-shell recursion relations for the integrand suggest to study the following combination of integrands,
\begin{equation}\label{eq:BCJloopintegr}
\sum_{i=1}^{n-1} k_\alpha \cdot (l+k_{n-1} + \sum_{j=1}^{i-1} k_j) I(1\ldots i-1,\alpha,i,\ldots, n-1)
\end{equation}
Here $I(\sigma)$ is the integrand of the one loop planar color ordered amplitude $A(\sigma)$. The loop momentum is chosen such that the propagator after the point where particle $n-1$ attaches to the loop is of the form $1/(l+k_{n-1})^2$.  A choice for the loop momentum  is needed to fix the relative normalization of the loop momentum dependent pre-factor compared to the amplitude integrand. This particular choice is suggested by on-shell recursion for the integrand. In the expression above the loop momentum can of course be shifted as required up to terms which vanish after integration. Note\begin{equation}
 \sum_{i=1}^{n-1} k_\alpha \cdot (l+k_{n-1}) I(1\ldots i-1,\alpha,i,\ldots, n-1) = I(1,\ldots,n-1;\alpha) 
 \end{equation}
holds where the quantity on the right hand side is a non-planar one loop integrand. With the indicated definitions it will be shown below
\boxit{\begin{equation}\label{eq:BCJatoneloop}
\sum_{i=1}^{n-1} k_\alpha \cdot (l+k_{n-1} + \sum_{j=1}^{i-1} k_j) I(1\ldots i-1,\alpha,i,\ldots, n-1) = 0
\end{equation}}
holds for the one loop integrand of any gauge theory with adjoint matter on the external lines, including pure Yang-Mills. The zero here is up to terms which vanish after integration.  Note that the left hand side of equation \eqref{eq:BCJatoneloop} scales as $\sim\!(z^{-1})$ under a BCFW shift of particles $1$ and $n-1$. 

\subsubsection*{Simple example}
As a quick sanity check, consider the four point one loop scattering amplitude in $\mathcal{N}=4$ super Yang-Mills theory. This amplitude can be written as
\begin{equation}
A^1_{4} \sim s t A^0_{4} \int d l^d \frac{1}{l^2 (l+k_4)^2 (l+k_4+k_1)^2 (l-k_3)^2}
\end{equation}
up to unimportant numerical factors. The factor $s t A^0_{4}$ is completely symmetric. Hence the sum in equation \eqref{eq:BCJatoneloop} boils down to 
\begin{multline}
\frac{k_1 \cdot (l+k_4) }{l^2 (l+k_4)^2 (l+k_4+k_1)^2 (l-k_3)^2} + \frac{k_1 \cdot (l+k_4+k_2) }{l^2 (l+k_4)^2 (l+k_4+k_2)^2 (l-k_3)^2}\\ + \frac{k_1 \cdot (l) }{l^2 (l+k_4)^2 (l+k_4+k_2)^2 (l-k_1)^2}
\end{multline}
Using $k_1 \cdot (l+k_4) = \frac{1}{2}((l+k_4 + k_1)^2 - (l + k_1)^2)$ and similar for all terms it is seen this particular sum vanishes up to terms which integrate to zero. This verifies \eqref{eq:BCJatoneloop} in this particular case. 

\subsection{Proof by standard unitarity cuts}
As in the tree level case one can consider all kinematic singularities of the expression in equation \eqref{eq:BCJatoneloop}. Focus first on standard two particle unitarity cuts of the quantity in equation \eqref{eq:BCJatoneloop}. By this the usual procedure of replacing two propagators in the integrand by delta functions is meant,
\begin{equation}
\frac{1}{(l + k_x + k_R)^2} \frac{1}{(l + k_x )^2} \rightarrow \delta^D(l + k_x + k_R)^2  \delta^D(l +k_x )^2 
\end{equation}
where the choice of the two propagators split the diagram into two halves. The incoming momentum on one half is $k_R$. The momentum $k_x$ appears since above a choice was made for the loop momentum $l$.  As was also done above a particular choice of which particles are on one side of the cut is referred to as a channel. A standard result obtained by simply considering Feynman graphs expresses the unitarity cut of the planar color-ordered integrand of an amplitude in a particular channel at one loop in terms of a product of tree level amplitudes, 
\begin{equation}
\textrm{Cut}_{\alpha_R } I^1(\alpha_R \alpha_L) = \sum_{\textrm{spin,species}} A^0(l_2, \alpha_L, -l_1) A^0(l_1, \alpha_R, - l_2)
\end{equation}
where the $l_i$ are the on-shell momenta
\begin{equation}
l_1 = l+k_x \qquad l_2 = l + k_x + k_R
\end{equation}
and the sum ranges over all particles in the loop which have propagators of this form as well as the spin or helicity quantum numbers of these particles. The unknown momentum $k_x$ is the sum over all momenta from one of the cut loop momenta leading up to particle $n-1$ in the convention for the loop momentum employed here. In this subsection it will be shown expression \eqref{eq:BCJatoneloop} does not have any non-vanishing two particle cuts. 

To cut down on calculational work it is useful to show equation \eqref{eq:BCJatoneloop} is cyclic. The only subtlety is the loop momentum: consider rotating the first $n-1$ integrands by one unit, and the last integrand by two units. The relation of equation \eqref{eq:BCJatoneloop} is invariant if the loop momentum is also shifted $l  \rightarrow l + k_{n-2}$. The latter shift preserves the above choice of loop momentum. Hence the study of all $2$ particle cut channels can be restricted to those parametrized by the choice of the set of particles $\{1, \ldots, m\}$ on one side of the cut. The special particle $n-1$ is on the side of the particle $\alpha$, except the case $m=n-1$. Consider first the case $1 \leq m < n-1$. In this case the momentum $l$ contracted into $k_\alpha$ can be expressed in terms of momentum $l_1$ as
\begin{equation}
k_\alpha \cdot (l+k_{n-1}) = k_{\alpha} \cdot (l_1 + \sum_{j=m+1}^{n-1} {k_j}) 
\end{equation}

Consider the double cut in the channel $(1, \ldots, m)$ of the sum over amplitudes in equation \eqref{eq:BCJatoneloop}. This yields
\begin{multline}
\sum_{\textrm{spin,species}} A^0(l_2, 1,\ldots, m, -l_1) \left(  k_{\alpha} \cdot (l_1 + \sum_{j=m+1}^{n-1} {k_j}) A^0(l_1,m+1,\ldots,n-1,\alpha, - l_2) \right. + \\ \left. \sum_{j=m+1}^{n-1} k_\alpha \cdot (l_1 + \sum_{i=m+1}^{j-1} k_i )A^0(l_1, m+1,\ldots,j-1,\alpha,j,\ldots,n-1, - l_2) \right)
\end{multline}
which can recognized as an expression proportional to the BCJ relation for the tree level amplitude in equation \eqref{eq:BCJtreelevel}. 

Now consider the exceptional case $m=n-1$. In this case every term in \eqref{eq:BCJatoneloop} contributes. Consider the first term ($i=1$) in equation \eqref{eq:BCJatoneloop}. The double cut of this term in the exceptional channel reads
\begin{multline}
\sum_{\textrm{spin,species}} k_{\alpha}\cdot \left(l_2 + k_{n-1} + \sum_{j=1}^{n-2} k_{i} \right)  A(l_2, 1,\ldots, n-1, -l_1)  A(l_1, \alpha, -l_2) = \\ \sum_{\textrm{spin,species}} k_{\alpha}\cdot \left(l_2 \right)  A(l_2, 1,\ldots, n-1, -l_1)  A(l_1, \alpha, -l_2) =0
\end{multline}
where the last zero follows by momentum conservation. The second term  ($i=2$) in equation \eqref{eq:BCJatoneloop} differs from this in minor ways as the particle $1$ now appears on the right of particle $n-1$. This influences the calculation of $l$ in terms of $l_2$ since momentum $1$ does not appear any more. However, it is reinserted explicitly in equation \eqref{eq:BCJatoneloop}. Hence the double cut in this channel of every term in equation \eqref{eq:BCJatoneloop} vanishes. 

In conclusion, all $D$-dimensional double cuts of the sum in  \eqref{eq:BCJatoneloop} vanish. By Cutkosky's observations \cite{Cutkosky:1960sp}, the fact that the quantity in equation  \eqref{eq:BCJatoneloop} does not have two particle cuts implies that after integration the resulting function does not have any branch cut singularities, assuming the only branch cuts of this expression are physical. Hence up to terms which vanish after integration, the sums of equation \eqref{eq:BCJatoneloop} must yield after integration a rational function of the external momenta and helicities.

Since the loop momenta and cuts are in $D$ dimensions, it will be impossible to construct a non-trivial rational function of this type. This can be made more precise as the rational function which remains can have no tree level pole singularities. For this consider one of the kinematic singularities as
\begin{multline}\label{eq:kinmasingBCJone}
\lim_{(k_1 + \ldots k_m)^2 \rightarrow 0} R \sim\\ \frac{A^0_L(1,\ldots,m, X)}{(k_1 + \ldots k_m)^2} \left[\sum_{\sigma \in OP({\alpha}\cup \{m+1, \ldots n-1 \})} (k_\alpha \cdot (l+k_{n-1} +k_X + k_{<\alpha})) I^1_R(X,\sigma)  \right]
\end{multline}
where the quantity on the right hand side is of course the sum of equation \eqref{eq:BCJatoneloop} for a one loop integrand with strictly less particles. Since the tree amplitude does not contain loop momenta, both left and right hand side of this equation can be integrated.  This therefore turns into a relation for the possible rational function on the right hand side of equation \eqref{eq:BCJatoneloop}. However, the three particle version for instance
\begin{equation}
\sim k_\alpha \cdot (l+k_2) \left(I(\alpha 1 2) + I(1 \alpha 2)\right)
\end{equation}
must vanish since in color ordered perturbation theory the three particle integrand is anti-symmetric. Hence the four particle version of equation \eqref{eq:BCJatoneloop} must sum to a polynomial function of the external momenta and helicities, assuming there are no non-physical singularities. By standard dimensional analysis, this cannot exist. Hence up to assumptions equation \eqref{eq:BCJatoneloop} holds for four particles. This can be iterated to show that up to the physically reasonable assumptions mentioned equation \eqref{eq:BCJatoneloop} holds for all multiplicity. 

\subsection{Minimal basis for integrands}

Leaving a general formula for future work it will be demonstrated here how to express $(n-1)!/2$ integrands in a basis of $(n-2)!$ integrands using the relation of equation \eqref{eq:BCJatoneloop}. Label all $n$ particles $1,\ldots, n$. First fix the position of the particle $n$ on the last position without loss of generality. In addition, this allows us to pick the same convention for the loop momentum for all integrands. The set of all integrands is given by
\begin{equation}\label{eq:nminusone}
I(\sigma,n) \qquad \sigma \in P(\{1,\ldots, n-1\}]
\end{equation} 
The basis sought for is the one where particle $1$ is adjacent to $n$,
\begin{equation}\label{eq:nminustwo}
I(1,\sigma,n) \qquad \sigma \in P(\{2,\ldots, n-1\}]
\end{equation} 
Note that $I(\sigma,1,n)$ is also an element of this basis by the reflection property in equation \eqref{eq:reflprop}. Every element in the set in equation \eqref{eq:nminusone} can be classified according to the distance between particles $1$ and $n$. In the set of equation \eqref{eq:nminustwo} this distance is zero. For a distance of one particle, one can use equation \eqref{eq:BCJatoneloop} directly to express everything into the set in equation \eqref{eq:nminustwo}. Consider the case of a distance of two particles which will be labelled, say, $2$ and $3$.  From equation \eqref{eq:BCJatoneloop} one can construct the following system of equations,
\begin{align}
k_2 \cdot (l+k_{n} ) I(2,3,1,\sigma,n) + k_2 \cdot (l+k_{n} + k_3) I(3,2,1,\sigma,n) & = f(3;2) \\
k_3 \cdot (l+k_{n} ) I(3,2,1,\sigma,n) + k_3 \cdot (l+k_{n} + k_2) I(2,3,1,\sigma,n) & = f(2;3)
\end{align}
where $\sigma$ stands for some permutation of $\{4,\ldots, n-1\}$. The function $f$ indicates the remaining sum generated in equation \eqref{eq:BCJatoneloop}. For our purposes here it suffices that this is a sum over integrands where the distance between particles $1$ and $n$ is only one particle. The determinant of the system is
\begin{equation}
\textrm{Det} = - k_2 \cdot k_3 \left((l +k_n)\cdot (k_2 + k_3) + k_2 \cdot k_3 \right)
\end{equation}
which vanishes only
\begin{equation} 
\textrm{if} \quad k_2 \cdot k_3 =0  \quad  \textrm{or}  \quad  (l +k_n)\cdot (k_2 + k_3) + k_2 \cdot k_3 = 0
\end{equation}
For generic kinematics this system of linear equations therefore admits a unique solution. This allows one to express all integrands with $2$ particles between $1$ and $n$ to be expressed in the basis given in equation \eqref{eq:nminustwo}. 

The previous argument for a distance of two particles between $1$ and $n$ can be generalized to distances of $i$ particles. Without loss of generality, label the particles between $1$ and $n$ $2,\ldots, i$ for some $i < n/2$. Now a system of equations can be constructed as all $i!$ permutations of the particles $\{2,\ldots, i\}$ in the following equation
\begin{multline}
k_2 \cdot (l+k_{n} ) I(2,3, \ldots, i,1,\sigma,n) + k_2 \cdot (l+k_{n} + k_3) I(3,2,\ldots,i,1,\sigma,n)+ \ldots \\ + k_2 \cdot (l+k_{n} + \sum_{j=3}^{i} k_j) I(3,\ldots,i,2,1,\sigma,n)  = f(\{3,\ldots,i\}; 2)
\end{multline}
The right hand side consists of integrands with a permutation of the set $\{3,\ldots,i\}$ between particles $1$ and $n$. This set contains one less particle than $i$. The system of equations contains $i!$ equations for $i!$ variables and can in generic kinematics be solved uniquely, assuming there are no accidental degeneracies. We have checked the latter numerically in four dimensions up to $i=7$. Up to this subtlety this shows one can in general express the $(n-1)!/2$ integrands of  equation \eqref{eq:nminusone} in the basis of $(n-2)!$ integrands given by equation \eqref{eq:nminustwo}. Finding general and effective expressions for this reduction is left to future work.

\subsection*{Comments}

The relation in  equation \eqref{eq:BCJatoneloop} is independent of dimensionality in principle. Furthermore, in the above the precise field content of the theories under study was left deliberately vague: equation \eqref{eq:BCJatoneloop}  is expected to apply to all  minimally coupled gauge theories with possible scalar potential and Yukawa terms in four and higher dimensions with adjoint matter on the external lines. Note that for instance fundamental matter in the loop can easily be related to adjoint matter in the loop. Special cases include pure Yang-Mills and maximally supersymmetric field theory. Similar relations are expected for fermions in the adjoint on the external lines with extra minus signs for all interchanges of fermions, see \cite{Sondergaard:2009za} for the tree level case of this. 

For practical (phenomenological) purposes equation \eqref{eq:BCJatoneloop} is not effective as given as it involves the loop momenta: most standard approaches to loop amplitudes involve reduction to a scalar integral basis such as that in equation \eqref{eq:massiveloopampexp}. For practical applications it would for instance be very interesting to derive consequences of equation \eqref{eq:BCJatoneloop} for the integrated amplitudes. 

The relation for the integrand in equation \eqref{eq:BCJatoneloop} can also be proven using on-shell recursion for the integrand of supersymmetric Yang-Mills in four dimensions \cite{ArkaniHamed:2010kv} and \cite{Boels:2010nw}. This is a straightforward generalization of the argument in subsection \eqref{eq:treelevelBCJfromBCFW}. For this particles $1$ and $(n-1)$ are shifted in equation \eqref{eq:BCJatoneloop} and the definition of the residue at the single-cut singularity of the integrand is taken from the suggestion in \cite{CaronHuot:2010zt}. Note the pole terms will work out courtesy of equation \eqref{eq:kinmasingBCJone}.  Crucial in the derivation is the result from section \ref{sec:BCJatoneloop} that the integrand scales better under a non-adjacent shift than under an adjacent shift. However, making the on-shell recursive argument precise leads too far beyond the present article.


\section{Generalizing non-adjacent shifts for tree level amplitudes}\label{sec:gennonadj}

Given the above results motivated from improved BCFW shift behavior for non-adjacent shifts, a natural question is then if even more improved BCFW shift behavior can be achieved. Two examples are known in the literature which involve such an improvement of shift behavior compared to the naive one. These are Einstein gravity and QED \cite{Badger:2008rn}. Note that both of these are un-ordered theories: there is no equivalent of color ordering for gravitons or photons so amplitudes involve sums over all orders of the bosonic particles. Below two related mechanisms are discussed in pure Yang-Mills theory  which can improve the large BCFW shift behavior. These will involve permutation sums closely related to the QED case as well as certain cyclic sums. 

\subsection{Improved large shift behavior from permutation sums}
For inspiration and orientation consider again the MHV amplitude at tree level,
\begin{equation}
A^{\textrm{MHV}} = \frac{\braket{i j}^4}{\braket{12} \ldots \braket{n 1}}
\end{equation}
up to unimportant numerical constants. Particle $i$ and $j$ are the opposite helicity gluons. Motivated by QED one can consider a partial permutation sum over all particles $2,3,\ldots n-1$ over this amplitude, e.g.
\begin{equation}\label{eq:examppermsumMHV}
\sum_{\textrm{P}(\{2,3,\ldots,n-1\})}  \,\, A^{\textrm{MHV}} = \braket{i j}^4 \sum_{\textrm{perms }
2,3,\ldots,n-1} \frac{1}{\braket{12} \ldots \braket{n 1}}
\end{equation}
The permutation sum on the right is the same as that used to obtain QED amplitudes from QCD ones and
has a known simple form,
\begin{equation}
\sum_{\textrm{P}(\{2,3,\ldots,n-1\})} \frac{1}{\braket{12} \ldots \braket{n-1 n}} = \frac{1}{\braket{n 1}} \prod_{i=2}^{n-1} \frac{\braket{1n}}{\braket{1i}\braket{i n}}
\end{equation}
This equation can be neatly proven by induction using on-shell recursion. For this one uses the facts that the left hand side obeys on-shell recursion relations and that left and right hand side have the same pole structure. Hence
\begin{equation}
\sum_{\textrm{P}(\{2,3,\ldots,n-1\})}   \,\, A^{\textrm{MHV}} = \frac{\braket{i j}^4 }{\braket{n 1}^2} \,\,\prod_{i=2}^{n-1} \frac{\braket{1n}}{\braket{1i}\braket{i n}}
\end{equation}
From this equation it can be seen this amplitude shifts as $\sim\!(z^{-n+2})$ under a BCFW shift  of legs $1$ and $n$ which boils down to a shift of the holomorphic spinors
\begin{equation}
 |1\rangle \rightarrow |1\rangle \quad |n\rangle \rightarrow |n\rangle + z|1\rangle
\end{equation}
Permutation sums over smaller sets can also be considered, for instance
\begin{multline} \label{eq:usefullsumforphot}
\sum_{P(\beta)} A^{\textrm{MHV}}(1,\alpha, n, \beta)  = \frac{ \braket{i j}^4 }{\braket{1 \alpha_1} \ldots \braket{\alpha_{\# \alpha} n}\braket{n 1}} \left(\prod_{i=1}^{\# \beta} \frac{\braket{1n}}{\braket{1\beta_i}\braket{\beta_i n}}\right) \\
= \sum_{P(\beta)} \sum_{\sigma \in OP(\alpha^T \cup \beta)} A^{\textrm{MHV}}(1,\sigma, n)
\end{multline}
where the last line follows by the Kleiss-Kuijf relations \eqref{eq:KKrel}. This amplitude shifts under a BCFW shift as $\sim\!(z^{-k-1})$ as long as the set $\alpha$ is not empty. 

These observations on MHV amplitudes admit a wide generalization. We have checked numerically using GGT \cite{Dixon:2010ik} up to eight particles that permutation sums over NMHV amplitudes show the same behavior. Further note that the non-adjacent shifts analyzed above in section \ref{sec:nonadjshftintegr} in $D\geq4$ can be interpreted as a permutation sum of one particle. This leads to the following
\begin{suspicion}\label{susp:permsumimprov}
The power of $z$ fall-off of a BCFW shift of two particles on either side of a permutation sum over
$k$ legs of a  color ordered  tree level Yang-Mills amplitude in $D \geq 4$ dimensions is suppressed
by $\sim\!(z^{-k+1})$ compared to same shift without the permutation sum. 
\end{suspicion}
There is a natural extension of this suspicion to correlation functions and integrands. Concretely, it implies that for a shift of particles $i$ and $j$ of a purely gluonic color ordered tree amplitude
\begin{equation}
\sum_{P(\{i+1,\ldots,j-1\})}  A(z)\sim \epsilon_i^{\mu}\epsilon_j^{\rho}\left(\frac{1}{z^{k}}\right)
\Bigg(z^1 \eta_{\mu\rho}f_1(1/z)+z^0 B_{\mu\rho}(1/z)+\mathcal{O}\left(\frac{1}{z^2}\right)\Bigg)
\label{eq:gennonadscale}
\end{equation}
is suspected to hold where $k$ is the number of particles in the permutation sum, as long as $i$ and $j$ are not adjacent. If they are adjacent, the pre-factor becomes 
\begin{equation}
\left(\frac{1}{z^{k}}\right) \rightarrow \left(\frac{1}{z^{k-1}}\right)
\end{equation}
The case where $|i-j-1|=1$ is of course the non-adjacent shift studied in the previous section for the more general case of integrands.

\subsection*{Interpretation of permutation sums as a choice of color basis and naive powercounting}
It is useful to note that the partial permutation sums discussed above have a physical interpretation in terms of the color quantum numbers for the group $U(N)$. This follows most easily in the following basis,
\begin{equation}\label{eq:explcolorbas}
\begin{array}{ll}
\left(h_i\right)_{kl} & =  \delta_{ik} \delta_{il} \equiv e_{ii}\\ 
\left(e_{ij}^+\right)_{kl} & = \delta_{ij} \delta_{jl} \qquad i>j  \\
\left(e_{ij}^-\right)_{kl} & = \delta_{ij} \delta_{jl} \qquad i<j  
\end{array}\end{equation} 
Since the $\pm$ labels are superfluous they are sometimes suppressed. Note that 
\begin{equation}\label{eq:conjbas}
\left(e_{ij} \right)^{\dagger} = e_{ji}
\end{equation}
In general
\begin{equation}\label{eq:multgenun}
e_{nk} e_{lm} = \delta_{kl} e_{nm}
\end{equation}
holds. In string theory the row and column labels in this basis have the interpretation of labeling the branes where a string starts and ends. One can assign all adjoint valued particles in a scattering process  a definite color quantum number (i.e. one definite matrix from the set in equation \eqref{eq:explcolorbas}). Consider the full amplitude for a scattering process defined as a sum over non-cyclic permutations of the color ordered amplitudes 
\begin{equation}
A^{\textrm{full}} = \sum_{\sigma \in P(n)/\mathbb{Z}_n} A^{\textrm{co}} \textrm{tr}
\left(T^{\sigma(1)} \ldots T^{\sigma(n)}  \right)
\end{equation}
For specific choices of quantum numbers this permutation sum simplifies dramatically, courtesy of equation \eqref{eq:multgenun}. Assigning for instance color quantum numbers
\begin{equation}\label{eq:exampcolorassign}
1 \leftrightarrow e_{ij} \qquad n \leftrightarrow e_{ji} \qquad \{2,\ldots,n-1\} \leftrightarrow h_j
\end{equation}
gives
\begin{equation}
A^{\textrm{full}} = \sum_{\sigma \in P(\{2,3,\ldots,n-1\})} A^{\textrm{co}}(1,\sigma,n)
\end{equation}
which one can recognize as the permutation sum in equation \eqref{eq:examppermsumMHV}. Other permutation sums can be generated by assigning different color charges. In general the Cartan sub-algebra elements $h_i$ generate the permutation sums over the particles with these quantum numbers. Note that this color basis is closely related to the more basis-independent approach of  \cite{Zeppenfeld:1988bz}, as well as the brane picture of scattering amplitudes. 

The above color structure can be neatly merged with color ordered perturbation theory to continue the analysis of suspicion \ref{susp:permsumimprov}. In color ordered perturbation theory the Yang-Mills three and four vertices are associated with color factors,
\begin{equation}
\textrm{tr} \left(T^1 T^2 T^3  \right) \qquad \textrm{and} \qquad \textrm{tr} \left(T^1 T^2 T^3 
T^4\right)
\end{equation}
Hence the only non-trivial couplings in the above color basis involve the generators
\begin{equation}
\{e_{ij}, e_{jk}, e_{ki} \}  \qquad \textrm{and} \qquad \{e_{ij}, e_{jk}, e_{kl}, e_{li} \} 
\end{equation}
where of course indices are allowed to coincide. However, there are no three particle couplings which involve minimally two elements of the Cartan sub-algebra, as well as no four particle vertices which involve minimally three elements of the same sub-algebra. Hence in the above diagrammatic analysis of large BCFW shifts \emph{the particles in the permutation sums have to couple directly to the hard line}. It is this observation which makes powercounting simpler. For $k>0$ particles in the permutation sum this immediately improves the leading power of the large shift behavior for $(k<n-2)$ from the naive $\sim\!(z)$ to $\sim\!(z^{- 1 - [k/2]})$ where the brackets indicate the largest integer smaller or equal to $k/2$. This can be easily seen since naively the leading graphs consist of the maximal number of four-vertices and minimal number of propagators. For the exact scaling and further cancellations more work is required. 

The improvement in power-counting can also be seen in a color-basis independent way at tree level. The hard-line graphs can readily be written down in color ordered perturbation theory. At tree level all lines attaching to the hard line will involve a one-leg off-shell current. Now note that the permutation sum over this current vanishes
\begin{equation}\label{eq:currdecoup}
\sum_{\sigma \in  P(\{1,2,\ldots n\})} J^{\mu}(P ; \sigma) = 0
\end{equation} 
The only exception to this is when the current contains only one leg so that the corresponding particle attaches directly to the hard line. Hence in the permutation sum of suspicion \ref{susp:permsumimprov} the permuted particles should all connect to the hard line directly just as was concluded earlier from the explicit color basis argument. Note equation \eqref{eq:currdecoup} is a special case of
\begin{equation}\label{eq:U1currentvan}
\sum_{\sigma \in  POP\{\alpha \} \cup\{1,2,\ldots n\}} J^{\mu}(P ; \sigma) = 0
\end{equation}
where the ordered product is over all unions of the indicated sets with the order of the particles in the second set preserved. This is the $U(1)$ decoupling relation for currents. The equation \eqref{eq:currdecoup} can be obtained from the latter by replacing all gluons by photons.

\subsection{Improved large shift behavior in the background field method}\label{subsec:largzpermlegs}
It can be readily verified from the calculations in section \ref{sec:nonadjshftintegr} that a completely off-shell analysis of all Feynman graphs will be complicated. For two permuted particles for instance one would need to calculate up to $\sim\!(z^{-2})$ contributions which involves up to eight point graphs. For the rest of this section only tree level amplitudes will be considered. This restriction enables the background field method of \cite{ArkaniHamed:2008yf} for analyzing large BCFW shifts reviewed  in section \ref{sec:intro}. 

This method is to study the hard line graphs generated by  background gauge Lagrangian of equation \eqref{bgflag} with the soft field in AHK gauge, see equation \eqref{ahkgauge}. A first simplification follows from the simple propagator for the hard fields which simply reads
\begin{equation}
: a_{\mu} a_{\nu} : =- i \frac{\eta_{\mu\nu} }{p^2}
\end{equation}
in the Feynman-'t Hooft background gauge choice employed to derive equation  \eqref{bgflag} . Hence graphs with $j$ propagators along the hard line start contributing at order $\sim\!(z^{-j})$. For non-color-adjacent shift graphs the color ordered perturbation theory further simplifications are that the anti-symmetric term cannot either connect to two permuted legs or connect a permuted to a non-permuted non-shifted external leg. The latter would be represented by a diagram where a line crosses the hard line directly. Both these results follow immediately by considering the color factors. To verify suspicion \ref{susp:permsumimprov} up to $\sim\!(z^{-2})$ terms one needs to consider all graphs with at most two propagators which will be done below. Since the needed calculations are straightforward but tedious, only salient point will be mentioned.

\subsubsection*{One permuted leg (non-adjacent shift)}
At leading order, $\sim\!(z^0)$, there is only one graph to consider which yields a metric. The antisymmetric four-vertex does not contribute since it does not produce a color-ordered Feynman rule which crosses the hard line. The sub-leading order  ($\sim\!(z^{-1})$) requires a small calculation as there could be a symmetric combination as a result of the use of two anti-symmetric vertices. It is easy to verify these diagrams contribute at order $\sim\!(z^{-2})$. The other contributions are either anti-symmetric or proportional to the metric at order $\sim\!(z^{-1})$, as expected.

\subsubsection*{Two permuted legs}
\begin{figure}[ht] 
  \begin{center}
 \includegraphics[scale=0.55]{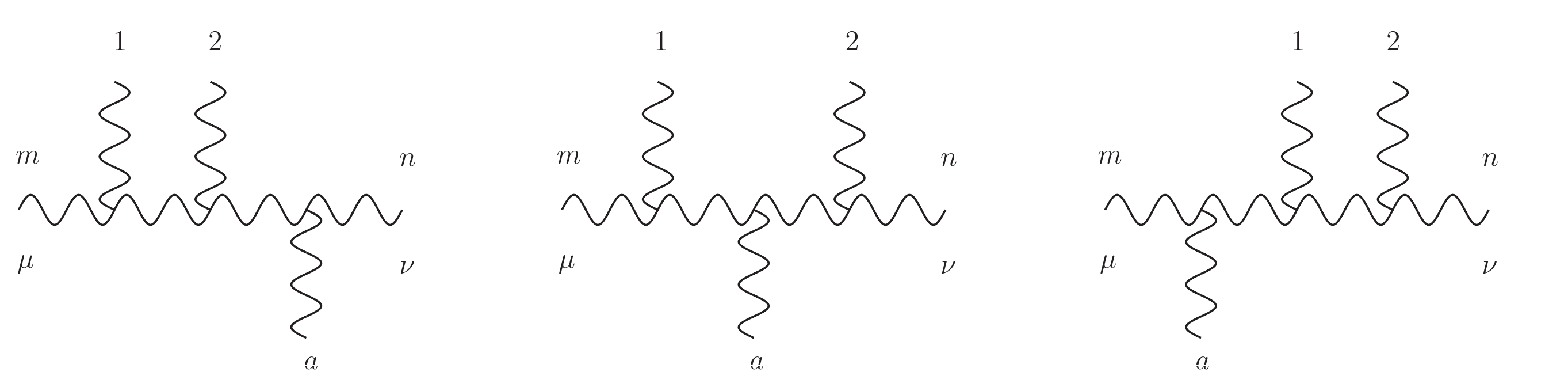}
\caption{Example diagrams of the diagrams contributing to the large shift limit of the permutation
sum over legs one and two, see equation \eqref{eq:diagramstwopermlegs}.}
\label{fig:diagramstwopermlegs}
  \end{center}
\end{figure}

At leading order, $\sim\!z^{-1}$, there are six graphs to consider. Since the anti-symmetric three vertices only depend on the momentum of the leg attached to the hard line all potentially anti-symmetric terms cancel at this order in $z$ in the permutation sum, leaving only terms proportional to the metric. At order $\sim\!(z^{-2})$ the only terms to be considered are those with two anti-symmetric vertices as the other contributions already fit the general scheme. Since the anti-symmetric three vertices only depend on the momentum of the leg attached to the hard line, all graphs with two anti-symmetric three point vertices are proportional to the same structure involving the currents attaching to these vertices. That leaves a comparatively simple computation which shows that the symmetric contribution at this order in $z$ vanishes in the permutation sum. To illustrate this point, consider the sum over the three graphs in figure
\ref{fig:diagramstwopermlegs}. 
The only term to be considered has two anti-symmetric vertices and one symmetric vertex. First choose the leg pointing down as the one contracting into the symmetric vertex. Summing the graphs gives
\begin{multline}\label{eq:diagramstwopermlegs}
(k^{\mu}_1 e_1^{\alpha} - k^{\alpha}_1 e_1^{\mu}    ) (k^{\nu}_2 e_{2,\alpha} - k_{2,\alpha} e_2^{\nu}    )\\
J(k_a)_\rho \left( \frac{2 (k^{\rho}_m + k^{\rho}_1) + k^{\rho}_a}{(k_m + k_1)^2 (k_m+k_1+k_a)^2} +  \frac{2 (k^{\rho}_m + k^{\rho}_1 + k^{\rho}_2) + k^{\rho}_a}{(k_m + k_1)^2 (k_m+k_1+k_2)^2} \right.\\ 
\left.+  \frac{2 (k^{\rho}_m ) + k^{\rho}_a}{(k_m + k_a)^2 (k_m+k_1+k_a)^2}\right)
\end{multline}
where the indices $\mu$ and $\nu$ contract with the shifted legs. Under a BCFW shift this graph scales as
\begin{multline}
\sim (k^{\mu}_1 e_1^{\alpha} - k^{\alpha}_1 e_1^{\mu}    ) (k^{\nu}_2 e_{2,\alpha} - k_{2,\alpha} e_2^{\nu}    )\\
 \frac{J(k_a)_\rho}{z^2 (q \cdot k_1) (q \cdot k_2) (q \cdot k_a)} \left((2 (k^{\rho}_m + k^{\rho}_1) + k^{\rho}_a) (q \cdot k_a) + (2 (k^{\rho}_m + k^{\rho}_1 + k^{\rho}_2) + k^{\rho}_a) (q \cdot k_2)\right.\\ \left.+  (2 (k^{\rho}_m ) + k^{\rho}_a) (q \cdot k_1)\right) + \mathcal{O}\left(\frac{1}{z^3} \right)
\end{multline}
which evaluates to
\begin{equation}
\sim   \frac{J(k_a)_\rho}{z^2 (q \cdot k_1) (q \cdot k_2) (q \cdot k_a)} \left( k_1^{\rho} (q \cdot k_1) - k_2^{\rho} (q \cdot k_2) \right) + \mathcal{O}\left(\frac{1}{z^3} \right) 
\end{equation}
When summed over permutations of legs $1$ and $2$ this quantity vanishes. The case where the symmetric three vertex is on one of the permuted legs follows from the same calculation by, symbolically, summing
\begin{equation}
\big[ (1 \leftrightarrow a) + (2 \leftrightarrow a)\big] + (1 \leftrightarrow 2)
\end{equation}
leading to the same conclusion: at order $\sim\!(z^{-2})$ the sum over these graphs is either proportional to the metric or anti-symmetric. A similar calculation can be repeated for all two propagator hard line graphs.

\subsubsection*{Three permuted legs}
There is a potential contribution from two graphs at order$\sim\!(z^{-1})$. It is easy to verify these two graphs combine to give an order $\sim\!(z^{-2})$ contribution proportional to the metric. That leaves quite some graphs at order $\sim\!(z^{-2})$ to consider. Most of these fall into the same category as studied at sub-leading order for the case of two permuted legs. The remainder is a simple addition of similar type graphs which leaves only terms proportional to the metric at order $\sim\!(z^{-2})$. The terms at order $\sim\!(z^{-3})$ will not be needed below.

\subsubsection*{Four and five permuted legs}
It can be checked that the first non-trivial order $\sim\!(z^{-2})$ terms cancel, leaving contributions starting at order $\sim\!(z^{-3})$.

\subsubsection*{More general shifts}
From the results just obtained other formulae may be derived by application of the Kleiss-Kuijf relations of equation \eqref{eq:KKrel}. Since
\begin{equation}\label{eq:apllKKshi}
\sum_{P(\alpha)}  A (1,\alpha,2,\beta) = \sum_{P(\alpha)}  \sum_{\sigma \in OP(\alpha^T \cup \beta)}
A(1, 2, \sigma)
\end{equation}
holds shifts of gluons $1$ and $2$ on the right hand side are directly related to shifts of these gluons on the left hand side. The latter have been studied above. Moreover, with one particle in set $\beta$ 
\begin{equation}
\sum_{P(\alpha)} A (1,\alpha,2,\beta_1) = \sum_{\sigma \in OP( \alpha \cup \beta)} A(1, 2, \sigma)
\end{equation}
Hence from the above results one can infer the behavior of shifts of amplitudes where the shifted particles next to the permutation sum are adjacent. Note that from this formula it is clear there is one power of $z$ less suppression in this case compared to the case where the shifted particles are non-adjacent. 

With the above calculations the suspicion \ref{susp:permsumimprov} has been proven up to and including $\sim\!(z^{-2})$ terms. In principle this proof could be pushed to more permuted legs with the help of computer-based algebra, but we were unable to see a way to a general proof of suspicion \ref{susp:permsumimprov} in the case of tree level amplitudes. A general proof, as well as the study of similar relations with more general matter couplings are interesting avenues for further research.

\subsection{Improved large shift behavior from cyclic sums}\label{sec:improvfromcyclic}
A second mechanism can be identified which improves the large shift behavior of Yang-Mills amplitudes. The main observation to motivate this is that the gluon current obeys a sub-cyclic identity:
\begin{equation}\label{eq:subcyclic}
\sum_{\sigma \in \mathbb{Z}(\{1,\ldots,n\})} J^{\mu}(\sigma) = 0 
\end{equation}
where the sum ranges over all cyclic permutations of all the gluons on the current. This is simply the $U(1)$ decoupling relation where the 'photon' is the off-shell leg. The leading terms of both the adjacent as well as the one-particle non-adjacent shift involve a single gluon current. Hence it is natural to study shifts of two particles adjacent to a cyclic sum. This leads to the following 
\begin{suspicion}\label{susp:cyclsumimprov} The power of $z$ fall-off of a BCFW shift of two particles on either side of a sub-cyclic sum of a 
color ordered  tree level Yang-Mills amplitude in $D \geq 4$ dimensions is suppressed by $(z^{-1})$
compared to same shift without the cyclic sum, unless the sum only contains one particle. 
\end{suspicion}
Again, there is a natural extension of this suspicion to correlation functions and integrands. Concretely, it implies that for a shift of particles $i$ and $j$ of a purely gluonic color ordered tree amplitudes 
\begin{equation}
\sum_{\mathbb{Z}(\{i+1,\ldots,j-1\})}  A(z)\sim \epsilon_i^{\mu}\epsilon_j^{\rho}\left(\frac{1}{z}\right) \Bigg( \eta_{\mu\rho}f_1(1/z)+ \frac{1}{z} B_{\mu\rho}(1/z)+\mathcal{O}\left(\frac{1}{z^2}\right)\Bigg)
\end{equation}
is suspected to hold  as long as $i$ and $j$ are not adjacent. If they are adjacent, the pre-factor becomes 
\begin{equation}
\left(\frac{1}{z^1}\right) \rightarrow \left(\frac{1}{z^{0}}\right) = 1
\end{equation}
This suspicion is obviously closely related to the previous one: for one and two particles in the cyclic sum it is equivalent. For five points one can use the KK relation to express the cyclic sum as
\begin{equation}
\sum_{\mathbb{Z}( \{1,2,3\})}  A(1,2,3,4,5) = - A(2,3,4,1,5) + A(1,3,4,2,5) - A(1,2,4,3,5)
\end{equation}
such that the $(4,5)$ shift conforms to \ref{susp:cyclsumimprov} by previous results. We have been unable to prove exact equivalence in general. 

Below the shifts of tree amplitudes where one side is a cyclic and the other a permutation sum will be needed. In examples below it will be seen that the improvements in BCFW shift behavior of suspicions \ref{susp:permsumimprov} and \ref{susp:cyclsumimprov} add, leading to further improved BCFW shift behavior. Hence one arrives at the following  
\begin{suspicion}\label{susp:sumofsusps}
The improvements of suspicions \ref{susp:permsumimprov} and \ref{susp:cyclsumimprov} can be obtained simultaneously for a BCFW shift of two particles adjacent to both a sub-cyclic as well as a permutation sum on a color ordered tree level Yang-Mills amplitude in $D \geq 4$ dimensions. 
\end{suspicion}
Just as above this particular suspicion will be checked up to and including $\sim\!(z^{-2})$ terms below for tree amplitudes using the background field approach. 

\subsubsection*{No permuted leg (adjacent shift)}
The first case to be considered has no permuted legs: the shifted legs are adjacent. Without loss of generality the BCFW shift of the following cyclic sum over color ordered amplitudes can be studied: 
\begin{equation}
\sum_{\sigma \in \mathbb{Z}(\{1,\ldots,n-2\})}  A(z)(\sigma, \widehat{n-1}, \widehat{n})
\end{equation}
Here legs $n-1$ and $n$ will be shifted. As observed above it is obvious that the leading $\sim\!(z)$ term vanishes by the sub-cyclic identity for gluon currents: this removes graphs where the sub-cyclic sum is on one leg. This removes the leading order graph for the adjacent shift case. At order $\sim\!(z^0)$ one graph has to be considered. This is either proportional to the metric or anti-symmetric. The anti-symmetric part, which does not confirm to suspicion \ref{susp:cyclsumimprov}, evaluates to 
\begin{equation}
\sim \epsilon^{\mu}_{\widehat{n-1}} \epsilon^{\nu}_{ \widehat{n}} \left(\sum_{\sigma \in \mathbb{Z}(\{1,\ldots,n-2\})} \sum_{\alpha \cup \beta = \sigma} \left[ J_{\mu}(\alpha) J_{\nu}(\beta) - J_{\nu}(\alpha) J_{\mu}(\beta)  \right] \right)
\end{equation}
where the sum is over all ways in which $\omega$ can be split into two ordered non-empty subsets. Combined with the cyclic sum this makes the expression cyclic under interchange of the sets $\alpha$ and $\beta$. Therefore this particular term is zero, validating suspicion \ref{susp:cyclsumimprov} in this particular case, to this particular order. Actually, since invariance under cyclic interchange of the sets $\alpha$ and $\beta$ is the same as invariance under permutation sums, many graphs with two cyclicly permuted legs have already been considered above.  

\begin{figure}[ht] 
  \begin{center}
 \includegraphics[scale=0.55]{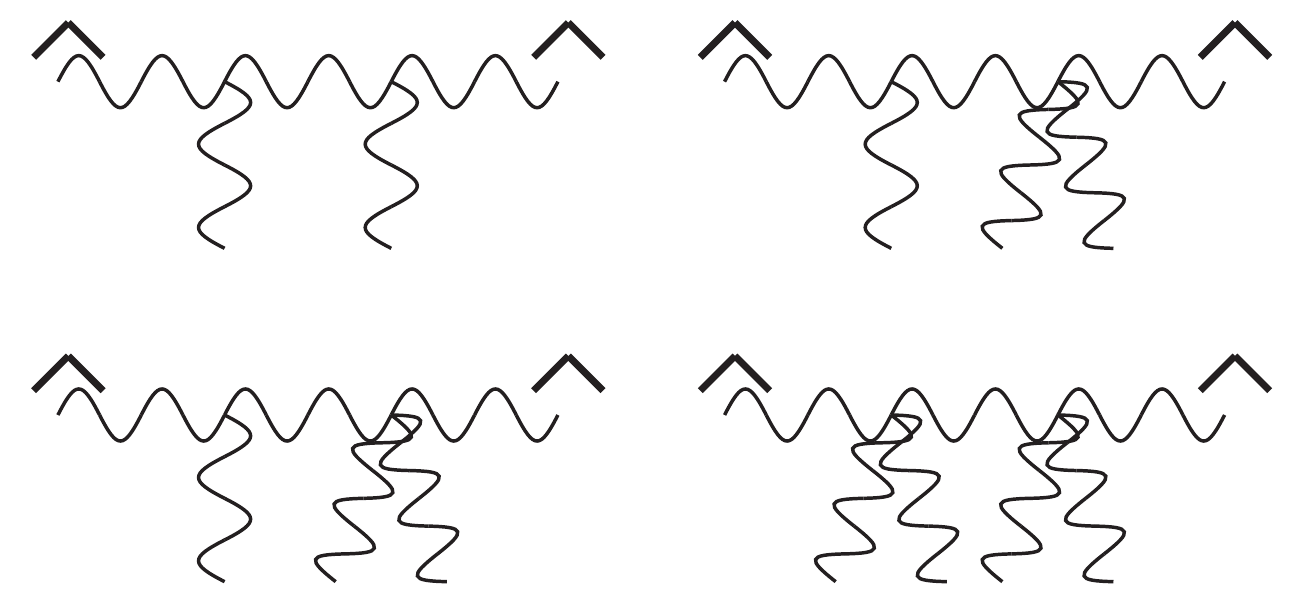}
\caption{Graphs contributing to the BCFW shift of two adjacent gluons flanking a cyclic sum at order $\sim\!(z^{-1})$.}
\label{fig:cyclicsumsadj}
  \end{center}
\end{figure}

At order $\sim\!(z^{-1})$ there are four graphs which at face value do not conform to suspicion \ref{susp:cyclsumimprov} as they are symmetric and not proportional to the metric. These have to be examined in turn. They either have two, three or four currents coupling to a one-propagator hard line graph and are depicted in figure \ref{fig:cyclicsumsadj}. 

The case of four currents coupling to the one-propagator graph evaluates to
\begin{multline}
\sim  \epsilon^{\mu}_{\widehat{n-1}} \epsilon^{\nu}_{ \widehat{n}} \left(\sum\sum_{\sigma \in \mathbb{Z}(\{1,\ldots,n-2\})}  \sum_{\alpha \cup \beta \cup \gamma \cup \delta = \sigma} \right. \\
\left. \frac{1}{z (q \cdot (k_{\alpha} + k_{\beta}))}  \left[ J_{\mu}(\alpha) J_{\nu}(\delta) (J(\beta) \cdot J(\gamma)) + J_{\mu}(\beta) J_{\nu}(\gamma) (J(\alpha) \cdot J(\delta)) \right. \right. \\ 
\left. \left. - J_{\mu}(\alpha) J_{\nu}(\gamma) (J(\beta) \cdot J(\delta)) - J_{\mu}(\beta) J_{\nu}(\delta) (J(\alpha) \cdot J(\gamma))  \right] \right)
\end{multline}
As before, the cyclic sum combined with the sum over all ordered partitions of the sets $\sigma$ make the expression under the sum cyclic under exchange of the sets $\alpha,\beta,\gamma,\delta$. Using this it is easy to show that the first two terms and the last two terms sum to zero under the sum taking into account the pre-factor and momentum conservation. 

The symmetric part of the two graphs with three currents not proportional to the metric can similarly be seen to shown to zero, while the graph with two currents has already been considered as argued above. Hence for adjacent shifts of a tree level gluon amplitude next to a cyclic sum suspicion \ref{susp:cyclsumimprov} holds. 

\subsubsection*{One permuted leg (non-adjacent shift)}
Without loss of generality the BCFW shift of the following cyclic sum over color ordered amplitudes can be studied: 
\begin{equation}
\sum_{\sigma \in \mathbb{Z}(\{1,\ldots,n-2\})} A(z)(\sigma, \widehat{n-2}, n-1, \widehat{n})
\end{equation}
Here legs $n-2$ and $n$ will be shifted. As observed above the leading $\sim\!(z^0)$ term vanishes by the sub-cyclic identity for gluon currents: this actually removes all four point graphs. Hence the first non-trivial graphs are at order $\sim\!(z^{-1})$. The graphs with two gluon currents in the cyclic sum have been analyzed already in the permutation sum section. With three gluon currents the only contributions from the two graphs possibly spoiling suspicion \ref{susp:cyclsumimprov} is anti-symmetric at order $\sim\!(z^{-1})$. Since the tensor structure will be the same for the two graphs, the calculations boils down to 
\begin{equation}
\frac{1}{(k_{\beta} + k_{\delta} + \hat{k}_{n})^2} + \frac{1}{(k_{\beta} + k_{\delta} + \hat{k}_{n-2})^2} = \mathcal{O}\left(\frac{1}{z^2}\right)
\end{equation}

\begin{figure}[t] 
  \begin{center}
 \includegraphics[scale=0.55]{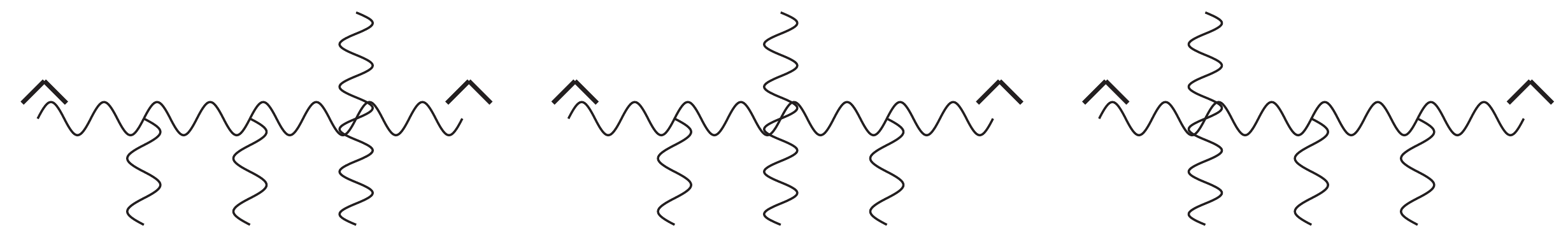}
\caption{Graphs contributing to the BCFW shift of two non-adjacent gluons flanking a cyclic sum at order $\sim\!(z^{-2})$.}
\label{fig:cyclicsumnonadj}
  \end{center}
\end{figure} 

This leaves possibly symmetric contributions at order $\sim\!(z^{-2})$ which are not proportional to the metric. Diagrams with three to five currents in the cyclic sum need to be considered. However, if the `permuted' leg connects to one of these currents by a Feynman graph which crosses the hard line, the two other vertices in the calculation have to be anti-symmetric. See figure \ref{fig:cyclicsumnonadj} for an example set of graphs in this set. Using cyclicity of the gluon current legs one arrives at
\begin{multline}
\frac{1}{(k_{\alpha}+ \hat{k}_{n-1})^2(k_{\alpha} + k_{\beta} + \hat{k}_{n-2})^2} + \frac{1}{(k_{beta}+ \hat{k}_{n-2})^2(k_{\alpha} + \hat{k}_{n})^2} \\ + \frac{1}{(k_{\beta}+ \hat{k}_{n})^2(k_{\alpha} + k_{\beta} + \hat{k}_{n})^2}  = \mathcal{O}\left(\frac{1}{z^3}\right)
\end{multline}
where $\alpha$ and $\beta$ indicate two consecutive gluons legs. This mechanism can be shown to hold for all graphs where the permuted leg attaches directly to a gluon current. This leaves the class of  $\sim\!(z^{-2})$ diagrams where the permuted leg connects only to the hard line. The six graphs with three currents can be shown to confirm suspicion \ref{susp:cyclsumimprov}: it's symmetric part at this order is proportional to the metric. The same holds for the three graphs with four currents.  As these calculations are simple and tedious they will be suppressed. 

\subsubsection*{Two permuted legs}
For two permuted legs there are four diagrams at order $\sim\!(z^{-1})$ with two gluon currents attached, two of which cancel among each other. The remaining contribution reads
\begin{equation}
\sim \sum_{\textrm{cyclic}} \sum_{\textrm{dist}} \frac{1}{z} \left[\frac{\left(e_1 \cdot J_\alpha\right)\left( e_2 \cdot  J_\beta \right)}{q (k_1 + k_\alpha)} - \frac{\left( e_1 \cdot  J_\beta \right)\left(e_2 \cdot J_\alpha\right)}{q (k_1 + k_\beta)}  \right]
\end{equation}
where the second term is the permuted version of the first. $J_\alpha$ and $J_\beta$ denote two gluon current legs. The sums are over the distribution of the gluonic particles over the two currents, as well as over all cyclic permutations of the gluonic legs. These terms cancel against each other: for every term in the sum for the first quotient there is a corresponding term in the sum for the second quotient. 

The symmetric part at order $\sim\!(z^{-2})$ can be shown to be proportional to the metric. This is basically the same as a part of the calculation in the previous case of one permuted leg: the two permuted legs can only couple to the hard line through a symmetric vertex. 

\subsubsection*{Three permuted legs}
\begin{figure}[t] 
  \begin{center}
 \includegraphics[scale=0.55]{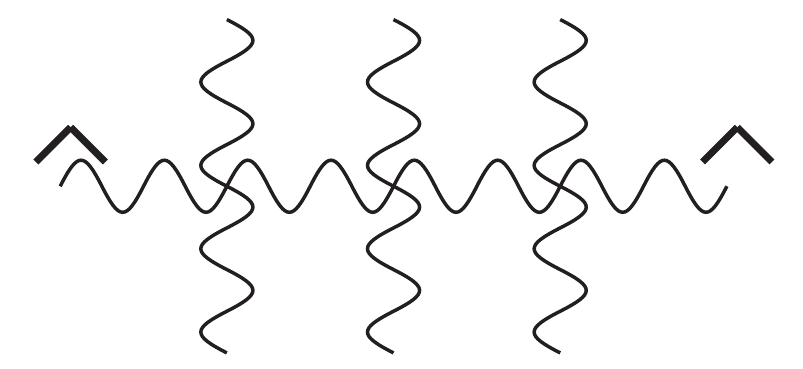}
\caption{Example graph contributing to the BCFW shift of two gluons flanking a cyclic sum and a permutation sum at order $\sim\!(z^{-2})$.}
\label{fig:threecrossesgraph}
  \end{center}
\end{figure}

For three permuted legs there are graphs at order $\sim\!(z^{-2})$ with either two or three gluon currents. Consider first the latter case. By the same argument as before, the indices on the gluon current can be treated as being cyclic. Before the permutation sum there are seven possible diagrams which can be drawn contributing at order $\sim\!(z^{-2})$, each containing four point vertices only. These can be classified according to the appearance of connections which cross the hard line. The single graph where the hard line is crossed three times, depicted in figure \ref{fig:threecrossesgraph} evaluates to 
\begin{equation}
 \sum_{\textrm{cyclic}} \sum_{\textrm{dist}} \sum_{perms}\frac{1}{z}  \left[\frac{(J_\alpha e_1)(J_\beta e_2)(J_\gamma e_3)}{q \cdot(k_\alpha + k_1) (q \cdot (k_3 + k_\gamma) )} \right]
\end{equation}

The sum over permutations can be written as
\begin{multline}
 \sum_{\textrm{cyclic}} \sum_{\textrm{dist}} \frac{1}{z} \left[ \frac{(J_\alpha e_1)(J_\beta e_2)(J_\gamma e_3)}{q \cdot(k_\alpha + k_1) (q \cdot (k_\gamma + k_3) } +  \frac{(J_\alpha e_3)(J_\beta e_3)(J_\gamma e_1)}{q \cdot(k_\alpha + k_3) (q \cdot (k_\gamma + k_2) } \right. \\ \left. +  \frac{(J_\alpha e_2)(J_\beta e_3)(J_\gamma e_1)}{q \cdot(k_\alpha + k_1) (q \cdot (k_\gamma + k_1)) }\right] + (\leftrightarrow)
\end{multline}
where $(\leftrightarrow)$ stands for the missing three terms which have a similar structure. Cyclicity in the indices $\alpha,\beta$ and $\gamma$ can now be used to rewrite this as 
\begin{multline}
 \sum_{\textrm{cyclic}} \sum_{\textrm{dist}} \frac{1}{z} \left[ \frac{(J_\alpha e_1)(J_\beta e_2)(J_\gamma e_3)}{q \cdot(k_\alpha + k_1) (q \cdot (k_\beta + k_2)) (q \cdot (k_\gamma + k_3) } (q \cdot (k_1 + k_2 + k_3 + k_\alpha + k_\beta + k_\gamma)) \right] \\ + (\leftrightarrow)
\end{multline}
which vanishes by momentum conservation. The ``$\leftrightarrow$'' graphs vanish similarly. By a similar computation the class of graphs containing six elements can be shown to sum to zero. To see this one first divides up the six graphs into three sets, distinguished by the appearance of a line crossing the hard line. These can be summed separately. Then it is seen the contribution of the graphs with the hard line crossing in the middle is minus the sum of the other two possibilities. 

The remaining diagrams involve two currents coupling to the hard line with two propagators. These have either one or two crosses of the hard line. In the case of one cross, they have the cyclicly summed legs either on the same or on a different vertex. This leads to three classes of graphs which can be shown to sum to zero separately. 

\subsubsection*{Conclusion}
This concludes the verification of suspicion  \ref{susp:sumofsusps} which itself is an extension of suspicion \ref{susp:cyclsumimprov} up to and including terms of order $\sim\!(z^{-2})$. As noted above in the permutation sum case, it is hard to see how to proof  \ref{susp:sumofsusps} in full generality, let alone possible extensions to the integrand. The achieved order is however enough for the purposes of the next section. We have explicitly verified the same improvements as for gluons hold for  shifts of massive scalar pairs. Other matter shifts can be analyzed along the same lines as above at least in principle. If shifts of two gluons coupled to arbitrary matter are known then as explained in appendix \ref{app:othermatters} this result can be used in a supersymmetric field theory to obtain shifts of certain fermions and scalar pairs.


\section{Relations for scalar integral basis coefficients at one loop}\label{sec:amplreloneloop}
In QED  \cite{Badger:2008rn} it has been shown that improved BCFW shift behavior in this theory is intimately related to the absence of so-called rational terms in light-by-light scattering amplitudes with more than four photons, as well as the absence of so-called bubble and triangle coefficients for amplitudes with more than six photons. In this section it will be demonstrated similar results can be obtained for leading color one loop amplitudes in pure Yang-Mills. 

In general the goal of generalized unitarity techniques is to calculate one loop amplitudes from tree level amplitudes. One-loop amplitudes in Yang-Mills theory in four dimensions coupled to a variety of massless scalar or spin-half matter can be captured in dimensional regularization in the four dimensional helicity-scheme \cite{Bern:1991aq} in a standard basis of scalar integrals,
\begin{equation}\label{eq:massiveloopampexp}
A^{\textrm{co}, 1-\textrm{loop}} = \sum a_b \left(\textrm{Boxes} \right) +  a_t  \left(\textrm{Triangles} \right) + a_{bb}  \left(\textrm{Bubbles} \right) +  \textrm{Rational} + \mathcal{O}(\epsilon)
\end{equation}
where the last terms are simply rational functions of polarizations and momenta. The sum ranges in principle over all ways to distribute the external particles over the legs of the integrals, disregarding the
order of the particles at the corners. One such particular choice will be referred to as a 'channel'. For a color ordered amplitude only the coefficients in those channels where all gluons on every corner are consecutive on the color-ordered amplitude are non-vanishing. 

As the scalar integrals can be integrated once and for all, the problem reduces to calculating the coefficients of the integrals which are functions of the external momenta and polarizations. Much effort has been devoted to determining the coefficients. By now all coefficients have a known expression in terms of tree level amplitudes, see \cite{Britto:2004nc} for the box coefficient, \cite{Forde:2007mi} for the triangle and bubble coefficients and \cite{Badger:2008cm} for the rational terms. For box, triangle and bubble terms the coefficients are functions of tree amplitudes with four-dimensional massless legs, including the `loop' legs. The box-integral coefficients are particularly simple while the others get progressively more involved with the rational terms the most complicated.

It is known in special theories that some of the coefficients of equation \eqref{eq:massiveloopampexp} vanish. Supersymmetric theories with massless matter generically do not have rational terms for instance. Maximally supersymmetric gauge and gravity theories in four dimensions have vanishing bubble and triangle coefficients\footnote{These statements are proven for supersymmetric gauge theories in \cite{Bern:1994zx}, while the analogous statements in maximal supergravity were the subject of the much more recent \cite{BjerrumBohr:2008ji}. A particularly simple explanation of the absence of triangles for maximal SUGRA can be found in \cite{ArkaniHamed:2008gz}.}. Both absence of rational terms as well as that of bubble and triangle coefficients are shared by theories with exotic matter content \cite{Lal:2009gn}, \cite{Lal:2010qq}. As recalled previously it has been shown photon amplitudes in QED show similar simplicity \cite{Badger:2008rn}. 

In this section relations between coefficients in the basis of equation \eqref{eq:massiveloopampexp} will be studied for the leading color amplitudes. To our knowledge the only known relations in pure Yang-Mills theory beyond the relation between planar and non-planar parts, equation \eqref{eq:sub-leadingfromleading}, are for finite one loop amplitudes. These are the amplitudes with all helicities equal \cite{Bern:1993qk} \cite{BjerrumBohr:2011xe} or one unequal \cite{BjerrumBohr:2011xe}. In \cite{Bern:1993qk} for instance it was shown that the  following permutation sum 
\begin{equation}\label{eq:vanamptemplate}
\sum_{\sigma \in POP(\alpha_3 \cup \beta) }A^{1-\textrm{loop}}(\sigma) = 0 \qquad \textrm{helicity equal}
\end{equation}
over the leading-color helicity equal amplitudes at one loop vanishes, where the subindex indicates the number of particles in the set $\alpha$. The sum ranges over all non-cyclic permutations of the union of the sets $\alpha_k$ and $\beta$ keeping the order of the set $\beta$ fixed. This  type of sum is generated when $k$ gluons of the planar loop amplitude are replaced by photons, i.e. their associated color matrices are replaced by the identity as
\begin{equation}
T^a \rightarrow \mathbb{I}
\end{equation}
Because of this interpretation equation \eqref{eq:vanamptemplate} will be referred to as the three photon decoupling identity. Below it is shown it generalizes to all rational terms.  

\subsection{Three photon decoupling for rational terms}
Since supersymmetric theories do not have rational terms it is useful to use a decomposition of the gluon contribution to the quantum effective action at one loop in terms of supersymmetric multiplets, e.g.
\begin{equation}
S^1_{\textrm{gluon}} = S^1_{\mathcal{N}=4} - 4 S^1_{\mathcal{N}=1} + S^1_{\textrm{scalar}}
\end{equation}
to reduce the calculation of rational terms to graphs with a `massive' complex scalar in the loop. The mass of the scalar is the $-2 \epsilon$ part of the momentum in dimensional regularization which gets integrated over to yield the loop amplitude. 

The rational terms in Yang-Mills theory have been expressed in terms of tree level amplitudes of gluons coupled to massive scalars in \cite{Badger:2008cm} using generalized unitarity techniques. Schematically,
\begin{equation}\label{eq:simonsays}
\textrm{Rational} = \sum \left(\textrm{massive Box} \right) +  \left(\textrm{massive Triangle} \right) + \left(\textrm{massive Bubbles} \right) 
\end{equation}
where the sum is over all channels of the massive box, triangle and bubble topologies. Note these are all rational functions: their names refer to their origin in terms of integral reduction, see \cite{Badger:2008cm} for details. The exact expressions of these coefficients will be given below.

\subsubsection*{Massive box contribution}
\begin{figure}[t!]
\centering
\includegraphics[scale=0.6]{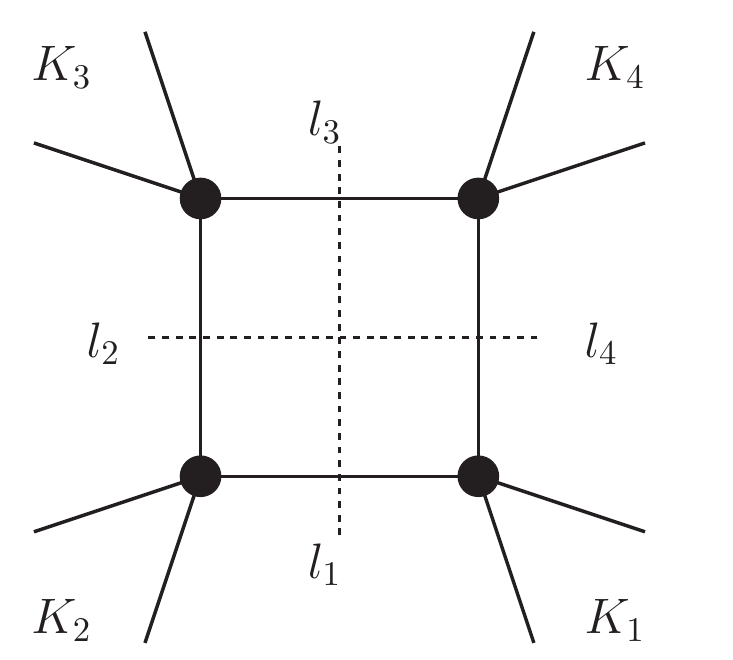}
\caption{\label{fig:quadruplecut} Momentum assignment of a massive box coefficient in a certain
channel. The $i$-th corner contain a number of gluons with total momentum $K_i$.  }
\end{figure}

The massive box contribution reads for a certain channel,
\begin{multline}\label{eq:cutmassivebox}
\left(\textrm{massive Box} \right) = \\ \sum_{\sigma} \textrm{Inf}_{\mu} \left(A_{\phi \bar{\phi}}(-l_1, K_1, l_2)A_{\phi \bar{\phi}}(-l_2, K_2, l_3) A_{\phi \bar{\phi}}(-l_3, K_3, l_4) A_{\phi \bar{\phi}}(-l_4, K_4, l_1) \right)_{\mu^4} 
\end{multline}
where the particle in the loop is the massive scalar with mass $\mu$ and $A_{\phi \bar{\phi}}$ are scattering amplitudes for a massive scalar pair coupled to gluons. The gluons on the amplitudes must appear in the same order as on the color ordered amplitude under study. In particular only coefficients in channels where all gluons on every corner are consecutive on the color-ordered amplitude are non-vanishing. See figure \ref{fig:quadruplecut} for a graphical illustration of the distribution of momenta, with momentum conservation on all corners. The sum is over the two solutions to the on-shell equations for the four dimensional loop momenta $l_i$,
\begin{equation}\label{eq:momsonshellbox}
l_i^2 = \mu^2 \qquad i \in \{ 1,2,3,4\}
\end{equation}
The symbol $\textrm{Inf}_x$ stands for taking the polynomial part of the Laurent series of a function $f(x)$ at $x \rightarrow \infty$ and the final instruction is to isolate the $\mu^4$ term in this polynomial. 

To make this more precise, consider the kinematics of the solution to the four equations \eqref{eq:momsonshellbox} along the lines of \cite{Badger:2008cm}. For two massive momenta $K_1$ and $K_2$ one can construct a light-like basis
of four vectors as
\begin{equation}
K_1^{\nu,\flat} \qquad  K_2^{\nu,\flat} \qquad n^{\nu} \qquad \bar{n}^{\nu}
\end{equation}
whose only two non-vanishing inner products are equal,
\begin{equation}
K_1^{\flat} \cdot K_2^{\flat} = n \cdot  \bar{n}
\end{equation}
In terms of the massive momenta $K_1$ and $K_2$  the massless vectors $K_1^\flat$ and $K_2^\flat$  are defined by
\begin{equation}
K_1 = K_1^{\flat} + \frac{K_1^2}{\gamma_{12}} K_2^{\flat} \qquad K_2 = K_2^{\flat} + \frac{K_2^2}{\gamma_{12}} K_1^{\flat} 
\end{equation}
with 
\begin{equation}
\gamma_{12} = 2 K_1^{\flat} \cdot K_2^{\flat}
\end{equation}
From the massless vectors  $K_1^\flat$ and $K_2^\flat$ it is easy to construct $n$ and $\bar{n}$ in terms of spinor helicity. The explicit expressions will not be needed here. In terms of the basis above the loop momentum $l_1$ is given by
\begin{equation}\label{eq:solsforthreeconstr} 
l^{\nu}_1 = a K_1^{\flat, \nu}  +  b  K_2^{\flat, \nu}  + t n^{\nu} + \frac{ \gamma_{12} a b - \mu^2}{t \gamma_{12}} \bar{n}^{\nu}
\end{equation}
which solves $l_1^2 = \mu^2$. The coefficients $a$ and $b$ follow from $l_2^2 = \mu^2$ and $l_4^2 = \mu^2$ as
\begin{equation}
a = - \frac{K_1^2 (K_2^2 + \gamma_{12})}{(K_1^2 K_2^2 - \gamma_{12}^2)}   \qquad b = \frac{K_2^2 (K_1^2 + \gamma_{12})}{(K_1^2 K_2^2 - \gamma_{12}^2)}
\end{equation}
The remaining equation
\begin{equation}
l_3^2 = (l_1 + K_1 + K_2)^2 = \mu^2
\end{equation}
can be written as
\begin{equation}\label{eq:twosolsmassbox}
a 2 (K_1^{\flat} K_3) + b 2 (K_1^{\flat} K_3) + t 2 (n \cdot K_3) +  \frac{ \gamma_{12} a b - \mu^2}{t \gamma_{12}} 2 (\bar{n} K_3) = - K_3^2 - 2 K_2 \cdot K_3
\end{equation}
This equation has two solutions for $t$. For our purposes only the limiting behavior $\mu\rightarrow \infty$ is important, which originates in the coefficient of $\frac{1}{t}$ in the above equation. Note the solution for $t$ scales as $\sim\!|\mu|$ in the large $\mu$ limit. 

More precisely in the limit $\mu \rightarrow \infty$ the loop momenta scale as
\begin{equation}\label{eq:massinfbox}
l^{\nu}_i \rightarrow \mu\, V^{\nu} + \mathcal{O}\left( \mu^0\right)
\end{equation}
for some space-like vector $V^{\nu}$ which is orthogonal to all external momenta $K_i$. In particular $V$ only depends on the choice of channel, not on the order of the gluons on the sub-amplitudes. 

The analysis will now parallel the BCFW shift closely. In particular it is natural to choose the gauge $V^{\nu} A_{\nu}=0$ to analyze the large $\mu$ behavior of the scalar amplitudes coupled to glue. The same powercounting argument as before shows that in the usual case this limit is dominated by one graph: the one in which the scalar couples to a glue current directly. Hence generically
\begin{equation}\label{eq:formoflargemasslimit}
\lim_{\mu \rightarrow \infty} A_{\phi \bar{\phi}}(-l_i, K_i, l_{i+1}) \rightarrow \sim \mu \, V_{\nu} J^{\nu}
\end{equation} 
with $J$ the one-leg off-shell color-ordered gluon current in this particular gauge which contains all the gluons coupling to this particular corner of the massive box. 

Now consider changing one of the external particles to a photon. For a particular channel of the massive box there will now be a sum over all ways of inserting the photon on the external line, keeping the scalars always color-adjacent. By the Kleiss-Kuijf relations in equation \eqref{eq:apllKKshi} this sum is related to the non-adjacent BCFW shift of a pair of massive scalars. Hence a generic massive box with a photon on one of the corners does not contribute to the rational terms.  The exception to this is when the photon couples to the corner directly through a three point amplitude. Hence 
\begin{equation}
\textrm{massive box} \left[\sum_{i=1}^{n+1} A(2,\ldots,i-2,1, i,\ldots,n) \right] = \sum_{\textrm{channels}'} \left(V_{\nu} e_1^{\nu}\right) \left(V^c_{\nu} J^{\nu} \right) \left(V^c_{\nu} J^{\nu} \right) \left(V^c_{\nu} J^{\nu}\right) 
\end{equation}
where the restricted sum is over all ways to distribute the $n-1$ particles $(2,\ldots,n)$ over the remaining three corners, keeping their order and summing over all cyclic permutations. This is already a simplification of the box coefficients. Now consider adding more photons, all of which have to couple through a three point amplitude. Hence for more than three photons the involved channels always have at least one power of $\mu$ suppression and hence do not contribute to the massive box coefficient. 

That leaves the boundary case of three photons. In this case three photons couple directly to the loop, while all gluons couple to the remaining free corner of the box. The gluons of this remaining corner are summed however over all cyclic permutations. This follows directly from the permutation sum in \eqref{eq:vanamptemplate}. By suspicion \ref{susp:cyclsumimprov} (which has been proven to the required order) this scales one order of $\mu$ down from the naive expectation. This can also be seen directly from equation \eqref{eq:formoflargemasslimit}: the large $\mu$ term vanishes in the cyclic sum over gluons on one corner.  

Hence there are no massive box contributions to the rational terms of pure Yang-Mills amplitudes for three photons and above. The argument removing the boundary case of three photons is quite generic: whenever all gluons couple to only one corner of a certain box,  bubble or triangle with photons on the other corners the cyclic sums generate an extra power of suppression.

\subsubsection*{Massive triangle contribution}
The massive triangle contribution reads
\begin{multline}
 \left(\textrm{massive Triangle} \right) = \\ \sum_{\sigma} \textrm{Inf}_{\mu^2} \left( \textrm{Inf}_{t}   \left(A_{\phi \bar{\phi}}(-l_1, K_1, l_2) A_{\phi \bar{\phi}}(-l_2, K_2, l_3)A_{\phi \bar{\phi}}(-l_3, K_3, l_1) \right)_{t=0} \right)_{\mu^2}
\end{multline}
where the loop momentum is parametrized by (\ref{eq:solsforthreeconstr}) or its conjugate. These possibilities are summed over, as indicated by the sum over $\sigma$. Here large $t$ behavior is easily seen to be closely related to a BCFW shift. However, a double expansion is needed. For this note that the analysis of the large $t$ behavior in the appropriate lightcone gauge is the same as for BCFW shifts. The difference is a $\mu^2/t$ term in the momentum. This is the only source of dependence on the mass $\mu$ in the Feynman graphs as the massive scalar propagators are on the mass shell. 

Therefore for almost all amplitudes the $\mu$ dependence starts two powers of $t$ beyond the leading naive scaling dimension, see equation (\ref{eq:solsforthreeconstr}). Since this is the naive expectation, care has to be taken when cancellations are involved. For our purposes, the only exception in a relevant channel is that with three photons coupling to gluons and a massive scalar pair. This scales as $\sim\!(t^{-2} + \mu t^{-3})$ by powercounting. It is easy to see that this channel does not contribute to the rational terms. In fact, in the generic case with three photons distributed over corners of the triangle the leading scaling dimension is $t^0$ and will hence not contribute to the rational terms as there is no $\mu^2$ dependence left. The class of channels with scaling dimension $t$ (where one or more of the photons couples directly to the triangle) will also not contribute to the rational terms. 

Hence the only channels which need to be considered in the three photon case are those where the product of amplitudes has leading scaling dimension $\sim\!(t^2)$. The two channels involved have either two or three photons coupling directly to the loop. However, by the same argument as for the massive box these channels have a single gluon current coupling to the loop which will be summed over all cyclic permutations. This reduces the leading scaling dimension to $\sim\!(t^{1})$ and it is easy to verify this also drop the $\mu^2$ dependence to terms of order $\sim\!(t^{-1})$. Hence the massive triangle does not contribute to the rational terms for three photons and above. 

\subsubsection*{Massive bubble contribution}
The contribution to the rational terms which remains to be analyzed is of massive bubble type. This actually contains \cite{Badger:2008cm} two different types of term, one of which arises as a variation on the above massive triangle. Since this has already been shown to vanish it can be neglected. What is left is a pure bubble contribution:
\begin{multline}
 \left(\textrm{massive Bubble} \right) = -i \, \\
 \textrm{Inf}_{\mu^2}\left(  \textrm{Inf}_{t} \left( \textrm{Inf}_{y} \left( A_{\phi \bar{\phi}}(-l_1, K_1, l_2) A_{\phi \bar{\phi}}(-l_2, K_2, l_1) \right)_{y^i \rightarrow Y_i} \right)_{t^0} \right)_{\mu^2}
\end{multline}
where the loop momentum is parametrized as
\begin{equation}\label{eq:purebubbleloops}
l_{1, \alpha \da} = y 1_{\alpha} 1_{\da}  + \frac{m_1^2}{\bar{\gamma}} (1-y) \xi_{\alpha} \xi_{\da} + t 1_{\alpha} \xi_{\da} + \frac{ y (1-y) m_1^2 - \mu^2}{t \bar{\gamma}} \xi_{\alpha} 1_{\da}
\end{equation}
in terms of an arbitrary light-like momentum $\xi$ with
\begin{equation}
\bar{\gamma} = 2 K_1 \cdot \xi
\end{equation}
The $Y_i$ are certain known functions of the external kinematics which will be unimportant in the following. It is advantageous to perform powercounting in the $ \xi_{\alpha} 1_{\da} A^{\alpha \da} = 0$ gauge to consider large $y$ contributions. Generic channels are suppressed by powers of $y$ as they will always involve at least two gluons on either tree amplitude. This leaves the special class of channels where the photons couple to the loop directly. We can discard all channels which involve a gluon current containing all the gluons. This leaves the channel with one gluon coupling to the loop and the other three photons on the other side. Note that in this case the parametrization of the loop momentum in equation \eqref{eq:purebubbleloops} is special as $m_1^2 = 0$. It is easy to verify that the amplitude with one single photon does not depend on $y$, while the other amplitude is $1/y$ suppressed. Hence there are no bubble contributions to the rational terms for three photons and above. 

This completes the proof of the three photon decoupling relation 
\begin{equation}
\textrm{Rational} \left[ \sum_{\gamma \in POP(\alpha_3 \cup \beta) } A^{1-\textrm{loop}}(\gamma) \right]= 0
 \end{equation}
for rational terms in pure Yang-Mills theory. This relation can be numerically cross-checked \cite{badgerprivate} up to seven points using NGluon \cite{Badger:2010nx}. The first non-trivial example of this particular relation is at six points: at four points a complete permutation sum arises which vanishes. At five points the permutation sum can be neatly grouped into pairs of amplitudes which involve the inverse ordering of particles. These sums vanish pairwise. 

The three photon decoupling relation can be re-written using equation \eqref{eq:sub-leadingfromleading}  as a relation for rational terms of the double trace one loop amplitude with the three photons on a separate trace from the gluons. There are two terms in this class, i.e. 
\begin{equation}
\textrm{Rational} \left[A^{1-\textrm{loop}}(\sigma; 1,2,3) +A^{1-\textrm{loop}}(\sigma; 1,3,2)  \right]= 0
\end{equation}
where the semi-colon indicates the separation between the particles on the different traces. 

\subsection{Five photon decoupling for bubbles and triangles}\label{sec:fivphdecop}
Based on the previous subsection and the results in \cite{Badger:2008rn} it is a natural question if there are photon decoupling relations for the ordinary bubble and triangle terms as well. These coefficients have been expressed in terms of four dimensional tree level amplitudes in \cite{Forde:2007mi}. Just as above for the rational-type terms, both bubble and triangle terms involve picking up a certain coefficient in a scaling limit which is closely related to BCFW. Since photons improve scaling in general, decoupling relations are to be expected from some number of photons onwards. In this subsection it is shown this number is five for triangle and bubble coefficients (three for ``purely bubble'' contribution to the bubbles).  

\subsubsection*{Absence of triangles}
The triangle coefficients can be expressed as \cite{Forde:2007mi}
\begin{equation}
 \left(\textrm{Triangle} \right) = \sum_{\sigma} \textrm{Inf}_{t} \sum_{\textrm{helicities}}  \left(A(-l_1, K_1, l_2) A(-l_2, K_2, l_3) A(-l_3, K_3, l_1) \right)_{t=0}
\end{equation}
with the loop momentum parametrized as in equation \eqref{eq:solsforthreeconstr} with $\mu^2$ set to zero. New here is the sum over helicities, which arises as the particle in the loop is a four dimensional on-shell gluon. To evaluate these sums one uses the completeness relation
\begin{equation}\label{eq:complrel}
 \sum_{\textrm{helicities}} e_{\mu}(l) e_{\nu}(l) = \eta_{\mu\nu} - \frac{\xi_{\mu} l_{\nu} + \xi_{\nu} l_{\mu}}{l \cdot \xi}
\end{equation}
for some gauge choice $\xi$ for which $\xi \cdot l \neq 0$. Note that in the limit $t \rightarrow \infty$ the completeness relation of equation \eqref{eq:complrel} scales as $\sim\!(t^0) + \mathcal{O}(\frac{1}{t})$ and has a universal limit for all three cut legs. Hence for a calculation of the leading scaling behavior this factor can be ignored.

\begin{figure}[ht]
  \begin{center}
 \includegraphics[scale=0.40]{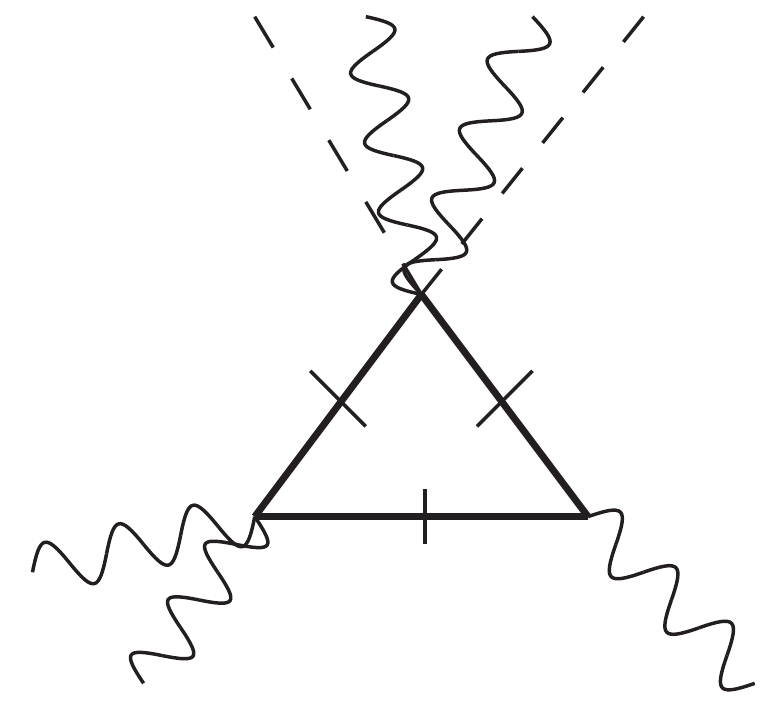}
\caption{Representative example of a triangle contribution in the five photon decoupling relation. The wiggly lines indicate photons. The gluons couple to the top amplitude, indicated by the dotted lines. }
 \label{fig:repfivephotoncontrib}  \end{center}
\end{figure}

Since the na\"ive scaling of the scattering amplitudes in this expression is $\sim\!(t^1)$, four powers of suppression are needed to obtain a vanishing result. A minor complication is the fact that purely photon corners shift one power of $t$ less well than photons mixing with gluons. This quickly leads to the identification of six photons as the case where the triangle coefficient surely vanishes, as long as there is also at least one gluon around. This leaves the boundary case of five photons. Taking into account permutation sums the only remaining channels which scale as $\sim\!(t^0)$ involve all gluons coupling to one corner of the triangle, with the photons distributed over all corners. A representative channel in this class is depicted in figure \ref{fig:repfivephotoncontrib}. 

Since all gluons will couple to one amplitude on the triangle all these channels will involve a cyclic sum over the gluons. From the results in subsection \ref{sec:improvfromcyclic} these channels scale one order of $t$ better than naively expected. Hence the triangle coefficients obey a five photon decoupling relation. 

\subsubsection*{Absence of bubbles}
The bubble coefficients are given by two different types of contribution: pure bubbles and terms originating in triangles. The latter vanish in the five photon sum as has just been shown. The pure bubble coefficients read
\begin{equation}
 \left(\textrm{pure bubble} \right) = \sum_{\sigma} \textrm{Inf}_{t} \sum_{\textrm{helicities}}    \left(A(-l_1, K_1, l_2) A(-l_2, K_2, l_1) \right)_{t=0}
\end{equation}
where the on-shell loop momentum is given by equation \eqref{eq:purebubbleloops}, with $\mu^2$ set to zero. It is easy to see that for more than two photons this term will always vanish, based on the results of section \ref{sec:gennonadj}. Note the pure bubble contribution to the bubble coefficients obey a three photon decoupling relation.

This completes the proof of the five photon decoupling relation
\begin{equation}
\textrm{Triangle}/\textrm{Bubble} \left[ \sum_{\gamma \in POP(\alpha_5 \cup \beta) } A^{1-\textrm{loop}}(\gamma) \right]= 0
 \end{equation}
for triangle and bubble terms in pure Yang-Mills theory.   This relation has been numerically cross-checked \cite{badgerprivate} for a specific helicity configuration at eight points using NGluon \cite{Badger:2010nx}. The first non-trivial example of this particular relation is at eight points: at six points a complete permutation sum arises which vanishes. At seven points the permutation sum can be neatly grouped into pairs of amplitudes which involve the inverse ordering of particles. These sums vanish pairwise. 

Phrased differently for five photons pure Yang-Mills amplitudes only have box contributions. Note that to calculate non-vanishing box coefficients for for instance MHV amplitudes in four dimensions \eqref{eq:usefullsumforphot} is quite useful. Beyond MHV amplitudes the Kleiss-Kuijf relations can be used to simplify the permutation sums somewhat. It should be noted that the five photon decoupling relation can be re-written using equation \eqref{eq:sub-leadingfromleading} as a relation for the rational terms of the double trace amplitude with the five photons on a separate trace from the gluons. There are $24$ terms in this sum.  

\subsection{Relations for massive box contributions to rational terms}
In \cite{BjerrumBohr:2011xe} new relations for the helicity equal and one un-equal Yang-Mills amplitudes at one loop have been conjectured. These were checked there to certain relatively high numbers of external legs, leaving little doubt about their general validity. It will be shown below that the conjectured  relations of \cite{BjerrumBohr:2011xe} for helicity-equal amplitudes have a helicity-independent extension to massive box coefficients in the expansion \eqref{eq:simonsays} which can be proven to all multiplicity.

\subsubsection*{Helicity equal amplitudes only have massive box coefficients}
To get an idea why this may be true, consider the fact that helicity equal amplitudes are not only purely rational one loop amplitudes but also only have massive box contributions in the expansion in equation  \eqref{eq:simonsays}. To see this consider BCFW-type shifts of a color-adjacent massive scalar pair of legs coupled to helicity equal and one helicity opposite gluons at tree level. Generically by argument reviewed several times above BCFW shifts of massive color-adjacent scalar legs coupled to glue scale as $\sim\!(z^1)$. However, the leading and sub-leading contribution in AHK gauge both come from the graph where the two scalar lines connect directly to a three vertex. This diagram contributes,
\begin{equation}
A(z) \sim z (q_{\mu}J^{\mu}_q) + (k_1 - k_n)_{\mu} J^{\mu}_q
\end{equation}
where $k_1$ and $k_n$ are the momenta of the massive scalar legs. Here $J^{\mu}_q$ is the gluon current calculated in AHK gauge. As pointed out for instance in \cite{Boels:2011zz} for helicity equal gluons one can show\footnote{Technically, this requires a gauge transformation from the Feynman-'t Hooft gauge employed in \cite{Berends:1987me} to the AHK gauge. It is easy to check this gauge transformation and the off-shell leg can be considered to be in the Cartan sub-algebra. The explicit transformation will then add a term $\sim\!p_{\mu} (q\cdot J)$ to the current, but this vanishes in the case we are interested in.} using the explicit expression in \cite{Berends:1987me} this current is proportional to the BCFW shift vector $q^{\mu}$, hence the amplitude scales as $\sim\!(z^{-1})$ for more than one attached gluon. 

For one helicity un-equal the current is proportional to the helicity-equal current with the opposite helicity leg removed, 
\begin{equation}
\begin{array}{rcl}
J_{\alpha \da}^{-} & 	\sim &	J_{\alpha \da}^+   \\
	& 	\sim &	q_{\alpha \da} + L k^-_{\alpha} q_{\da}
\end{array} 
\end{equation}
from the explicit solution for some function $L$. The difference in form to the helicity equal case is caused by a difference in the momentum conserving delta function. $k^-_{\alpha}$ is one of the spinors of the on-shell momentum of the single opposite helicity leg. Hence a BCFW shift of the massive scalar legs of an amplitude of a massive scalar pair of particles coupled to gluons with one helicity opposite gluon will scale as $\sim\!(z^0)$. Following the same steps as above this analysis can be used to show that the massive triangle and bubble coefficients are zero for the helicity equal loop amplitude. These coefficients are not zero for any other helicity amplitude, although they are comparatively simple for the one-helicity opposite contribution. In conclusion, helicity equal amplitudes at one loop in pure Yang-Mills only have massive box contributions. Therefore it is not unreasonable to suspect that the relations conjectured in  \cite{BjerrumBohr:2011xe} for these amplitudes generalize to relations for massive box coefficients. This will be shown next. 

\subsubsection*{Relations for massive box coefficients}
From the analysis of the massive box reviewed briefly above around equation \eqref{eq:cutmassivebox} it should be clear that in the analog of the AHK gauge a box coefficient in a particular channel is proportional to 
\begin{equation}\label{eq:explmassboxpartchannel}
\textrm{box coefficient} \sim \prod_{i=1}^4 V^c_{\mu} J^{\mu}_i
\end{equation}
for a space-like vector $V^c$ discussed above in equation \eqref{eq:massinfbox}. The gluon currents contain the gluons on the specified corner of the box, indexed by the index $i$.  The total massive box contribution to the rational terms of a particular amplitude is
\begin{equation}\label{eq:massboxcoef}
\textrm{mb} \left[A(1,2,\ldots n)\right] \sim \sum_{\textrm{channels}} \left[ \prod_{i=1}^4 V^c_{\mu} J^{\mu}_i \right]
\end{equation}
where the sum ranges over all channels of the box topology. The vector $V^c$ depends on the choice of channel, but not on the order of the gluons on the corners of the box. Hence in the permutation sums considered below all contributions to a particular channel can be added directly. 

As a warm-up consider the massive box coefficients of the four and five point amplitudes. At four points there is only one possible channel. From equation \eqref{eq:explmassboxpartchannel} it is clear that the complete massive box contribution is completely symmetric,
\begin{equation}
\textrm{mb} \left[A(1,2,3,4)\right]  = \textrm{mb} \left[A(1,3,2,4)\right]  = \textrm{mb} \left[A(1,3,4,2)\right] 
\end{equation}
This is a well-known property of the four point helicity equal amplitude. The observation here is that the massive box coefficient shares this particular property for all helicities. As an aside, it is easy to verify that the massive triangle contributions are \emph{not} completely symmetric for four point rational terms. Therefore the one helicity opposite amplitude for instance is not completely symmetric. 

At five points, taking a hint from equation (3.10) from  \cite{BjerrumBohr:2011xe}, consider the permutation sum
\begin{multline}\label{eq:eq310fivept}
\textrm{mb} \left[A(1,2,3,4,5)+A(1,2,3,5,4)+A(1,4,3,5,2) \right. \\ + \left. A(1,4,3,2,5)+A(1,5,3,4,2)+A(1,5,3,2,4)\right]
\end{multline}
for a generic helicity configuration five point amplitude. It can be checked that massive box coefficients vanish in every channel separately. Consider for instance the channel,
\begin{equation}
\{1,2,3,4,5\} \rightarrow \{1,4\} \oplus \{3\} \oplus \{2\} \oplus \{5\}
\end{equation}
Not all amplitudes in equation \eqref{eq:eq310fivept} contribute to this particular channel as particles one and four need to be adjacent. Collecting contributions and keeping track of the order of the gluons involved one arrives at a particular sum over currents for this channel from equation \eqref{eq:explmassboxpartchannel},
\begin{equation}
\sim J(2) J(3) J(5) \left(J(41) + J(14) + J(14) + J(41) \right) = 0
\end{equation}
Here and in the following the vector $V^c$ will be suppressed for notational convenience. This calculation can be repeated in all channels. Note that this is independent of helicity, showing \eqref{eq:eq310fivept} holds for massive box coefficients. An equivalent way of seeing this is to relate the large $\mu$ limit in equation \eqref{eq:cutmassivebox} to a BCFW shift and to note that the type of permutation sum over gluons which arises in this particular channel of \eqref{eq:eq310fivept} results in sums over massive scalar pair amplitudes coupled to glue which scale suppressed by one order of $\mu$ in the large $\mu$ limit. This remark applies to all calculations in this subsection and will not be repeated. 

In general the massive box contribution will vanish for every set of permutations of the particles in the amplitude such that in every channel a sum emerges over orderings of particles on one of the currents which vanishes. This reduces the problem of finding relations for the massive box contributions in principle to combinatorics. Leaving a general solution of this problem to future work (apart from a comment on the number of independent constituents below), here the focus will be on two particular cases: equations (3.12) and (3.16) of  \cite{BjerrumBohr:2011xe}. 

\subsubsection{Proof of equation (3.12) from \cite{BjerrumBohr:2011xe} for massive box coefficients}
Slightly rewritten and extended to the massive box contribution, this series of relations reads
\begin{equation}\label{eq:312bjb}
0= \textrm{mb} \sum_{P(4,\ldots n)} \left[ A(1 4 2 3 5 \ldots n) +  A(1 2 4 3 5 \ldots n) + (n-6) A(1 2 3 4 5 \ldots n) \right]
\end{equation}
The case of $n=5$ of this equation is equivalent to equation \eqref{eq:eq310fivept}. The slightly exceptional case of $n=6$ can also be proven by considering all cuts. This leaves the cases $n>6$.

The proof of equation \ref{eq:312bjb} for all multiplicity follows by considering all channels for the massive box coefficients. From equation \eqref{eq:312bjb} it follows that the massive box coefficient of a generic channel will be zero as there will be a permutation sum of all gluonic particles on one of the currents. The exception to this is when either the current contains only one particle, or if the current contains the exceptional particles $1$, $2$ or $3$. First consider the case where the $1,2,3$ particles are on the same corner of the massive box. For a non-vanishing result there must be at least $n-6$ particles from the set $\{4,\ldots n\}$ on this corner to get a non-vanishing result. Also, $n$ should be larger than $6$ for a non-vanishing result. Since the sum ranges over all permutations of this set, $n-6$ particles $\{4, \ldots, n-6\}$ can be picked without loss of generality. Then this particular channel will yield a permutation sum over currents of the form
\begin{multline}
\sum_{P(4,\ldots n-6)} \left[J(1,2,4,3,5,\ldots n-6) + J(5,\ldots n-6,1,2,4,3\ldots) + J(1,4,2,3,5,\ldots n-6)\right. \\ \left.+ J(5,\ldots n-6,1,4,2,3\ldots) +(n-6)J(1,2,3,4,5,\ldots n-6) \right. \\ \left.+ (n-6)J(5,\ldots n-6,1,2,3,4\ldots) \right] 
\end{multline}
This can be rearranged to be a permutation sum of:
\begin{equation}
\sum_{P(4,\ldots n-6)} \left[\sum_{\beta \in OP {4} \cup \{1,2,3,5,\ldots n-6\}}J(\beta) \right] = 0
\end{equation}
which vanishes. Hence there are no contributions with all three particles $1,2,3$ on the same corner of the massive box. In the following this particular reasoning will be used many times. 

Now consider the two cases where there are two of the exceptional particles on the same corner. For the case particles $1,2$ are on the same corner several possibilities arise, distinguished by how many more particles appear with particles 1 and 2. Let $(k-4)$ be this number with $k\geq 4$. Without loss of generality particles $n$ and $n-1$ can be placed on the single gluon corners of the quadruple cut. The sum over massive box coefficients which arises contains products of four gluon currents, but two of them always involve particles $n$ and $n-1$ so can be disregarded. The sum reduces to a permutation sum over the set $\{4,\ldots, k\}$ and $\{k+1, \ldots, n-2\}$ of
\begin{multline}
\sim J(4, \ldots, k, 1, 2) \left[(n-5) J(3, k+1 \ldots n-2) + J(k+1,3, k+2,  \ldots n-2)  \right] \\ 
\left[J(5, \ldots, k, 1, 2, 4) + J(5, \ldots, k, 1, 4, 2) \right]\,J(3, k+1 \ldots n-2) 
\end{multline}
where $k\geq 4$. Due to the permutation sum these currents can equivalently be treated as being symmetric in $\{4,\ldots, k\}$ and $\{k+1, \ldots, n-2\}$. Using 
\begin{equation}
(n-5) J(3, k+1 \ldots n-2) + J(k+1,3, k+2,  \ldots n-2) =  (k - 3) J(3, k+1 \ldots n-2) 
\end{equation}
the following sum,
\begin{equation}
\sim (k-3) J(4, \ldots, k, 1, 2) + J(5, \ldots, k, 1, 2, 4) + J(5, \ldots, k, 1, 4, 2) 
\end{equation}
arises which vanishes because it can, again, using rotational symmetry be arranged to spell the photon decoupling relation for the gluon current. Hence there are no contributions to the massive box coefficient with particles $1$ and $2$ on the same corner of the box. The same applies directly to the case of $2,3$ on the same corner of the box. Particles $1$ and $3$ cannot appear on the same corner of a box without particle $2$, a case already considered. 

This leaves the case with particles $1$, $2$ and $3$ each on a different corner of the box. This case has two possibilities: either particle $2$ appears alone on one corner, or with one additional particle. In the latter case in this particular channel the combination of currents
\begin{equation}
\sim J(2,4) + J(4,2)
\end{equation}
appears which vanishes. This leaves the case where particle $2$ is isolated on one corner of the box. Without loss of generality pick particle $n$ to be alone on a corner of the quadruple cut. The calculation now reduces to a permutation sum over the sets $\{4,\ldots,k\}$ and $\{k+1, \ldots, n-1\}$ over
\begin{multline}
\sim (n-6) J(k+1,\ldots, n-1, 1) J(3, 4, \ldots k) +  J(k+1,\ldots, n-2, 1, n-1) J(3, 4, \ldots k)\\ +  J(k+1,\ldots, n-1, 1) J(4, 3, 5, \ldots k)
\end{multline}
Using the photon decoupling relation for currents twice this is seen to vanish as well. This concludes the proof of equation \eqref{eq:312bjb}. 

\subsubsection{Proof of equation (3.16) from \cite{BjerrumBohr:2011xe} for massive box coefficients}
The series of relations in equation (3.16) of \cite{BjerrumBohr:2011xe}  suitably extended to massive box coefficients reads for an $n$ gluon amplitude
\begin{equation}\label{eq:316bjb}
0= \textrm{mb} \left[ 6 A(1,2,3,\ldots, n) - \sum_{k=2}^{n-1} \sum_{\sigma \in OP(\alpha_k \cup \beta_k)} \left[ A(1, \sigma) \right] \right]
\end{equation}
with the sets $\alpha_k$ and $\beta_k$ defined as ordered sequences,
\begin{equation}
\alpha_k = \{2,\ldots, k\} \qquad \beta_k = \{k+1,\ldots, n\}
\end{equation}
The five and six gluon cases of equation \eqref{eq:316bjb} can be verified by hand by considering all possible channels as was done above.

For the proof of equation \eqref{eq:316bjb} for general multiplicity can be split into two broad cases: the one where all channels involve sets of consecutively labelled gluons and the one where this is not the case. Consider the latter case first. This can again be split into those channels where the not-consecutively-labelled gluons involve particle one and those which don't. Suppose this is true. Some experimentation shows that this case can only arise for channels involving two consecutive subsets, i.e.
\begin{equation}
(1,\ldots, i, j, \ldots l)
\end{equation}
The reason for the restriction to two subsets is that other possibilities will not be generated by the permutation sums in equation \eqref{eq:316bjb}. The channel with two subsets arises when the consecutive gluons $(i+1, \ldots, j-1)$ are cycled through the $(j \ldots k)$ gluons. Hence the label $k$ on the set $\alpha_k$ in equation \eqref{eq:316bjb} can only be either $j$ or $l$. The latter possibility is when the entire set $(j, \ldots l)$ is moved to the other side of the amplitude. Generically these terms will contain several currents summed over insertions of gluons which will not contribute to the massive box coefficient. For the non-vanishing possibilities the following sum over currents arises in this channel:
\begin{equation}
J(j\ldots l, 1, \ldots i) + \sum_{\sigma \in OP (\{j\ldots l\} \cup \{ j, \ldots l\}} J(1,\sigma)= 0
\end{equation}
which vanishes by application of the Kleiss-Kuijf relations \eqref{eq:KKrel} for currents. 

Next consider channels involving non-consecutive gluons which do not involve particle $1$. By  the same reasoning as before only channels which contain two consecutive gluon subsets need be considered, say 
\begin{equation}
(m,\ldots, i, j, \ldots l)
\end{equation}
These channels are generated by those permutation sums for which the label $k$ on the set $\alpha_k$ in equation \eqref{eq:316bjb} is either $i$ or $j-1$. This leads to the sum 
\begin{equation}
J(m\ldots i, j, \ldots l) + \sum_{\sigma \in OP (\{m\ldots i\} \cup \{ j, \ldots l\}} J(1,\sigma) = 0
\end{equation}
again by application of the Kleiss-Kuijf relations \eqref{eq:KKrel}. This reduces the proof of equation \eqref{eq:316bjb} to the consideration of channels which only involve consecutive gluon subsets. 

First consider the channel in the remaining consecutive gluons case where one of the subsets start with particle $1$, 
\begin{equation}
\{1,\ldots, i\} \oplus \{i+1, \ldots j\} \oplus \{ j+1, \ldots l\} \oplus \{l, \ldots n\}
\end{equation}
Generically, the permutation sums of equation \eqref{eq:316bjb} will kill the box coefficients since they generate $U(1)$ decoupling relations for the gluon current. This leaves exceptional cases. First consider the case of $k=j$ and $k=l$ on the set $\alpha_k$ in equation \eqref{eq:316bjb}. Both of these contribute three terms to the channel under consideration, all six of which are the same as the massive box coefficient of the ordered amplitude in equation \eqref{eq:316bjb}. These then sum to zero. That leaves the permutation sums in  \eqref{eq:316bjb} with $2\leq k < i$. It is fairly easy to see that these can be summed as
\begin{equation}
J(\alpha, 1, \beta) + \sum_{\sigma \in OP (\{1\ldots k \} \cup \{k+1, \ldots i\}} J(1,\sigma)= 0
\end{equation}
so that these channels do not contribute to the massive box coefficient. The channels where particle one is joined by particles $n$ or more on the left work out exactly similarly. 

This concludes the proof of equation \eqref{eq:316bjb} for massive box coefficient contributions to the rational terms of planar pure Yang-Mills one loop amplitudes for general helicity.

\subsubsection{The number of independent massive box contributions}
From the above it is seen that the proof of equations \eqref{eq:312bjb} and \eqref{eq:316bjb} only rely on the $U(1)$ decoupling identity for gluon currents. This observation can be used to calculate the number of independent box coefficients identified under the relations generated by the $U(1)$ decoupling-type identities for currents. This number is important as an indication how much mileage can be gained from relations of this type with a view of speeding up numerical calculations. 

First consider all orderings of particles $1$ through $n$, disregarding cyclic symmetry. The box coefficients can be expressed in terms of currents through equation \eqref{eq:massboxcoef}. Now the $U(1)$ decoupling relations can be solved to express any ordering of the particles in the current in terms of a current with the position of one particle fixed, i.e. for two sets of particles $\alpha$ and $\beta$,
\begin{equation}
J^{\mu}(\alpha, 1, \beta) = \sum_{\sigma \in OP(\alpha^T \cup \beta)} J^{\mu}(1,\sigma)
\end{equation}
In other words there is an explicit basis for currents with $n$ on-shell particles with $(n-1)!$ elements. For counting purposes (only) this is the same as if the gluon current were cyclic in the gluons. 

Hence to count the number of box coefficients taking into account the decoupling identities is simple: this is simply the number of ways $n$ particles can be distributed non-trivially over four currents taking into account cyclic symmetry. This is a particular unsigned Stirling number of the first kind,
\begin{equation}
|S^{n}_4|
\end{equation}
Some of its values are given in table \ref{tab:somevalues}. From this table it is clear there are many more box coefficients than there are color-ordered one-loop amplitudes from six particles onward. Any amplitude can be expressed as a sum over all channels of the massive box coefficients. The question is how many of these coefficients are independent. 

\begin{table}[t] 
\begin{center}
\begin{tabular}{|c|c|c|c|c|c|}
\hline
 	qty $\backslash$ \#		 &	4	& 	5	&	6	&	7	&	8\\ \hline
$1/2 (n-1)! $	 & 	3	&	12	&	60	& 	360	&	2520\\
$|S^{n}_4|  $   	&	1	&	10	&	85	&	735	&	6769\\
$|S^{n-1}_3|$	 &	1	&	6	&	35	&	225	&	1624\\
\hline  
\end{tabular}
\caption{Representative values of several expressions as a function of particle number.  $|S^{n}_i|$ are the unsigned Stirling numbers of the first kind. \label{tab:somevalues}}
\end{center}
\end{table}

First of all note that the quadruple cuts can be classified according to the minimal number of gluons on a corner. For each of these classes one can count the number of quadruple cut coefficients which can be identified under the $U(1)$ decoupling relations for the currents. First consider the class which contains minimally one corner with one gluon. There are 
\begin{equation}
|S^{n-1}_{3}|\, \frac{n}{n} = |S^{n-1}_{3}| 
\end{equation}
different box coefficients in this particular class, identified under the $U(1)$ decoupling relations.  The fraction $\frac{n}{n}$ arises from the $n$ different gluons which can be put on the single corner, while a factor of $n$ has to be divided out because of cyclic symmetry. Now consider the class of box coefficients with minimally two gluons in a corner. There are $|S^{n-2}_{2}|$ different possibilities here, which is strictly less than $|S^{n-1}_{3}|$. Hence the minimal basis for this class is smaller than for the class with just one gluon. It follows that the class of one-gluon box-coefficients therefore sets the bound on the number of independent massive box contributions to the rational terms to $|S^{n-1}_{3}|$. This observation extends the counting of  \cite{BjerrumBohr:2011xe}  for independent helicity equal amplitudes up to eight points to all multiplicity.

It would be very interesting to construct an explicit basis for the massive box contributions to the rational terms as a solution to the identities given above. Beyond this obtaining similar relations for the massive triangle and bubble coefficients is of prime concern as this is a key to understanding and extending the relations for the one helicity opposite finite amplitudes conjectured in  \cite{BjerrumBohr:2011xe}.


\section{Discussion}
In this article non-adjacent BCFW shifts for amplitudes in Yang-Mills theory at tree and the integrand at loop level have been investigated along with applications to derive relations for (mostly one) loop amplitudes. The outcome that integrands scale just as tree level amplitudes better under non-adjacent shifts than under adjacent shifts is perhaps natural given the adjacent shift result \cite{Boels:2010nw}. On the other hand it is surprising that such generic all-loop statements can be found at all, in particular for such a wide variety of minimally coupled Yang-Mills theories with possible scalar potential and Yukawa terms. In particular, our results hold for pure Yang-Mills theory. 

One reason improved scaling for non-adjacent shifts of the integrand is interesting is that it suggests generalizations of the BCJ relations exist at all loop orders for said integrand. The same result is suggested by the extension of the decomposition of \cite{Bern:2008qj} at tree level which originally inspired the BCJ relations to the loop level integrand in \cite{Bern:2010ue}. In section \ref{sec:BCJatoneloop} an explicit BCJ type recursion relation has been presented. Clearly, it will be interesting to study this object further as well as determine its interaction with the standard scalar integral basis.

For tree amplitudes several mechanisms to improve BCFW scaling beyond generic non-adjacent shifts have been discussed. These mechanisms play a central role in the proof of several new relations for coefficients of the standard integral basis. These relations for the coefficients deserve further study, especially with a view toward speeding up numerical code to calculate cross-sections. Especially for the rational terms there is scope for improvements in existing numerical code for amplitude calculations. One step in this direction would be to study minimal basis type solutions of the relations presented. Some counting results have been presented above. Furthermore, relations for massive triangle and bubbles should also be studied: the results in \cite{BjerrumBohr:2011xe} for the one helicity-unequal amplitudes can serve as a guide here. 

There are various interesting avenues along which the results in this article can be extended. One such avenue are amplitudes and integrands at one loop with external fermionic legs which transform in the fundamental. QCD is an example of such a theory after all. More theoretically, the results above for generalizations of non-adjacent shifts certainly also translate to expectations for more supersymmetric shifts in supersymmetric theories such as $\mathcal{N}=4$ in four dimensions and the superstring in ten. What role these generalized shifts play in these theories is an interesting question open to wide speculation. 

\acknowledgments
It is a pleasure to thank Simon Badger, Zvi Bern, Emil Bjerrum-Bohr and Michael Kiermaier for discussions and/or correspondence. All Feynman graphs in this paper have been drawn using Jaxodraw \cite{Binosi:2008ig}, which is based on the Axodraw package \cite{Vermaseren:1994je}. This work was supported by the German Science Foundation (DFG) within the Collaborative Research Center 676 "Particles, Strings and the Early Universe". 
 \vspace{1cm}


\newpage

\begin{appendices}

\section{Color-ordered Feynman rules}
\label{app:feynrules}

This appendix contain an overview of color-ordered Feynman rules of Yang-Mills theory minimally coupled to scalar or spin one half matter in AHK lightcone gauge. 

\begin{figure}[h!]
\centering
\includegraphics[scale=0.57]{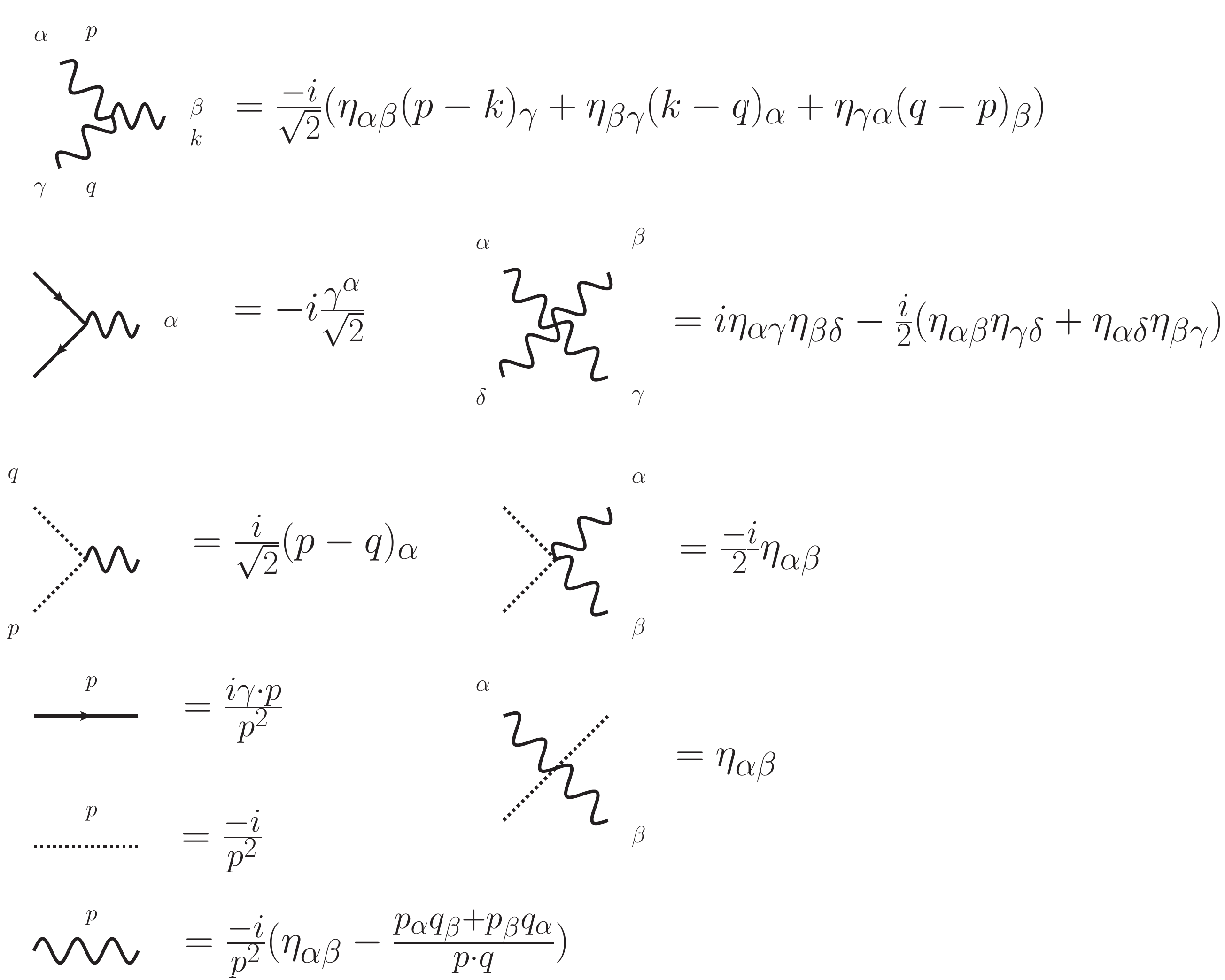}
\caption{\label{fig:feynrules} Color-ordered Feynman rules for gluons (wiggly lines), scalars (dotted lines) and fermions (straight lines).}
\end{figure}


\newpage

\section{List of Feynman graphs used in section \ref{sec:nonadjshftintegr}}
\label{app:diagrams} 
This section lists the diagrams used in the derivation of the non-adjacent large-$z$ scaling of equation \eqref{r_gennonadscale} in section \ref{sec:gennonadj}.

\subsection*{Gluonic diagrams}
The convention for labeling the external gluonic legs and indices is as follows: $(1_\mu)$, $(2_\nu)$, $(3_\rho)$, $(4_\sigma)$, $(5_\tau)$, $(6_\lambda)$. Hats denote the shifted legs.

\begin{figure}[h!]
\centering
\includegraphics[scale=0.55]{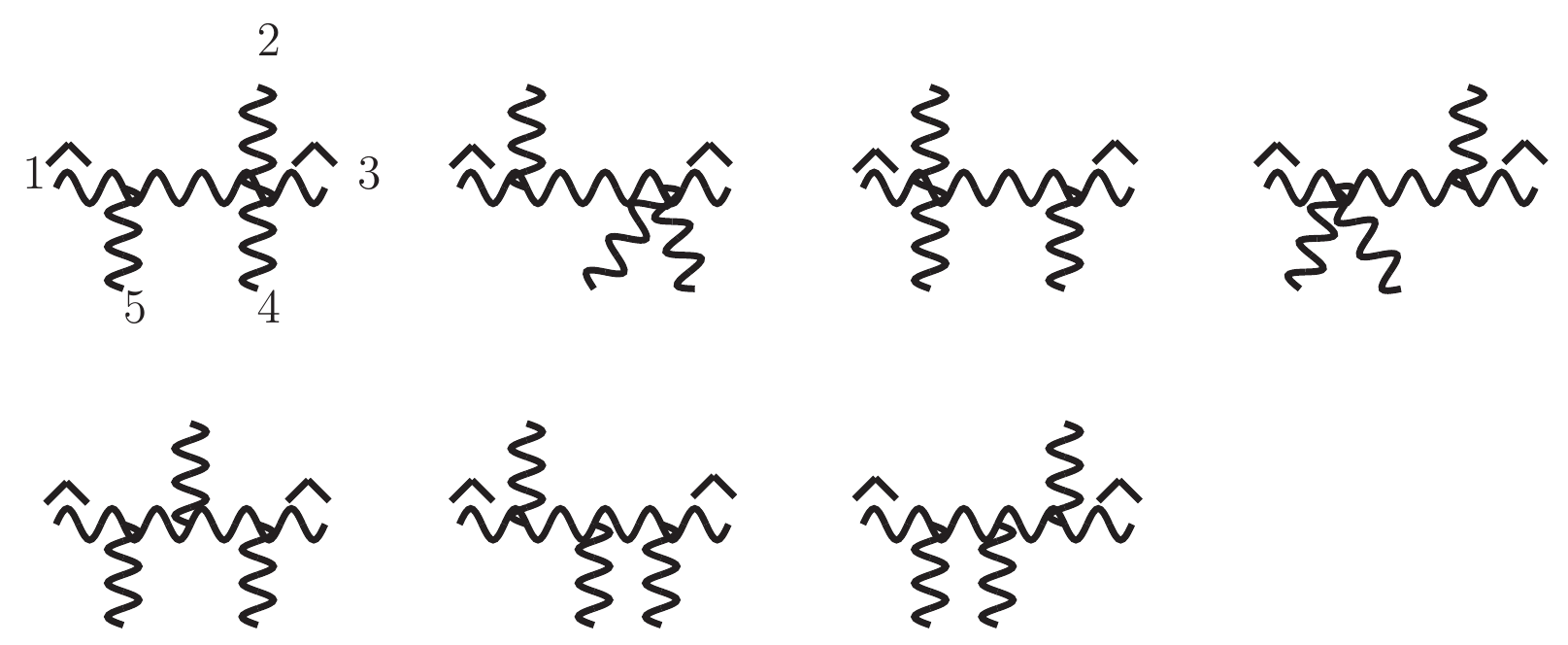}
\caption{\label{fig:5ptglue} Color-ordered gluon diagrams at five points. Legs 1 and 3 have been shifted.}
\end{figure}
\begin{figure}[h!]
\centering
\includegraphics[scale=0.55]{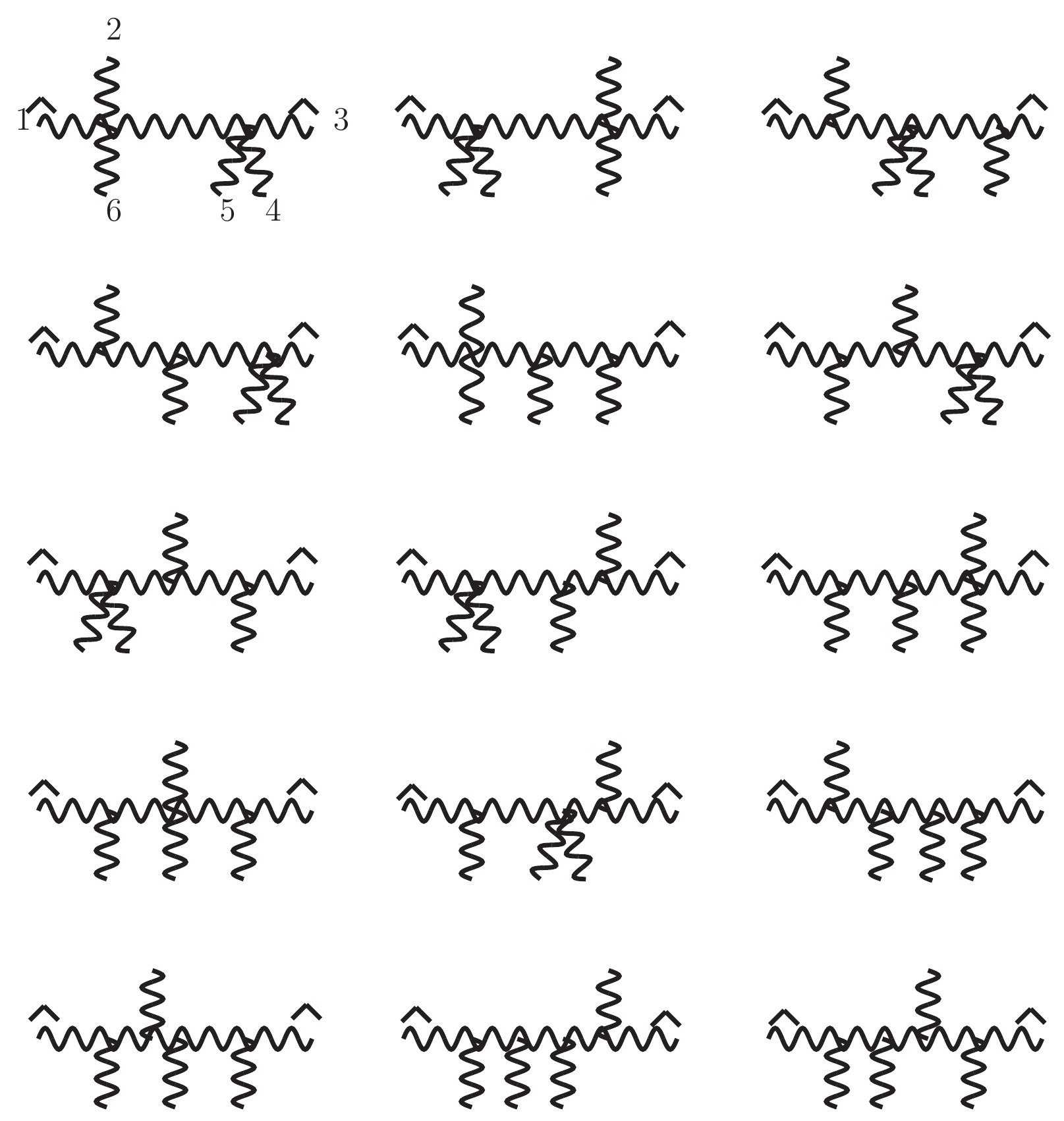}
\caption{\label{fig:6ptglue} Diagrams contributing to the shift (1, 3) at six legs. The diagrams for the shift (1, 4) are the same as in the 6 point scalar-gluon case with scalars exchanged to gluons (fig \ref{fig:6pt2scalar}).}
\end{figure}
\newpage
\subsection*{Scalar contribution diagrams}

\begin{figure}[h!]
\centering
\includegraphics[scale=0.5]{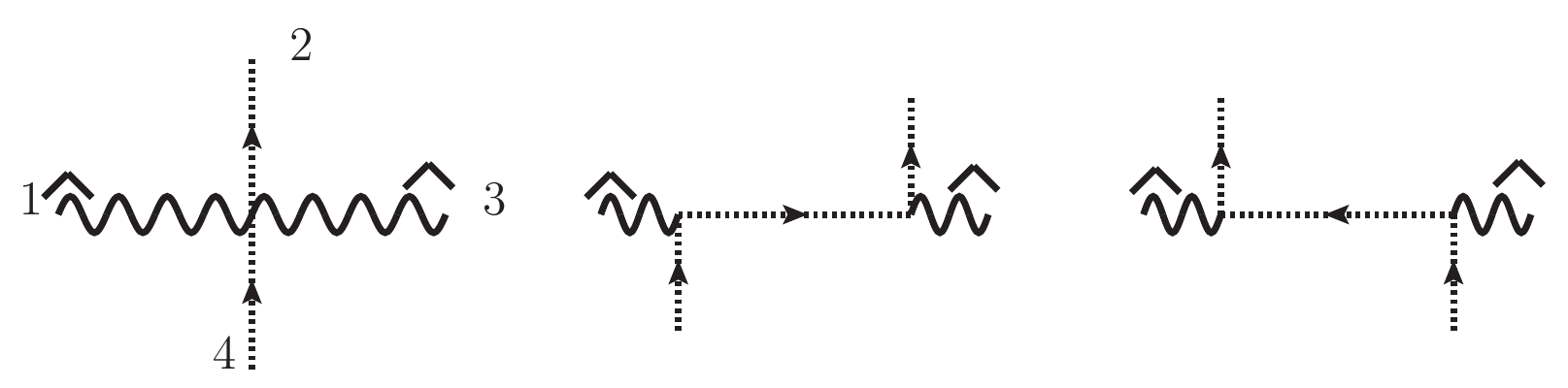}
\caption{\label{fig:4ptscalar} Scalar-glue diagrams contributing to the shift $(\hat{1},\mu),\;(\hat{3},\rho)$ at four points.}
\centering
\includegraphics[scale=0.5]{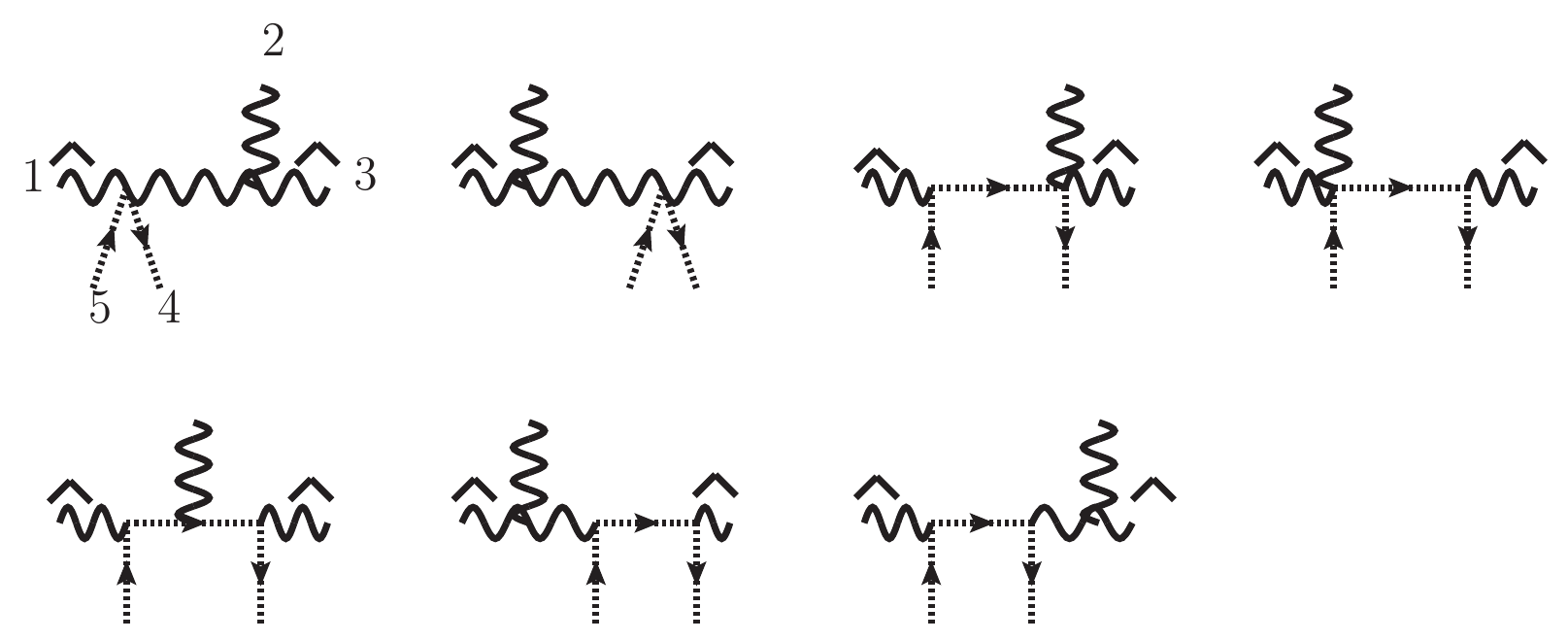}
\caption{\label{fig:5ptscalar} Scalar-glue diagrams contributing to the shift $(\hat{1},\mu),\;(\hat{3},\rho)$ at five points for the particle configuration $g_1,g_2,g_3,s_4,s_5$.}
\end{figure}

\begin{figure}[h!]
\centering
\includegraphics[scale=0.5]{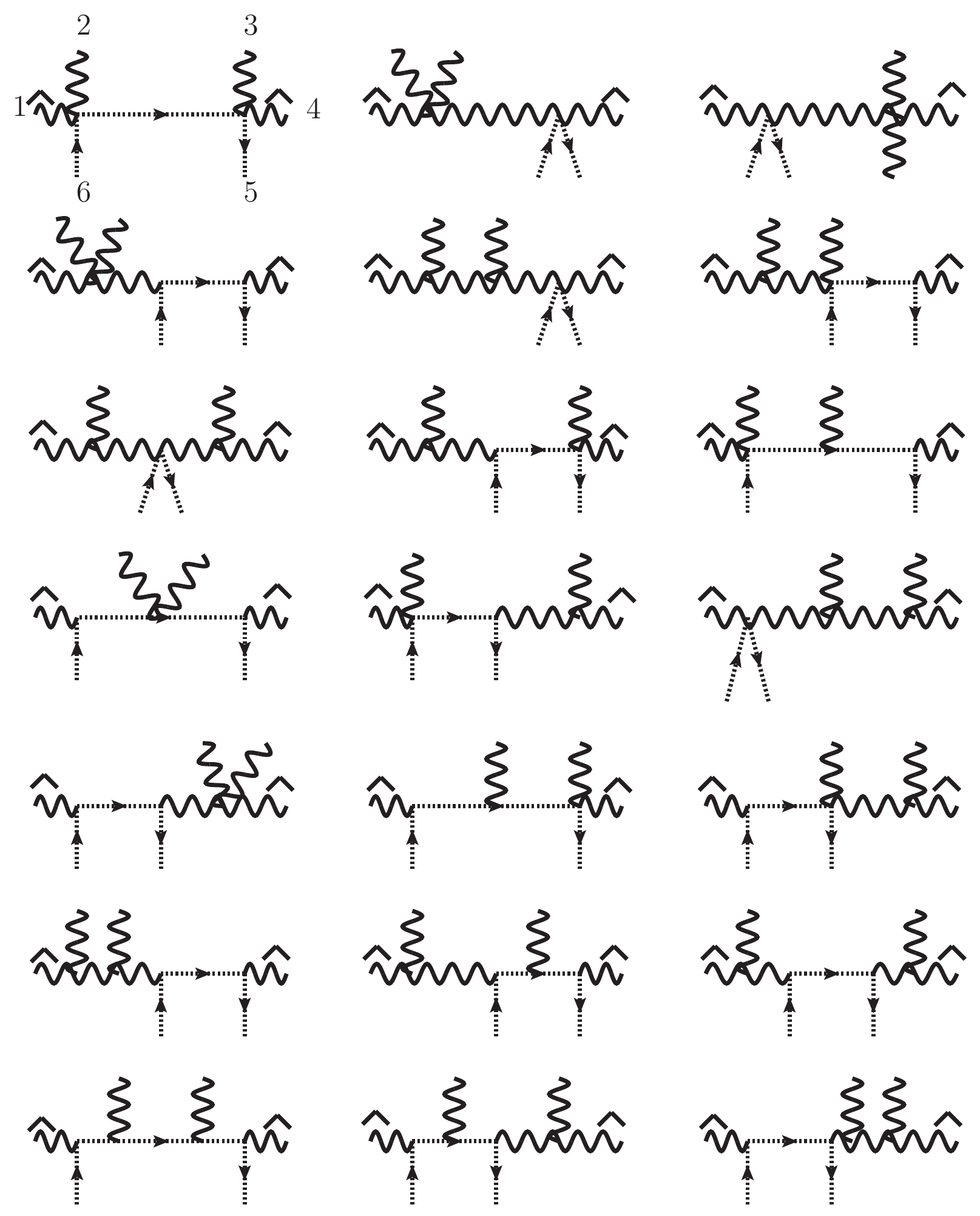}
\caption{\label{fig:6pt2scalar} Scalar-glue diagrams contributing to the shift $(\hat{1},\mu),\;(\hat{4},\rho)$ at six points for the particle configuration $g_1,g_2,g_3,g_4,s_5,s_6$.}
\end{figure}

\subsection*{Fermion contribution diagrams}

\begin{figure}[h!]
\centering
\includegraphics[scale=0.59]{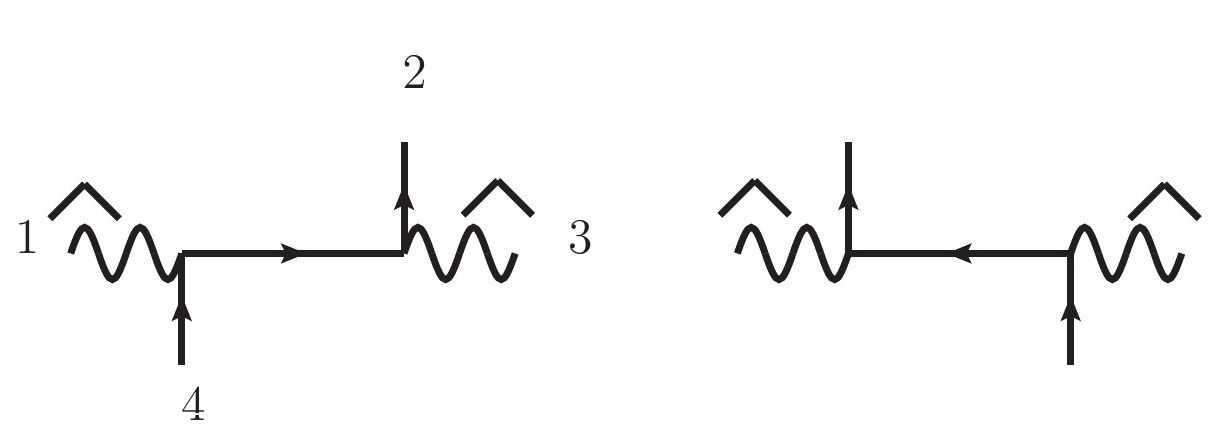}
\caption{\label{fig:4ptferm} Fermion-glue diagrams contributing to the shift $(\hat{1},\mu),\;(\hat{3},\rho)$ at four points.}
\centering
\includegraphics[scale=0.59]{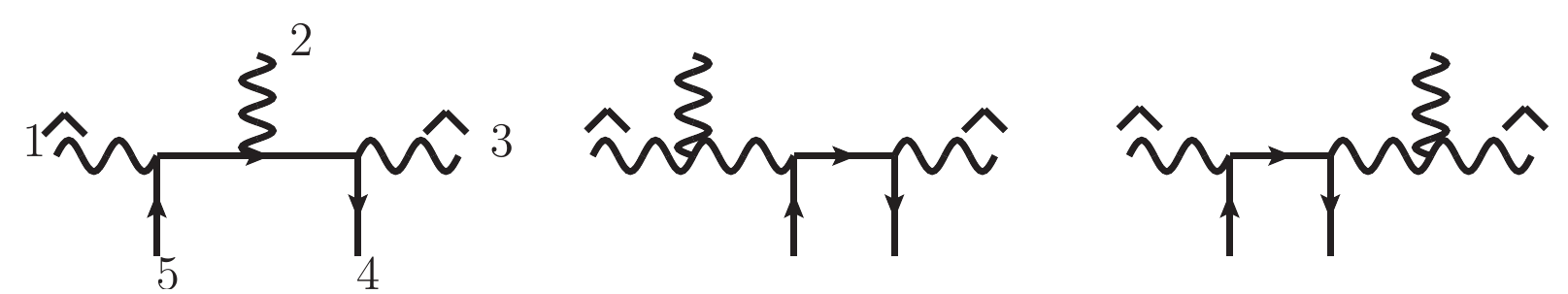}
\caption{\label{fig:5ptferm} Fermion-glue diagrams contributing to the shift $(\hat{1},\mu),\;(\hat{3},\rho)$ at five points for the particle configuration $g_1,g_2,g_3,f_4,f_5$.}
\end{figure}

\begin{figure}[h!]
\centering
\includegraphics[scale=0.59]{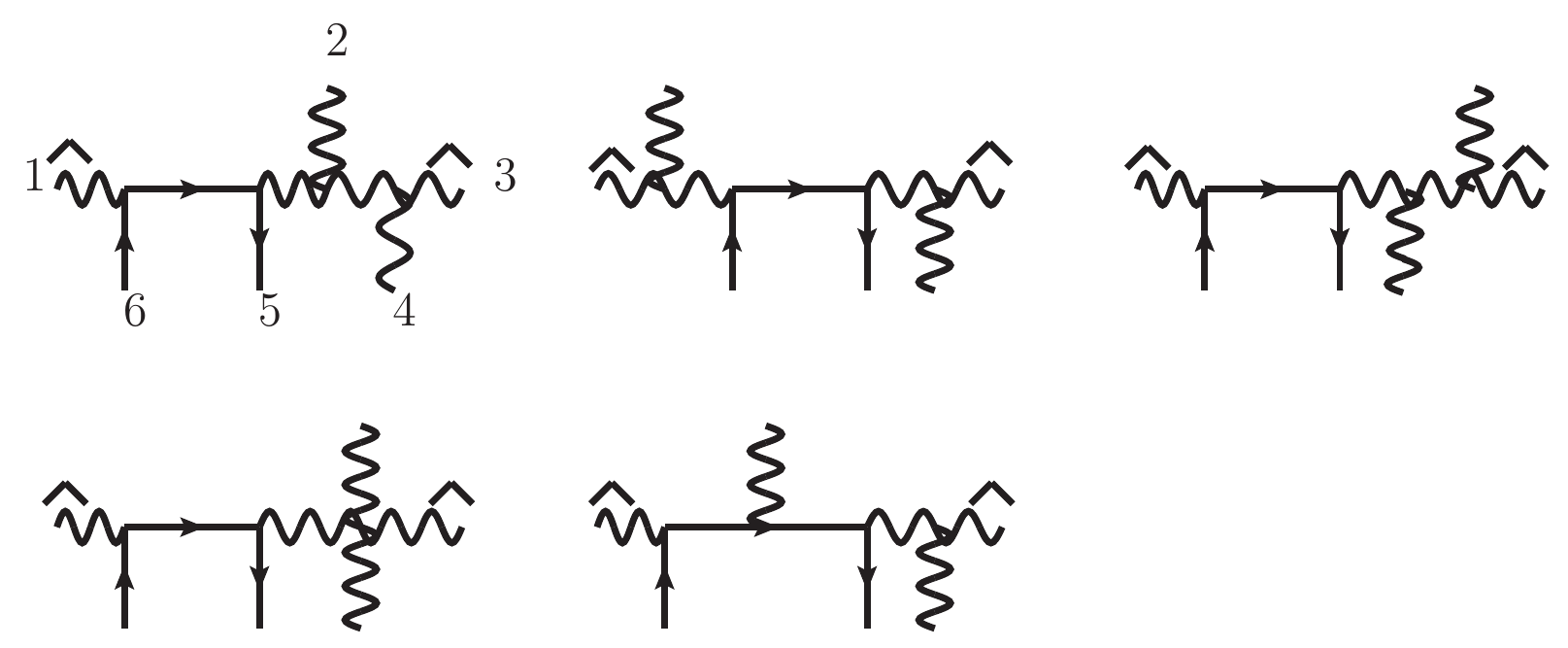}
\caption{\label{fig:6ptfermother} Fermion-glue diagrams contributing to the shift $(\hat{1},\mu),\;(\hat{3},\mu)$ at six points for the particle configuration $g_1,g_2,g_3,g_4,f_5,f_6$.}
\end{figure}

\begin{figure}[h!]
\centering
\includegraphics[scale=0.59]{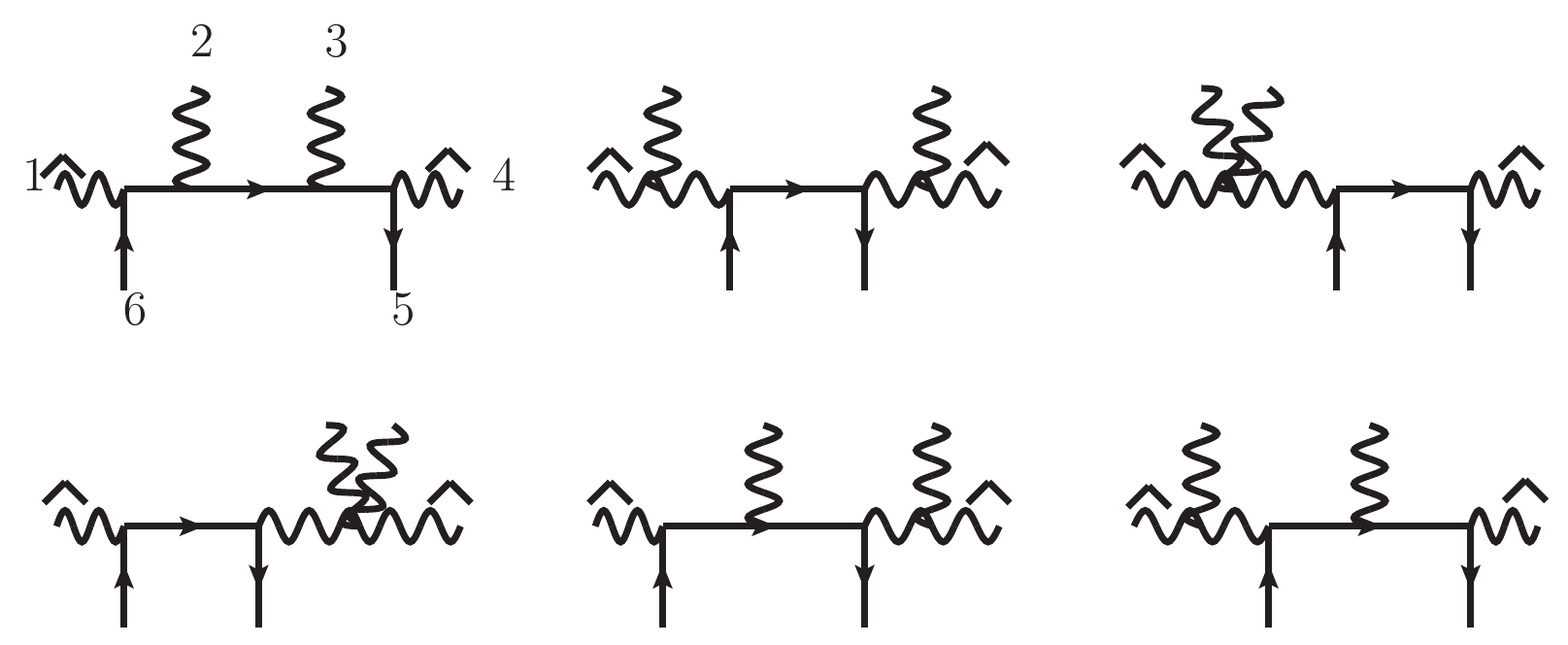}
\caption{\label{fig:6ptferm} Fermion-glue diagrams contributing to the shift $(\hat{1},\mu),\;(\hat{4},\rho)$ at six points for the particle configuration $g_1,g_2,g_3,g_4,f_5,f_6$. Note resemblance to dog of bottom left figure.}
\end{figure}

\newpage

\section{BCFW shifts of particles other than gluons from supersymmetry}\label{app:othermatters}

In principle the same analysis as in the main text in various places for BCFW shifts of two gluons can be performed for shifts of any two types of matter particle, either using the background field method or the more direct AHK gauge analysis. For amplitudes and integrands in supersymmetric field theories for which shifts of gluons to arbitrary matter are known this appendix presents a simple way to obtain some of these shifts using on-shell super-fields. 

For simple four dimensional supersymmetry the supersymmetry algebra can be represented on-shell by fields on an on-shell superspace which has one fermionic coordinate for every massless leg, 
\begin{equation}\label{eq:superfield}
\phi(\eta) = \phi^0 + \eta \phi^1
\end{equation}	
The supersymmetry algebra can be represented in terms of the on-shell spinors of the massless legs as
\begin{equation}\label{eq:susyalgonsh}
Q^{\alpha} = \sum_i \lambda^{\alpha}_i \eta_i \qquad \bar{Q}^{\da} = \sum_i \lambda^{\da}_i \frac{\delta}{\delta \eta_i} 
\end{equation}
where $i$ labels the legs of the amplitude. These should be thought of as fermionic partners to the momentum. The BCFW shift in terms of the spinors of, say, legs one and two is
\begin{align}
1_{\alpha} & \rightarrow \hat{1}_{\alpha} \equiv 1_{\alpha} + z 2_{\alpha} \\
2_{\da} & \rightarrow \hat{2}_{\da} \equiv 2_{\da} - z 1_{\da} 
\end{align}
This shift keeps bosonic momentum, but does not leave the supersymmetry charges of equation \eqref{eq:susyalgonsh} invariant. In order to achieve this also shift
\begin{equation}
\eta_2 \rightarrow \hat{\eta}_2 \equiv \eta_2 - z \eta_1
\end{equation}
Together with the shift of the bosonic spinors this is a BCFW super-shift \cite{Brandhuber:2008pf}. It can be shown \cite{ArkaniHamed:2008gz} that the amplitudes as a function on superspace shifts under this shift as
\begin{equation}
\lim_{z \rightarrow \infty} A\left(\{\eta_1, \hat{1}_{\alpha}, 1_{\da} \}, \{\hat{\eta}_2, 2_{\alpha}, \hat{2}_{\da} \}, \phantom{\tilde{X}} \!\!\!\!\! X \right) = \lim_{z \rightarrow \infty} A\left(\{0, \hat{1}_{\alpha}, 1_{\da} \}, \{0, 2_{\alpha}, \hat{2}_{\da} \}, \tilde{X} \right)  
\end{equation}
where $\tilde{X}$ is a transformed set of supersymmetric variables which is independent of $z$. Hence on a given superspace the behavior under the super-shift is determined by the ordinary BCFW shift of the particles on the amplitude which do not depend on $\eta$ in equation \eqref{eq:superfield}. These are the top-states of the supersymmetry multiplet. 

This result can be applied directly in $\mathcal{N}=4$ super Yang-Mills for instance. From the known results on shifts of color adjacent gluons for the integrand for instance one shows that the adjacent super-shift scales as $\sim(z^{-1})$ in the superspace where the top-states are one of the two gluon polarization states. Note for this to work one the gluon shift has to be known for all possible external states in principle. Similarly, from the above results on shifts of non-color-adjacent gluons for the integrand coupled to various forms of matter one obtains the non-adjacent super-shift of two particles as $\sim(z^{-2})$ on the same type of superspace as in the color-adjacent case. 

Furthermore, a fermionic Fourier transform can be used to choose any type of state in the multiplet as the top-state. More precisely, transforming only legs one and two yields
\begin{equation}
 A\left(\{\hat{\bar{\eta}}_1, \hat{1}_{\alpha}, 1_{\da} \}, \{\bar{\eta}_2, 2_{\alpha}, \hat{2}_{\da} \}, \phantom{\tilde{X}} \!\!\!\!\! X \right) = \int d\eta_1 d\hat{\eta}_2 e^{\hat{\bar{\eta}}_1 \eta_1 + \bar{\eta}_2 \hat{\eta}_2} A\left(\{\eta_1, \hat{1}_{\alpha}, 1_{\da} \}, \{\hat{\eta}_2, 2_{\alpha}, \hat{2}_{\da} \}, \phantom{\tilde{X}} \!\!\!\!\! X \right)
\end{equation}
where 
\begin{equation}
\hat{\bar{\eta}}_1 \equiv \bar{\eta}_1 + z \bar{\eta}_1
\end{equation}
The integration measure can be written in a $z$-independent way since
\begin{equation}
\int d\eta_1 d\hat{\eta}_2 e^{\hat{\bar{\eta}}_1 \eta_1 + \bar{\eta}_2 \hat{\eta}_2} \ldots = \int d\eta_1 d\eta_2 e^{\bar{\eta}_1 \eta_1 + \bar{\eta}_2 \eta_2} \ldots
\end{equation}
In an $\mathcal{N}=1$ supersymmetric field theory in four dimensions for instance, this relates the bosonic BCFW shift of $g^+, g^+$ to the shift of $\psi^+, \psi^+$. So for every case in which the shift of these gluons is known by the results in this article, the shift of the associated fermions follows. From this one shows that the bosonic BCFW shift of the following particles  in $\mathcal{N}=4$ supersymmetric Yang-Mills theory on the integrand scales as $\sim\!(z^{-2})$ 
\begin{equation}
A\left(\widehat{g^\pm},Y,\widehat{g^\pm}, \phantom{\tilde{X}} \!\!\!\!\! X \right), \quad A\left(\widehat{\psi^{I,\pm}},Y,\widehat{\psi^{I,\pm}},  \phantom{\tilde{X}} \!\!\!\!\! X \right), \quad A\left(\widehat{\phi^{IJ}},Y,\widehat{\phi^{IJ}}, \phantom{\tilde{X}} \!\!\!\!\! X \right)
\end{equation}
up to terms which integrate to zero.

Extension of this analysis to theories with higher dimensional supersymmetry is immediate as the above analysis basically only requires the two shifted legs to be in a four dimensional sub-space spanned by the momenta in the shifted legs, the BCFW shift vector and its conjugate. For the other legs it is then only necessary for a superspace to exist. An argument directly in terms of higher dimensional superspaces \cite{Boels:2009bv} should also be possible.


\section{Non-adjacent and non-planar shifts of one loop amplitudes}\label{sec:shiftoneloop}
This appendix presents a partial analysis of non-adjacent and non-planar shifts of integrated integrands: of the complete one loop amplitudes. 

\subsection*{Shifts of helicity equal amplitudes}
A particular simple class of one-loop amplitudes in pure Yang-Mills is formed by those amplitudes which have all helicities on the external legs equal. This makes these amplitudes ideally suited to study questions of BCFW shift dependence in this particular case. For the readers convenience, the leading color all-plus amplitude first obtained in \cite{Bern:1993qk} is given up to an overall constant by
\begin{equation}
A(1,\ldots, n) = \frac{\sum_{1 \leq i<j<k<l\leq n} \braket{i j} \sbraket{j k} \braket{k l} \sbraket{l i}    }{\braket{12} \ldots \braket{n 1}}
\end{equation}
The BCFW shifts of this amplitude and more importantly on-shell recursion relations for this amplitude have been studied first in \cite{Bern:2005hs}. Both adjacent and non-adjacent shifts of this leading color amplitude scale as $z^0$. In general the non-planar amplitudes are generated from this by formula \eqref{eq:sub-leadingfromleading}. For instance,
\begin{equation}
A(1;2, 3, 4) = A(1,2,3,4) + A(2,1,3,4) + A(2,3,1,4)
\end{equation}

For orientation, let us calculate BCFW shifts of some four point non-planar amplitudes first. Shifting particles $3$ and $4$ and some help from S@M \cite{Maitre:2007jq} gives
\begin{align}
A(1; 2, 3, 4) & \rightarrow - 3 \frac{\sbraket{14}\sbraket{24}}{\braket{1 3}\braket{2 3}} + \mathcal{O}\left(\frac{1}{z^{1}}\right) \\
A(1, 2; 3, 4) & \rightarrow 2 \frac{\sbraket{12}\sbraket{34}}{\braket{1 2}\braket{3 4}} -  4 \frac{\sbraket{14}\sbraket{24}}{\braket{1 3}\braket{2 3}} + \mathcal{O}\left(\frac{1}{z^{1}}\right) \\
A(1, 2, 3; 4) & \rightarrow  \frac{\sbraket{12}\sbraket{34}}{\braket{1 2}\braket{3 4}} -  2 \frac{\sbraket{14}\sbraket{24}}{\braket{1 3}\braket{2 3}} +\mathcal{O}\left(\frac{1}{z^{1}}\right)
\end{align}
as the distinct three possibilities for ordering the shifted particles. All of these are evidently order $\sim\!(z^0)$, just as the leading color part of the all-plus amplitude. This would seem to be the end of the story. 

However, one can go quite a bit further using (partial) numerics. This will circumvents possible obstructions to obtaining manifestly vanishing results, such as Schouten identities and momentum conservation. The following results have been obtained using Mathematica and the S@M package for a set of randomly generated momenta, leaving the shift parameter $z$ arbitrary. Coefficients smaller than $10^{-8}$ have been taken to be zero; all non-zero coefficients are typically order $1$.  For shifts of color adjacent gluons
\begin{equation}
A(1; 2, \ldots, \widehat{n-1},\widehat{n}) \rightarrow \quad \sim z^0 \qquad n \leq 10
\end{equation}
is obtained. Shifting two gluons on separate color traces yields
\begin{equation}
A(\widehat{1}; 2, \ldots, \widehat{n}) \rightarrow \quad \sim z^0 \qquad n \leq 10
\end{equation}
Again, these results simply mirror the leading color result.  For non-adjacent shifts of two gluons still on the same color trace with this color configuration we obtain
\begin{equation}
A(1; 2, \ldots, \widehat{i},\ldots, \widehat{n}) \rightarrow \quad \sim \frac{1}{z} \qquad n \leq 10
\end{equation}

The next class of examples are sub-leading amplitudes with two gluons in one trace and the rest in the other. Adjacent shifts on the same color trace give
\begin{equation}
A(1, 2; \ldots, \widehat{n-1},\widehat{n}) \rightarrow \quad \sim \frac{1}{z} \qquad n \leq 10
\end{equation}
and
\begin{equation}
A(\widehat{1},\widehat{2}; \ldots, n-1,n) \rightarrow \quad \sim z^0 \qquad n \leq 10
\end{equation}
while non-adjacent shifts give
\begin{equation}
A(1, 2; \ldots, \widehat{i}, \ldots, \widehat{n}) \rightarrow \quad \sim \frac{1}{z^2} \qquad n \leq 10
\end{equation}
Shifts of particles on two different traces yield
\begin{equation}
A(\widehat{1}, 2; \ldots, \widehat{n}) \rightarrow \quad \sim z^0 \qquad n \leq 10
\end{equation}
The pattern of the previous example persists for all other color traces and shift combinations of the all plus amplitude up to ten gluons. These results can be summarized by comparing the scaling of a certain shift to the adjacent shift of the leading color amplitude:
\begin{itemize}
\item shifts of two gluons on different color traces scale the same
\item shifts of two color adjacent gluons scale
	\begin{itemize}
	\item the same if the trace of the shifted gluons contains two gluons
	\item suppressed by an additional $\sim\!(z^{-1})$ else
	\end{itemize}
\item shifts of two non-color-adjacent gluons on the same trace scale  by an additional $\sim\!(z^{-2})$ if the other trace contains more than one gluon, and by $\sim\!(z^{-1})$ if there is one gluon.
\end{itemize}

There is one obstacle to drawing a generic conclusion from these results. This obstacle is the fact that an analysis of one loop diagrams as in \cite{Boels:2010nw} shows that the generic leading contribution of a BCFW shift of a one-loop amplitude is of order $z$, while that of the all-plus amplitude shifts as $\sim\!(z^0)$ to leading order. It can be seen from \cite{Boels:2010nw} that, barring special kinematics, all-plus is the only helicity configuration for which this happens. To account for this effect we have studied in addition shifts of the `one-minus' four and five point amplitudes using expressions found for instance in \cite{Bern:2005ji}. 

This leads to the pattern for non-adjacent shifts of one-loop amplitudes in
\begin{suspicion}\label{sus:nonplanshift}
For all multiplicities at one loop shifts of the sub-leading color amplitudes compare to the generic adjacent shift of particles with the same quantum numbers of the leading color amplitude as:
\begin{itemize}
\item shifts of two gluons on different color traces are suppressed by $\sim\!(z^{-1})$
\item shifts of two color-adjacent gluons scale are suppressed by $\sim\!(z^{-1})$ (unless the other color
trace is empty)
\item shifts of two non-color-adjacent gluons on the same trace  are $\sim\!(z^{-2})$ suppressed by $\sim\!(z^{-2})$ (unless
the other color trace is empty)
\end{itemize}
\end{suspicion}
Again, the evidence for this is numerical checks for up to $10$ particles for helicity equal amplitudes and up to $5$ particles for the one unequal ones at one loop. 

\subsection*{Partial analysis of non-planar shifts via the background field method }
It would be interesting to prove the above suspicion for all the different assertions made. Evidence for these can be obtained by considering the analysis of \cite{Boels:2010nw} for one-loop amplitudes in pure Yang-Mills theory. We refer to that paper for an in-depth discussion of the analysis and its various drawbacks. In brief, one can try to classify all diagrams which contribute to the large-$z$ shift of a general one-loop amplitude in background field Feynman-'t Hooft gauge keeping the tree diagrams in AHK gauge. The UV divergent structure of the diagrams follows from standard powercounting which can be translated into large-$z$ shift behavior. One finds that there are possible ambiguities at sub-leading ($\sim\!(z^{0})$) order (both diagrams unaccounted for as well as missing full analysis of IR divergences). These issues will, just as in \cite{Boels:2010nw}, be ignored here and postponed to future work. The upshot is for color-adjacent shifts of the leading color diagrams that there is only a limited class of diagrams to be considered. It was shown that the leading behavior for large $z$ is due to the trivalent triangle-loop graph. This mirrors the known results at tree level where the leading large $z$ behavior is caused by the three-vertex. Unintegrated this graph is given by
\begin{align}
\nonumber \epsilon_1^\mu \epsilon_2^\nu A^{triangle,planar}_{\mu\nu\rho} = \sqrt{2}\epsilon_1^\mu \epsilon_2^\nu \int \frac{d^D l}{(2\pi)^D}&\frac{1}{l^2(l-\hat{p}_1)^2 (l+ \hat{p}_2)^2}\Big((D-2)(l_\mu l_\nu (2l -\hat{p}_1+\hat{p}_2)_\rho)\\& + 4(l_\mu \eta_{\rho\nu}+l_\nu \eta_{\rho\mu}+l_\rho\eta_{\mu\nu})\Big) + \mathcal{O}\left(\frac{1}{z}\right)
\end{align}
where the order $\frac{1}{z}$ indicate the order after integration. It can be integrated using standard means to
\begin{equation}
 \epsilon_1^\mu \epsilon_2^\nu A^{triangle,planar}_{\mu\nu\rho} \sim \epsilon_1^\mu \epsilon_2^\nu \Big(\frac{1}{2 p_1\cdot p_2}\Big)^\epsilon \Gamma(\epsilon)\Big([z](\eta_{\mu\nu}f_0^1+\epsilon \frac{P_\mu P_\nu}{2p_1\cdot p_2}f_1^1)+f^1_{2\mu\nu}[z^0]+ \mathcal{O}\left(\frac{1}{z}\right) \Big)
\end{equation}
where $P$ is the sum of $p_1$ and $p_2$ and $f_i$ are functions close to unity for small $\epsilon$. Moreover $f_{2,\mu\nu}$ is antisymmetric.

In this section we indicate briefly how those results may be used to study shifts of two gluons on different color traces and the improved scaling behavior stated in the suspicions will be traced back to the absence / vanishing of trivalent triangle graphs for sub-leading color contributions.

\subsection*{Shifts of two gluons on different color traces}
\begin{figure}[h!]
\centering
\includegraphics[scale=0.15]{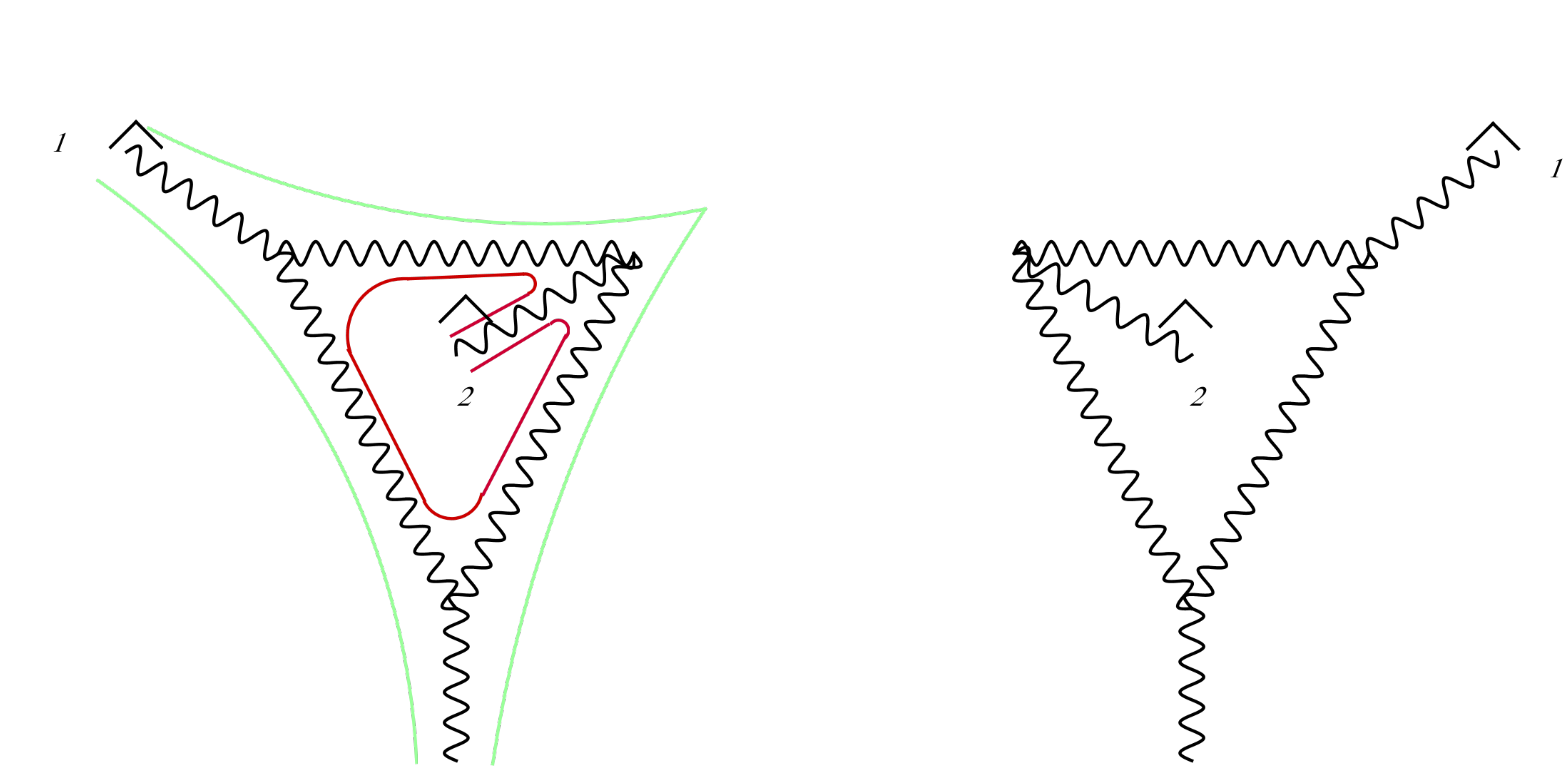}
\caption{\label{fig:diffcolortrace} The non-planar triangle graphs with particle one and two on distinct color traces. The two color traces are
given by the red and green line.}
\end{figure}

The non-planar versions of the trivalent triangle graph will have one or more legs inside the loop. One can see from picture \ref{fig:diffcolortrace} that the non-planar version differs from the planar version only by a minus sign arising from turning one leg inside the diagram thereby effectively flipping one three-vertex. Since there are two ways in which the two shifted legs can be distributed among the two color traces there are two distinct classes of diagrams. One class has the shifted leg $1$ on the outer color trace and the other class has leg $2$ on the outer trace. The diagrams belonging to the former case is depicted in figure \ref{fig:diffcolortrace}.
Pick for instance leg one on the outer color trace and particle two on the inner color trace. Within this class the relative position of the shifted legs is unimportant, i.e. there are  two diagrams for this configuration (the other diagrams being the mirror image of this one, see above) and one has to sum over them. Since the diagrams themselves are antisymmetric under the exchange of the position of the two shifted legs it follows that the sum over the two triangle diagrams vanishes at this order in $z$. In this way the overall scaling of the one-loop graphs is reduced from $\sim\!(z^{1-\epsilon})$ to $\sim\!(z^{0-\epsilon})$. This provides evidence for the first item in suspicion \ref{sus:nonplanshift}.

\begin{figure}[h!]
\centering
\includegraphics[scale=0.15]{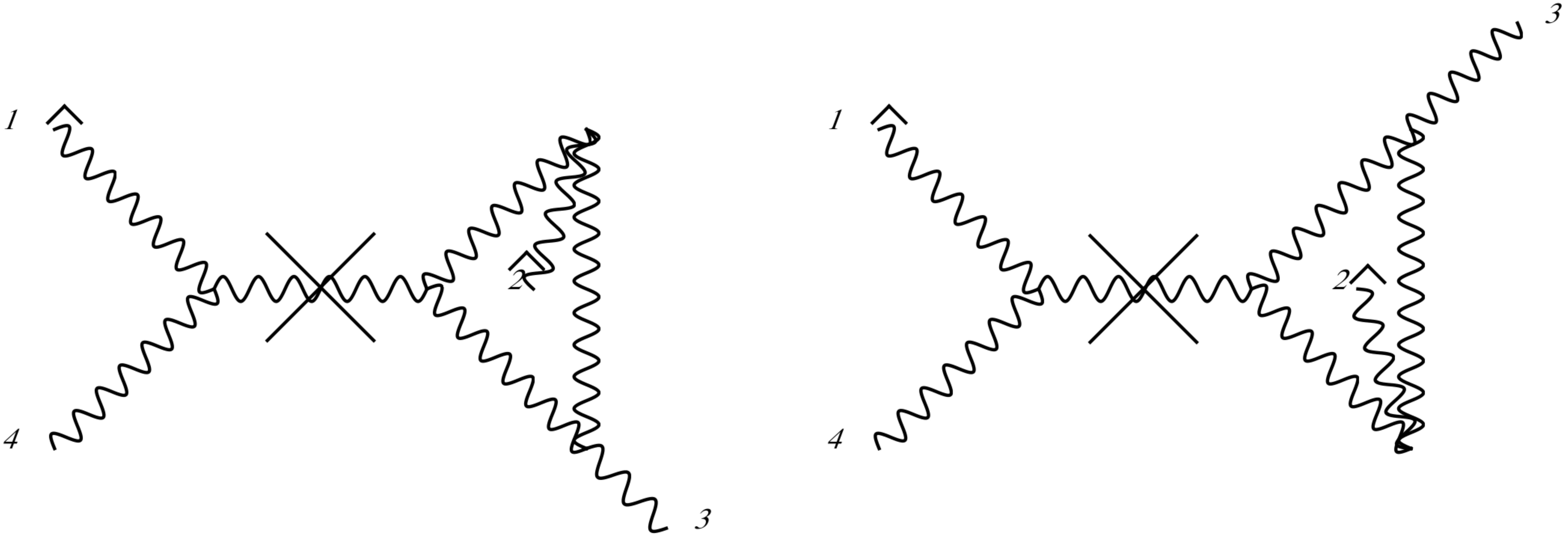}
\caption{\label{fig:ahktri} The non-planar triangle with an AHK propagator attached. The cross indicates that only the $\sim\!(z^{0})$ part of the propagator is considered (see equation \eqref{ahkhardline}) .}
\end{figure}

One can show in a similar fashion that the non-planar versions of the triangle graphs with an AHK propagator attached from \cite{Boels:2010nw} (fig. \ref{fig:ahktri}) cancel as well. The reasoning for these cancellations is exactly the same as above. 

\subsection*{Shifts of two gluons on the same trace}
The strategy is again to show that the non-planar siblings of triangle diagram either vanish or are absent. Consider first the case of non-adjacent gluon shifts on the same color trace. By definition there have to be more than three particles on one trace therefore explicitly ruling out the trivalent triangle diagram as a contribution. Consequently the scaling of color-non-adjacent gluons on the same trace will be suppressed by one power in $z$ starting at $\sim\!(z^{0-\epsilon})$. This mirrors the tree-level behavior again. To show that there is additional power of suppression requires more work and will be postponed to the future.

\begin{figure}[h!]
\centering
\includegraphics[scale=0.15]{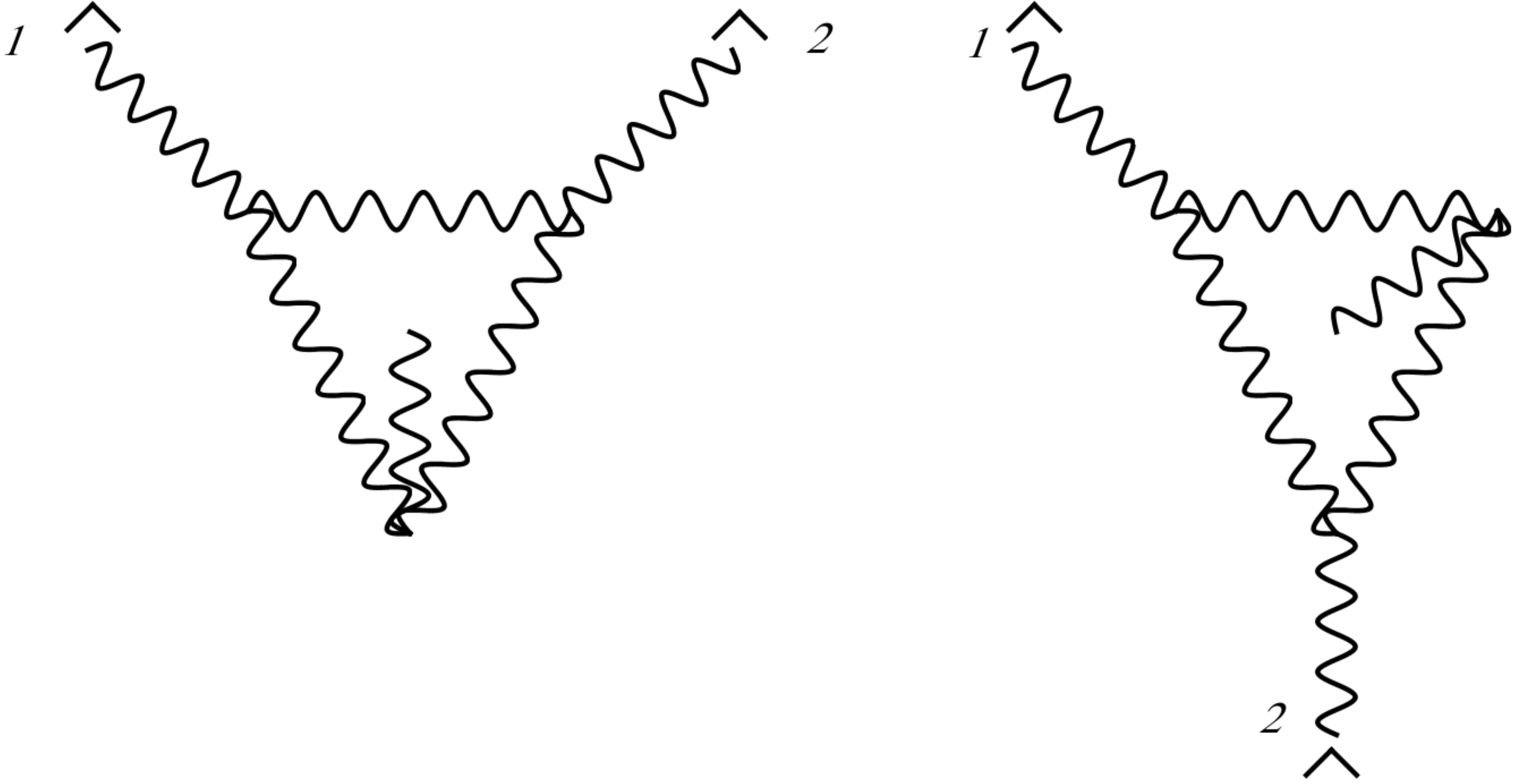}
\caption{\label{fig:tri_nonpl_adjacent} Adjacent shift on same color trace.}
\end{figure}

The case of color-adjacent shifts on the same trace is similar to the discussion of shifts on different color traces. The possible $\sim\!(z^{1-\epsilon})$ contribution stems again from the three legged triangle diagram. Shift particles $1$ and $2$ then particle $3$ will be on the second color trace. The two diagrams this results in are given in the figure \ref{fig:tri_nonpl_adjacent} but as before the sum of these two graphs vanishes due to antisymmetry giving a $\sim\!(z^{-1})$ suppression.

\subsection*{Conclusion}
The results in this appendix indicate that BCFW shifts of one-loop non-planar amplitudes are generically better behaved than their planar counterparts.


\end{appendices}

\bibliographystyle{JHEP}

\bibliography{nonadjbib}

\end{document}